\documentclass[universe,article,submit,moreauthors,pdftex,a4paper]{Definitions/mdpi}
 
\firstpage{1} 
\pubvolume{xx}
\issuenum{1}
\articlenumber{5}
\pubyear{2023}
\copyrightyear{2020}

\pdfoutput=1

\usepackage{xcolor}
\usepackage{float}
\usepackage{graphicx}
\usepackage{amsmath}
\usepackage{amssymb}
\usepackage{epstopdf}
\usepackage{comment}

\newcommand{\GeV}{\text{ GeV}}
\newcommand{\MeV}{\text{ MeV}}
\newcommand{\keV}{\text{ keV}}
\newcommand{\eV}{\text{ eV}}

\newcommand{\beqn}{\begin{equation}}
\newcommand{\eeqn}{\end{equation}}
\newcommand{\req}[1]{Eq.\,(\ref{#1})}
\newcommand*{\rf}[1]{Fig.~{\ref{#1}}}
\newcommand*{\rsec}[1]{Sect.\,{\ref{#1}}}

\newcommand*{\xblue}{\color{black}}
\preto{\abstractkeywords}{\nolinenumbers}

\Title{A SHORT SURVEY OF MATTER-ANTIMATTER EVOLUTION IN THE PRIMORDIAL UNIVERSE}

\Author{Johann Rafelski\orcidA{}, Jeremiah Birrell\orcidD, Andrew Steinmetz\orcidC{}, and Cheng Tao Yang\orcidB{}}
\AuthorNames{Johann Rafelski, Jeremiah Birrell, Andrew Steinmetz, and Cheng Tao Yang}

\address[1]{%
Department of Physics, The University of Arizona, Tucson, AZ 85721, USA} 

\corres{Correspondence: JohannR@arizona.edu}

\abstract{We offer a survey of the matter-antimatter evolution within the primordial Universe. While the origin of the tiny matter-antimatter asymmetry has remained one of the big questions in modern cosmology, antimatter itself has played a large role for much of the Universe's early history. In our study of the evolution of the Universe we adopt the position of the standard model Lambda-CDM Universe implementing the known baryonic asymmetry. We present the composition of the Universe across its temperature history while emphasizing the epochs where antimatter content is essential to our understanding. Special topics we address include the heavy quarks in quark-gluon plasma (QGP), the creation of matter from QGP, the free-streaming of the neutrinos, the vanishing of the muons, the magnetism in the electron-positron cosmos, and a better understanding of the environment of the Big Bang Nucleosynthesis (BBN) producing the light elements. We suggest but do not explore further that the methods used in exploring the early Universe may also provide new insights in the study of exotic stellar cores, magnetars, as well as gamma-ray burst (GRB) events. We describe future investigations required in pushing known physics to its extremes in the unique laboratory of the matter-antimatter early Universe.}

\begin{document}

Published: 27 June 2023 Universe 2023, 9(7), 309; https://doi.org/10.3390/universe9070309\\[0.2cm]
\centerline{\bf This article belongs to the Special Issue Remo Ruffini Festschrift, see Acknowledgements}
\centerline{https://www.mdpi.com/journal/universe/special\_issues/J0M337731D}
\hrule
\tableofcontents

\section{Timeline of Particles and Plasmas in the Universe}\label{sec:Intro}
\subsection{Guide to \texorpdfstring{$130\GeV>T>20\keV$}{130\GeV>T>20\keV}}\label{sec:Guide}
\noindent This survey of the early Universe begins with quark-gluon plasma (QGP) at a temperature of $T=130\GeV$. It then ends at a temperature of $T=20\keV$ with the electron-positron epoch which was the final phase of the Universe to contain significant quantities of antimatter. This defines the \lq\lq short\rq\rq\ $t\approx1/2$ hour time-span that will be covered. This work presumes that the Universe is homogeneous and that in our casual domain, the Universe's baryon content is matter dominated. Our work is rooted in the Universe as presented by Lizhi Fang and Remo Ruffini~\cite{fang1984cosmology,fang1985galaxies,fang1987quantum}. Within the realm of the Standard Model, we coherently connect the differing matter-antimatter plasmas as each transforms from one phase into another.

A more detailed description of particles and plasmas follows in \rsec{sec:Timeline}. We have adopted the standard $\Lambda$CDM model of a cosmological constant ($\Lambda$) and cold dark matter (CDM) where the Universe undergoes dynamical expansion as described in the Friedmann-Lema\^itre-Robertson-Walker (FLRW) metric. The contemporary history of the Universe in terms of energy density as a function of time and temperature is shown in \rf{CosmicDensity}. The Universe's past is obtained from integrating backwards the proposed modern composition of the Universe which contains $69\%$ dark energy, $26\%$ dark matter, $5\%$ baryons, and $<1\%$ photons and neutrinos in terms of energy density. The method used to obtain these results are found in \rsec{sec:Cosmo}.

After the general overview, we take the opportunity to enlarge in some detail our more recent work in special topics. In \rsec{sec:QGP}, we describe the chemical potentials of the QGP plasma species leading up to hadronization, Hubble expansion of the QGP plasma, and the abundances of heavy quarks. In \rsec{sec:Hadrons} we discuss the formation of matter during hadronization, the role of strangeness, and the unique circumstances which led to pions remaining abundant well after all other hadrons were diluted or decayed. We review the roles of muons and neutrinos in the leptonic epoch in \rsec{sec:Leptonic}. The $e^{\pm}$ plasma epoch is described in \rsec{sec:ElectronPositron} which is the final stage of the Universe where antimatter played an important role. Here we introduce the statistical physics description of electrons and positron gasses, their relation to the baryon density, and the magnetization of the $e^{\pm}$ plasma prior to the disappearance of the positrons shortly after Big Bang Nucleosynthesis (BBN). A more careful look at the effect of the dense $e^{\pm}$ plasma on BBN is underway. One interesting feature of having an abundant $e^{\pm}$ plasma is the possibility of magnetization in the early Universe {\xblue which we consider in \rsec{sec:Energy}. We introduce in this work the spin magnetic moment polarization for the first time in the context of cosmology.} We address this using spin-magnetization and mean-field theory where all the spins respond to the collective bulk magnetism self generated by the plasma. We stop our survey at a temperature of $T=20\keV$ with the disappearance of the positrons signifying the end of antimatter dynamics at cosmological scales.

\begin{figure}[ht]
 \centerline{\includegraphics[trim=20 90 30 90,clip,width=\textwidth]{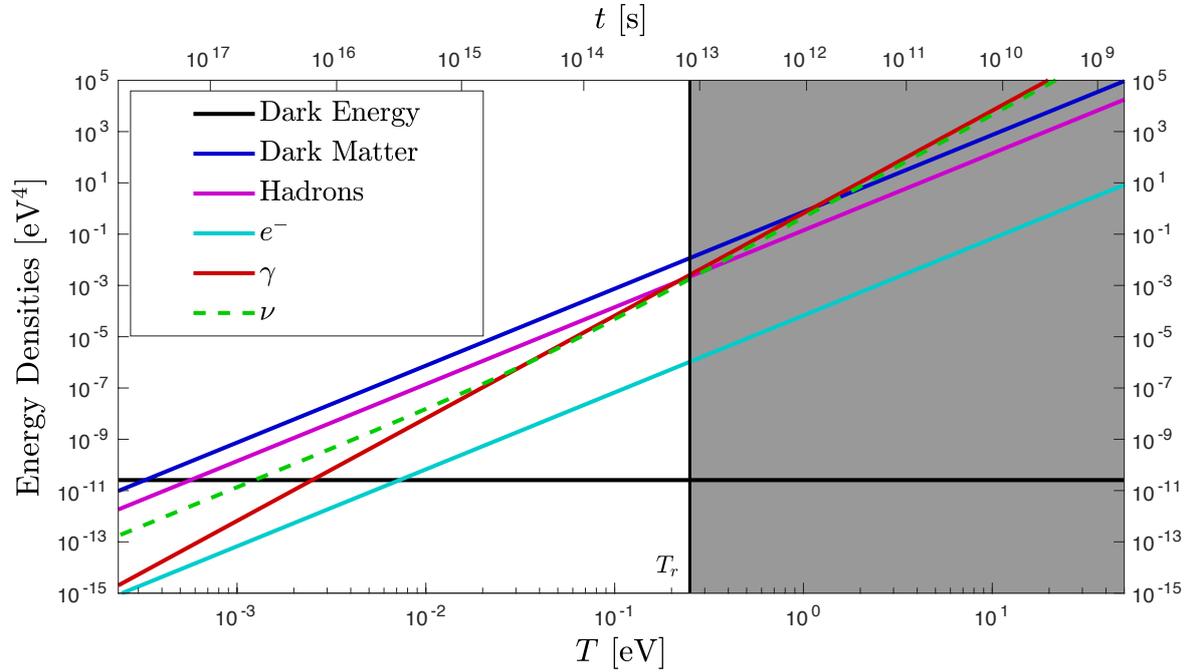}}
 \caption{Contemporary and recent Universe composition: In this example we assumed present day composition to be $69\%$ dark energy, $26\%$ dark matter, $5\%$ baryons, $<1\%$ photons and neutrinos. The dashed line shows how introduction of $2\times 0.1\eV$ mass in two of the three neutrinos impacts the energy density evolution (Neutrino mass choice is just for illustration. Other values are possible). The recombination temperature $T_{r}\approx0.25\eV$ delimits the era when the Universe was opaque shown as the shaded region. \label{CosmicDensity}}
\end{figure}

This primordial Universe is a plasma physics laboratory with unique properties not found in terrestrial laboratories or stellar environments due to the high amount of antimatter present. We suggest in \rsec{sec:Summary} areas requiring further exploration including astrophysical systems where positron content is considerable and the possibility for novel compact objects with persistent positron content is discussed. While the disappearance of baryonic matter is well described in the literature, it has not always been appreciated how long the leptonic ($\bar{\mu}=\mu^{+}$ and $\bar{e}=e^{+}$) antimatter remains a significant presence in the Universe's evolutionary history. We show that the $e^{\pm}$ epoch is a prime candidate to resolve several related cosmic mysteries such as early Universe matter in-homogeneity and the origin of cosmic magnetic fields. While the plasma epochs of the early Universe are in our long gone past, plasmas which share features with the primordial Universe might possibly exist in the contemporary Universe today. Such extraordinary stellar objects could poses properties dynamics relevant to gamma-ray burst (GRB)~\cite{Ruffini:2001fe,Aksenov:2008ze,Aksenov:2010vi,Ruffini:2012it}, black holes~\cite{Ruffini:2003yt,Ruffini:2009hg,Ruffini:2000yu} and neutron stars (magnetars)~\cite{Han:2011er,Belvedere:2012uc}.

\subsection{The five plasma epochs}\label{sec:Timeline}
\noindent At an early time in the standard cosmological model, the Universe began as a fireball, filling all space, with extremely high temperature and energy density~\cite{Rafelski:2015cxa}. {\xblue Our domain of the present day Universe originated from an ultra-relativistic plasma which contained almost a perfect symmetry between matter and antimatter except for a small discrepancy of one part in $10^{9}$ which remains a mystery today. There are two general solutions of this problem both of which suppose that the Universe's initial conditions were baryon-antibaryon number symmetric in order to avoid \lq fine-tuning\rq\ to a specific value:
\begin{itemize}
\item[A] Case of baryonic number (charge) conservation: In order to separate space domains in which either matter or antimatter is albeit very slightly dominant we  need a \lq force\rq\ capable of dynamically creating this matter-antimatter separation. This requires that two of the three Sakharov~\cite{Sakharov:1967dj,Sakharov:1988vdp} conditions be fulfilled:
\begin{itemize}
 \item[1.] Violation of CP-invariance allowing to distinguish matter from antimatter
 \item[2.] Non-stationary conditions in absence of local thermodynamic equilibrium
\end{itemize}
Other than very distant antimatter domains~\cite{Cohen:1997ac} the missing antimatter could be perhaps \lq stored\rq\ in a compact structure~\cite{Khlopov:2000as,Blinnikov:2014nea,Khlopov:2023dbg}. 
\item[B] There is no known cause for baryon charge conservation. Therefore it is possible to consider the full Sakharov model with 
\begin{itemize}
 \item[3.] Absence of baryonic charge conservation 
\end{itemize}
allowing the dynamical formation of the uniform matter-antimatter asymmetry typically occurring prior to the epoch governed by physics confirmed by current experiment to which environs we restrict this short survey. A well studied example is the Affleck-Dine
mechanism~\cite{Affleck:1984fy}.
\end{itemize} 
Very early formation of baryon asymmetry is further supported by the finding that the known CP-violation in the Standard Model's weak sector is insufficient to explain in quantitative terms the baryon asymmetry~\cite{Rubakov:1996vz}. However,  baryon asymmetry could  develop at a later stage in Universe evolution. We show in this review that this remains a topic deserving further investigation. In this work we take a homogeneous prescribed baryon asymmetry obtained from observed baryon to photon ratio in the Universe. Additional comments on the situation in the context of non-equilibria processes are made in \rsec{sec:BottomCharm}, at the end of \rsec{sec:Muons}, and in \rsec{sec:Summary}. 

The primordial hot Universe fireball underwent several practically adiabatic phase changes which dramatically evolved its bulk properties as it expanded and cooled.} We present an overview \rf{CosmicFraction} of particle families across all epochs in the Universe, as a function of temperature and thus time. The comic plasma, after the electroweak symmetry breaking epoch and presumably inflation, occurred in the early Universe in the following sequence:

\begin{figure}[ht]
 \centerline{\includegraphics[trim=70 120 60 100,clip,width=\textwidth,width=\linewidth]{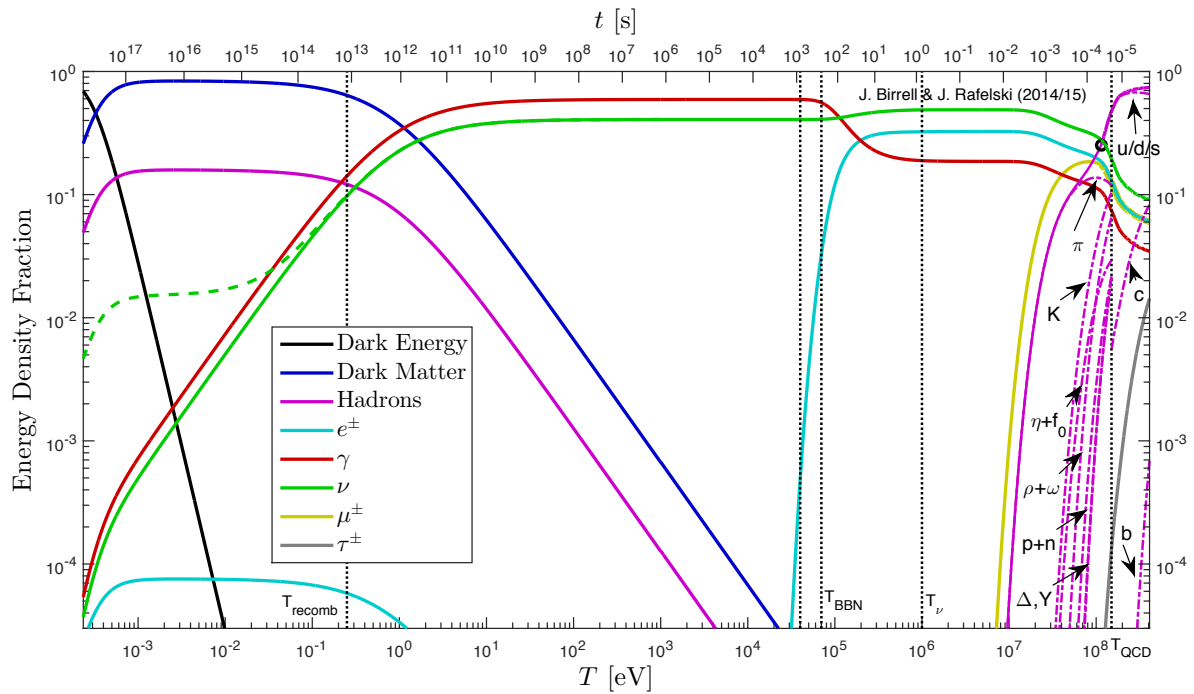}}
 \caption{Normalized  Universe constituent matter and radiation components $\Omega_i$ are evolved over cosmological timescales (top scale, bottom scale is temperature $T$) from contemporary observational cosmology to the QGP epoch of the Universe. Vertical lines denote transitions between distinct epochs. Solid neutrino (green) line shows contribution of massless neutrinos, while the dashed line shows $1$ massless and $2\times 0.1$ eV neutrinos (Neutrino mass choice is just for illustration. Other values are possible). \label{CosmicFraction}}
\end{figure}

\begin{enumerate}
    \item \textbf{Primordial quark-gluon plasma}: At early times when the temperature was between $130\GeV>T>150\MeV$ we have the building blocks of the Universe as we know them today, including the leptons, vector bosons, and all three families of deconfined quarks and gluons which propagated freely. As all hadrons are dissolved into their constituents during this time, strongly interacting particles $u,d,s,t,b,c,g$ controlled the fate of the Universe. Here we will only look at the late-stage evolution at around $150\MeV$.
    \item \textbf{Hadronic epoch}: Around the hadronization temperature $T_h\approx150\MeV$, a phase transformation occurred forcing the strongly interacting particles such as quarks and gluons to condense into confined states~\cite{Letessier:2005qe}. It is here where matter as we know it today forms and the Universe becomes hadronic-matter dominated. In the temperature range $ 150\MeV>T>20\MeV$ the Universe is rich in physics phenomena involving strange mesons and (anti)baryons including (anti)hyperon abundances~\cite{Fromerth:2012fe,Yang:2021bko}.
    \item \textbf{Lepton-photon epoch}: For temperature $10\MeV>T>2\MeV$, the Universe contained relativistic electrons, positrons, photons, and three species of (anti)neutrinos. Muons vanish partway through this temperature scale. In this range, neutrinos were still coupled to the charged leptons via the weak interaction~\cite{Birrell:2012gg,Birrell:2014ona}. During this time the expansion of the Universe is controlled by leptons and photons almost on equal footing.
    \item \textbf{Final antimatter epoch}: After neutrinos decoupled and become free-streaming, referred to as neutrino freeze-out, from the cosmic plasma at $T=2\MeV$, the cosmic plasma was dominated by electrons, positrons, and photons. We have shown in~\cite{Chris:2023abc} that this plasma existed until $T\approx0.02\MeV$ such that BBN occurred within a rich electron-positron plasma. This is the last time the Universe will contain a significant fraction of its content in antimatter.
    \item \textbf{Moving towards a matter dominated Universe}: The final major plasma stage in the Universe began after the annihilation of the majority of $e^{\pm}$ pairs leaving behind a residual amount of electrons determined by the baryon asymmetry in the Universe and charge conservation. The Universe was still opaque to photons at this point and remained so until the recombination period at $T\approx0.25\eV$ starting the era of observational cosmology with the CMB. This final epoch of the primordial Universe will not be described in detail here, but is well covered in~\cite{Planck:2018vyg}.
\end{enumerate}

\begin{figure}[ht]
 \centering
 \includegraphics[width=\textwidth]{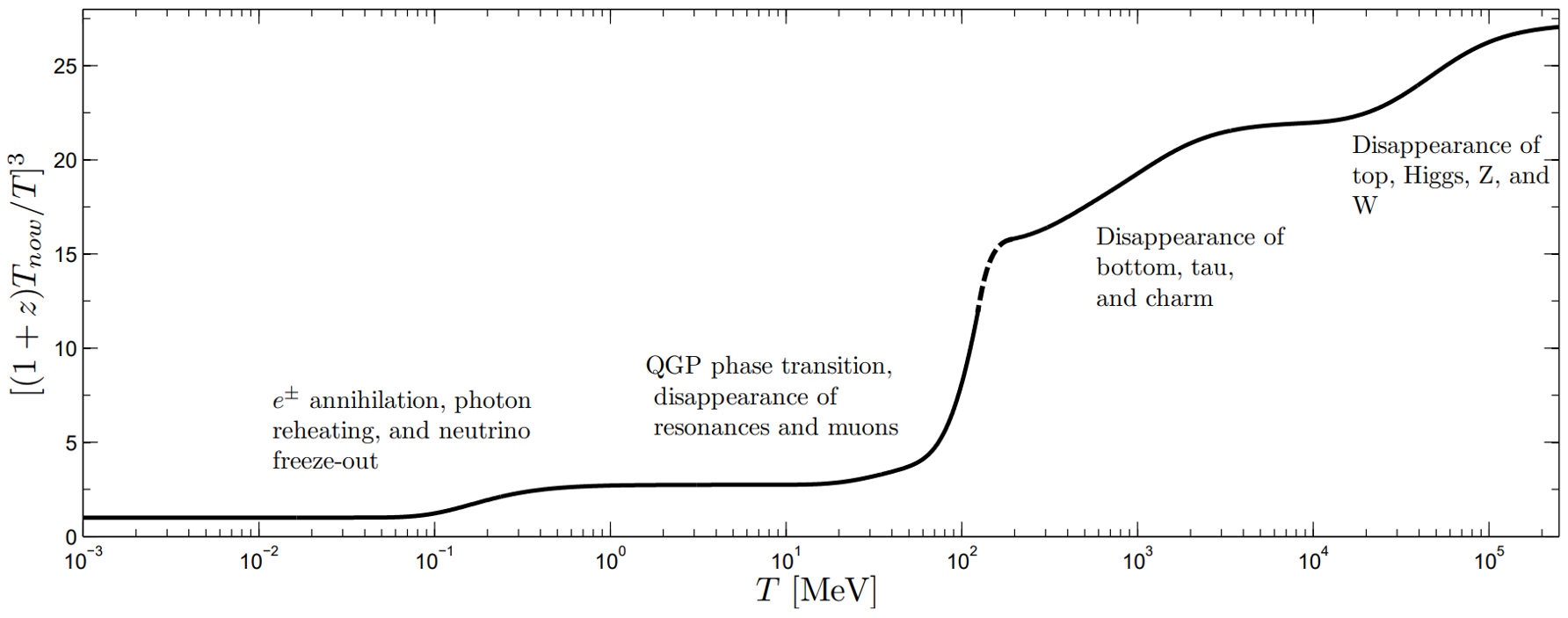}
 \caption{The evolution of the photon reheating (black line) process in terms of fractional temperature change in the Universe. Figure adapted from~\cite{Rafelski:2013yka}. The dashed portion is a qualitative description subject to the exact model of QGP hadronization.}
 \label{degrees_of_freedom} 
\end{figure}

Each plasma outlined above contributes to the thermal behavior of the Universe over time. This is illustrated in \rf{degrees_of_freedom} where the fractional drop in temperature during each plasma transformation is plotted. Each subsequent plasma lowers the available degrees of freedom (as the particle inventory is whittled away) as the Universe cools~\cite{Wantz:2009it,Rafelski:2013yka}. Each drop in degrees of freedom represents entropy being pumped into the photons as entropy is conserved (up until local gravitational processes become relevant) in an expanding Universe. As there are no longer degrees of freedom to consume, thereby reheating the photon field further, the fractional temperature remains constant today.

In \rf{CosmicFraction} we begin on the right at the end of the QGP era. The first dotted vertical line shows the QGP phase transition and hadronization, near $T=150\MeV$. The hadron era proceeds with the disappearance of muons, pions, and heavier hadrons. This constitutes a reheating period, with energy and entropy from these particles being transferred to the remaining $e^\pm$, photon, neutrino plasma. The black circle near $T=115\MeV$ denotes our change from $2+1$-flavor lattice QCD~\cite{Kronfeld:2012ym, DElia:2012ifm, Bonati:2013hsa} data for the hadron energy density, taken from Borsanyi et al.~\cite{Borsanyi:2012rr,Borsanyi:2013bia}, to an ideal gas model~\cite{Bernstein:1988bw} at lower temperature. We note that the hadron ideal gas energy density matches the lattice results to less than a percent at $T=115\MeV$~\cite{Philipsen:2012nu}. 

To the right of the QGP transition region, the solid hadron line shows the total energy density of quarks and gluons. From top to bottom, the dot-dashed hadron lines to the right of the transition show the energy density fractions of $2+1$-flavor (u,d,s) lattice QCD matter (almost indistinguishable from the total energy density), charm, and bottom (both in the ideal gas approximation). To the left of the transition the dot-dashed lines show the pion, kaon, $\eta+f_0$, $\rho+\omega$, nucleon, $\Delta$, and Y contributions to the energy fraction.

Continuing to the second vertical line at $T=\mathcal{O}(1\, \MeV)$, we come to the annihilation of $e^\pm$ and the photon reheating period. Notice that only the photon energy density fraction increases, as we assume that neutrinos are already decoupled at this time and hence do not share in the reheating process, leading to a difference in photon and neutrino temperatures. This is not strictly correct but it is a reasonable simplifying assumption for the current purpose; see~\cite{Mangano:2005cc,Fornengo:1997wa,Mangano:2001iu,Birrell:2012gg}. We next pass through a long period, from $T=\mathcal{O}(1\, \MeV)$ until $T=\mathcal{O}(1\, \eV)$, where the energy density is dominated by photons and free-streaming neutrinos. BBN occurs in the approximate range $T=40-70\keV$ and is indicated by the next two vertical lines in \rf{CosmicFraction}. It is interesting to note that, while the hadron fraction is insignificant at this time, there is still a substantial background of $e^\pm$ pairs during BBN (see \rsec{sec:ElectronPositronDensity}). 

We then come to the beginning of the matter dominated regime, where the energy density is dominated by the combination of dark matter and baryonic matter. This transition is the result of the redshifting of the photon and neutrino energy, $\rho\propto a^{-4} \propto T^4$, whereas for non-relativistic matter $\rho\propto a^{-3}\propto T^3$. Recombination and photon decoupling occurs near the transition to the matter dominated regime, denoted by the (\rf{CosmicFraction}) vertical line at $T=0.25\eV$.

Finally, as we move towards the present day CMB temperature of $T_{\gamma,0}=0.235$ meV on the left hand side, we have entered the dark energy dominated regime. For the present day values, we have used the energy densities proscribed by the Planck parameters~\cite{Planck:2013pxb} using \req{Planck_params} and zero Universe spatial curvature. The photon energy density is fixed by the CMB temperature $T_{\gamma,0}$ and the neutrino energy density is fixed by $T_{\gamma,0}$ along with the photon to neutrino temperature ratio and neutrino masses. Both constitute $<1\%$ of the current energy budget.

The Universe evolution and total energy densities were computed using massless neutrinos, but for comparison we show the energy density of massive neutrinos in the dashed green line. For the dashed line we used two neutrino flavors with masses $m_\nu=0.1\eV$ and one massless flavor. Note that the inclusion of neutrino mass causes the leveling out of the neutrino energy density fraction during the matter dominated period, as compared to the continued redshifting of the photon energy.

\subsection{The Lambda-CDM Universe}\label{sec:Cosmo}
\noindent Here we provide background on the standard $\Lambda$CDM cosmological (FLRW-Universe) model that is used in the computation of the composition of the Universe over time. We use the spacetime metric with metric signature $(+1,-1,-1,-1)$ in spherical coordinates
\beqn\label{metric}
ds^2=c^2dt^2-a^2(t)\left[ \frac{dr^2}{1-kr^2}+r^2(d\theta^2+\sin^2(\theta)d\phi^2)\right]
\eeqn
characterized by the scale parameter $a(t)$ of a spatially homogeneous Universe. The geometric parameter $k$ identifies the Gaussian geometry of the spacial hyper-surfaces defined by co-moving observers. Space is a Euclidean flat-sheet for the observationally preferred value $k=0$~\cite{Planck:2013pxb,Planck:2015fie,Planck:2018vyg}. In this case it can be more convenient to write the metric in rectangular coordinates
\beqn\label{metric2}
ds^2=c^2dt^2-a^2(t)\left[ dx^2+dy^2+dz^2\right].
\eeqn
We will work in units where $\hbar=1,\ c=1$.

The global Universe dynamics can be characterized by two quantities: the Hubble parameter $H$, a strongly time dependent quantity on cosmological time scales, and the deceleration parameter $q$:
\begin{align}
 \label{Hubble} H(t)^{2}&\equiv\left(\frac{\dot a}{a}\right)^2=\frac{8\pi G_{N}}{3}\rho_{tot}\,,\\
 \label{Deceleration} \frac{\ddot a}{a}=-qH^2,\qquad\qquad q&\equiv -\frac{a\ddot a}{\dot a^2},\qquad\qquad \dot H=-H^2(1+q)\,, 
\end{align}
where $G_{N}$ is the Newtonian gravitational constant and $\rho_{tot}$ is the energy density of the Universe and composed of the various energy densities in the Universe. The deceleration parameter $q$ is defined in terms of the second derivative of the scale parameter. 

In \rf{deceleration_evolution} (left) we illustrate the late stage evolution of the parameters $H$ and $q$ given in \req{Hubble} and \req{Deceleration} compared to temperature. This illustrates how the Universe evolves according to the Friedmann equations \req{Hubble} and \req{Deceleration} above. The deceleration begins radiation dominated with $q=1$ and then transitions to matter dominated $q=1/2$.  {\xblue Within the $\Lambda$CDM model} the contemporary Universe is undergoing a transition from matter dominated to dark energy dominated, where the deceleration would  settle on the asymptotic value of $q=-1$~\cite{Rafelski:2013yka}. {\xblue However, several alternate models:  phantom energy~\cite{Caldwell:2003vq}, Chaplygin gas~\cite{Bilic:2001cg}, or more generally dynamic (spatially and/or time dependent) dark energy~\cite{Benevento:2020fev} cannot be excluded in absence of strong evidence for the constancy of dark energy.}

{\xblue Within the $\Lambda$CDM model only usual forms of energy are relevant before recombination epoch, see \rf{CosmicFraction}. Any alternate model can be thus constrained by understanding precisely the evolution of the Universe prior to this epoch. Part of the program of this survey is to connect the late stage evolution to the very early Universe during and prior to BBN  accounting for the unexpectedly considerable antimatter content. 

The current tension in Hubble parameter measurements~\cite{Perivolaropoulos:2021jda,DiValentino:2021izs,Aluri:2022hzs} might benefit from closer inspection of these earlier denser periods should these contribute to modification of the conventional model of Universe expansion. We further note that} the JWST has recently discovered that galaxy formation began earlier than predicted which requires reevaluation of early Universe matter inhomogeneities~\cite{Yan:2022sxd}. \rf{deceleration_evolution} (right) shows the close relationship between the redshift $z$ and the Hubble parameter. Deviations separating the two occur from the transitions which changed the deceleration value.

\begin{figure}[ht]
 \centering
 \includegraphics[width=\textwidth]{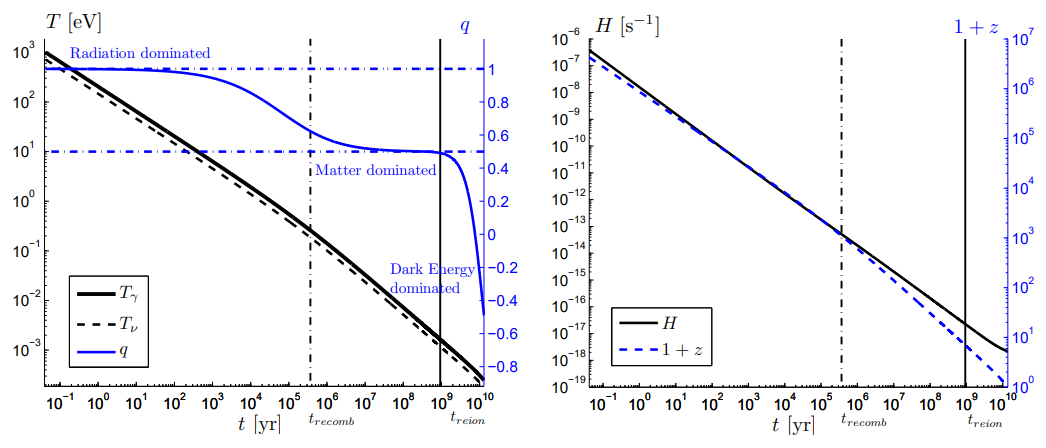}
 \caption{ (left) The numerically solved later $t>10^{-1}\ \mathrm{yr}$ evolution of photon and neutrino background temperatures $T_{\gamma},\ T_{\nu}$ (black and black dashed lines) and the deceleration parameter $q$ (thin blue line) over the lifespan of the Universe. (right) The evolution of the Hubble parameter $1/H$ (black line) and redshift $z$ (blue dashed line) which is related to the scale parameter $a(t)$. Figure adapted from~\cite{Rafelski:2013yka}.}
 \label{deceleration_evolution} 
\end{figure}

The Einstein equations with a cosmological constant $\Lambda$ corresponding to dark energy are:
\beqn\label{Einstein}
G^{\mu\nu}=R^{\mu\nu}-\left(\frac R 2 +\Lambda\right) g^{\mu\nu}=8\pi G_N T^{\mu\nu}, 
\quad R= g_{\mu\nu}R^{\mu\nu}.
\eeqn
The homogeneous and isotropic symmetry considerations imply that the stress energy tensor is determined by an energy density and an isotropic pressure
\begin{align}
 T^\mu_\nu =\mathrm{diag}(\rho, -P, -P, -P).
\end{align}
It is common to absorb the Einstein cosmological constant $\Lambda$ into the energy and pressure
\beqn\label{EpsLam}
\rho_\Lambda=\frac{\Lambda}{8\pi G_N}, \qquad P_\Lambda=-\frac{\Lambda}{8\pi G_N}
\eeqn
and we implicitly consider this done from now on.

Two dynamically independent Friedmann equations~\cite{weinberg1972gravitation} arise using the metric \req{metric} in \req{Einstein}:
\beqn\label{hubble}
\frac{8\pi G_N}{3} \rho = \frac{\dot a^2+k}{a^2}
=H^2\left( 1+\frac { k }{\dot a^2}\right),
\qquad
\frac{4\pi G_N}{3} (\rho+3P) =-\frac{\ddot a}{a}=qH^2.
\eeqn
We can eliminate the strength of the interaction, $G_N$, solving both these equations for ${8\pi G_N}/{3}$, and equating the result to find a relatively simple constraint for the deceleration parameter:
\beqn\label{qparam}
q=\frac 1 2 \left(1+3\frac{P}{\rho}\right)\left(1+\frac{k}{\dot a^2}\right).
\eeqn
For a spatially flat Universe, $k=0$, note that in a matter-dominated era where $P/\rho<<1$ we have $q\simeq 1/2$; for a radiative Universe where $3P=\rho$ we find $q= 1 $; and in a dark energy Universe in which $P=-\rho$ we find $q=-1$. Spatial flatness is equivalent to the assertion that the energy density of the Universe equals the critical density
\begin{equation}\label{crit_density}
\rho=\rho_{\text{crit}}\equiv \frac{3H^2}{8\pi G_N}.
\end{equation}

The CMB power spectrum is sensitive to the deceleration parameter and the presence of spatial curvature modifies $q$. The Planck results~\cite{Planck:2013pxb,Planck:2015fie,Planck:2018vyg} constrain the effective curvature energy density fraction,
\begin{equation}
\Omega_K\equiv1-\rho/\rho_{\text{crit}},
\end{equation}
to
\begin{equation}
|\Omega_K|<0.005.
\end{equation}
This indicates a nearly flat Universe which is spatially Euclidean. We will work within an exactly spatially flat cosmological model, $k=0$. 
As must be the case for any solution of Einstein's equations, \req{hubble} implies that the energy momentum tensor of matter is divergence free:
\beqn\label{divTmn}
T^{\mu\nu};_\nu =0 \Rightarrow -\frac{\dot\rho}{\rho+P}=3\frac{\dot a}{a}=3H.
\eeqn
A dynamical evolution equation for $\rho(t)$ arises once we combine \req{divTmn} with \req{hubble}, eliminating $H$. Given an equation of state $P(\rho)$, solutions of this equation describes the dynamical evolution of matter in the Universe. In practice, we evolve the system in both directions in time. On one side, we start in the present era with the energy density fractions fit by the central values found in Planck data~\cite{Planck:2013pxb}
\begin{equation}\label{Planck_params}
H_0=67.4\,\text{km/s/Mpc},\hspace{2mm} \Omega_b=0.05,\hspace{2mm} \Omega_c=0.26, \hspace{2mm}\Omega_\Lambda=0.69,
\end{equation}
and integrate backward in time. On the other hand, we start in the QGP era with an equation of state determined by an ideal gas of SM particles, combined with a perturbative QCD equation of state for quarks and gluons~\cite{Borsanyi:2013bia}, and integrate forward in time. As the Universe continues to dilute from dark energy in the future, the cosmic equation of state will become well approximated by the de Sitter inflationary metric which is a special case of FLRW.

\section{QGP Epoch}\label{sec:QGP}
\subsection{Conservation laws in QGP}\label{sec:Conservation}
\noindent During the first $\Delta t\approx30\ \mu$sec after the Big Bang, the early Universe is a hot soup that containing the elementary primordial building blocks of matter and antimatter~\cite{Rafelski:2015cxa}. In particular it contained the light quarks which are now hidden in protons and neutrons. Beyond this there were also electrons, photons, neutrinos, and massive strange and charm quarks. These interacting particle species were kept in chemical and thermal equilibrium with one another. Gluons which mediated the color interaction are very abundant as well. This primordial phase lasted as long as the temperature of the Universe was more than 110,000 times than the expected temperature $T_{\odot}=1.36\keV\ (1.58\times10^{7}\ \mathrm{K})$ at the center of the Sun~\cite{Castellani:1996cm}.

\begin{figure}[ht]
 \centering
 \includegraphics[width=\textwidth]{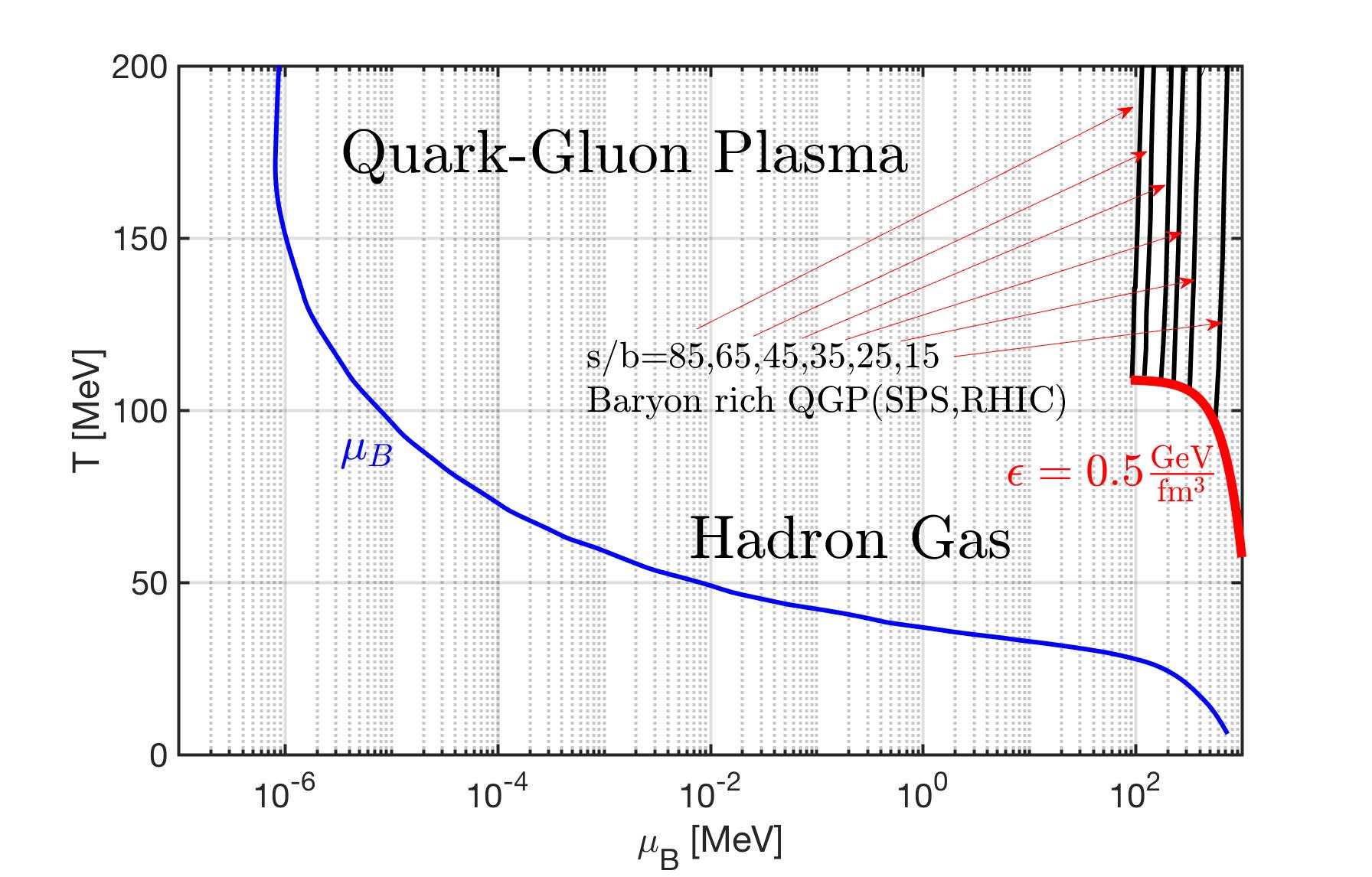}
 \caption{ The evolution of the cosmic baryon chemical potential $\mu_{B}$ after hadronization (blue line). Curves for QGP (thin black line) created in terrestrial accelerators for differing entropy-per-baryon $s/B$ values are included~\cite{Rafelski:1987nv}. The boundary (red line) where QGP condenses into hadrons is illustrated at an energy density of $0.5\GeV/\mathrm{fm}^{3}$ as determined through lattice computation~\cite{HotQCD:2014kol}.}
 \label{phaseQGP} 
\end{figure}

The conditions in the early Universe and those created in relativistic collisions of heavy atomic nuclei differ somewhat: whereas the primordial quark-gluon plasma survives for about 25 $\mu$sec in the Big Bang, the comparable extreme conditions created in ultra-relativistic nuclear collisions are extremely short-lived~\cite{Rafelski:2001hp} on order of $10^{-23}$ seconds. As a consequence of the short lifespan of laboratory QGP in heavy-ion collisions~\cite{Ollitrault:1992bk,Petran:2013lja}, they are not subject to the same weak interaction dynamics~\cite{Ryu:2015vwa} as the characteristic times for weak processes are too lengthy~\cite{Rafelski:1982ii}. Therefore our ability to recreate the conditions of the primordial QGP are limited due to the relativistic explosive disintegration of the extremely hot dense relativistic `fireballs' created in modern accelerators. This disparity is seen in \rf{phaseQGP} where the chemical potential of QGP $\mu_{q}=\mu_{B}/3$~\cite{Rafelski:1987nv} for various values of entropy-per-baryon $s/b$ relevant to relativistic particle accelerators are plotted alongside the evolution of the cosmic hadronic plasma chemical potential. The confinement transition boundary (red line in \rf{phaseQGP}) was calculated using a parameters obtained from~\cite{Letessier:2002ony} in agreement with lattice results~\cite{HotQCD:2014kol}. The QGP precipitates hadrons in the cosmic fluid at a far higher entropy ratio than those accessible by terrestrial means and the two manifestations of QGP live far away from each other on the QCD phase diagram~\cite{Jacak:2012dx}.

The work of Fromerth et. al.~\cite{Fromerth:2012fe} allows us to parameterize the chemical potentials $\mu_d$, $\mu_e$, and $\mu_\nu$ during this epoch as they are the lightest particles in each main thermal category: quarks, charged leptons, and neutral leptons. The quark chemical potential is determined by the following three constraints~\cite{Fromerth:2012fe}:

\begin{figure}[ht]
 \centering
 \includegraphics[width=\textwidth]{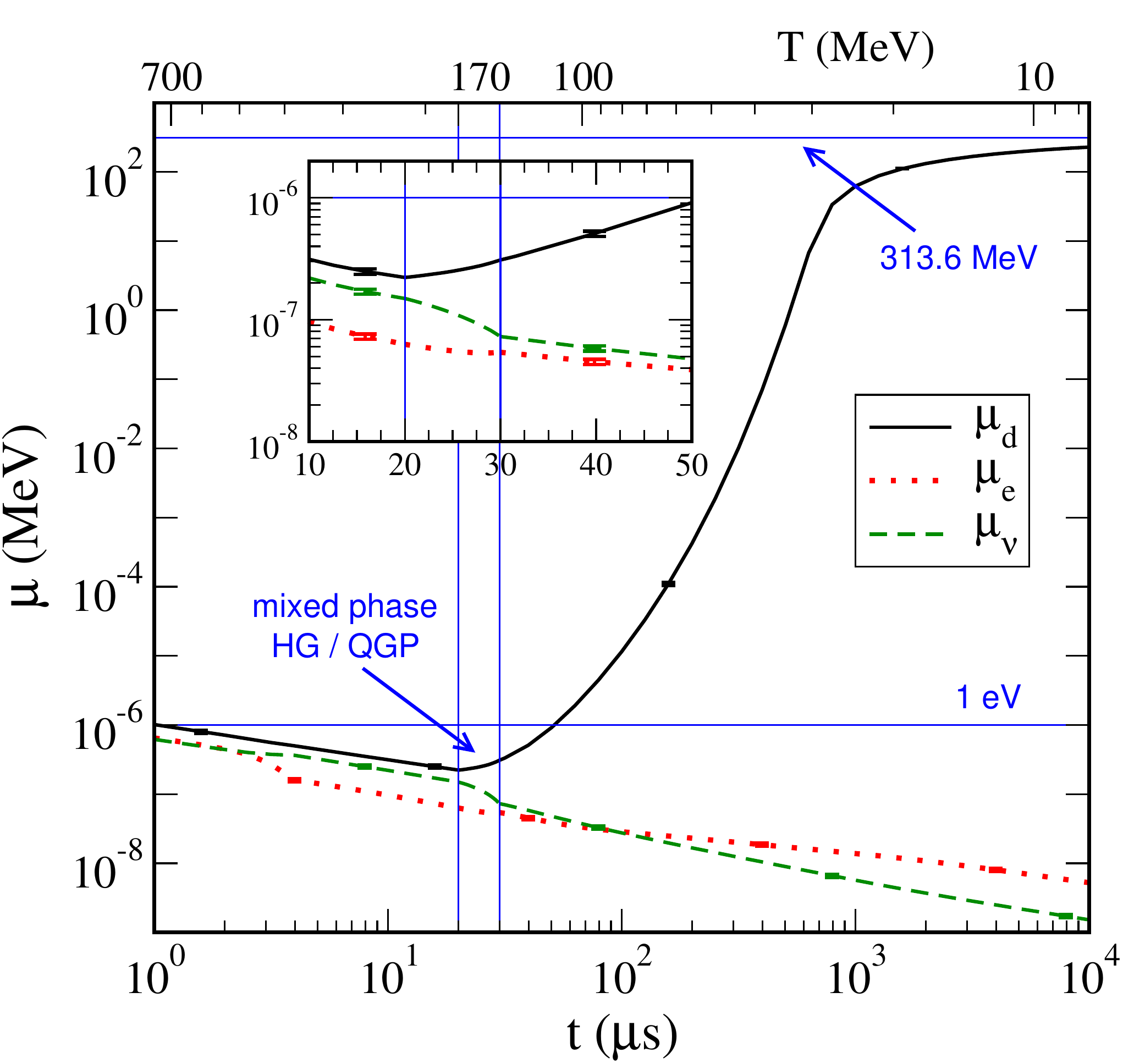}
 \caption{Plot of the down quark chemical potential (black), electron chemical potential (dotted red) and neutrino chemical potential (dashed green) as a function of time. These are 2003 unpublished results of Fromerth \& Rafelski~\cite{Fromerth:2002wb}; also presented in Ref.\,\cite{Rafelski:2019twp}}
 \label{QGPchem1} 
\end{figure}

\begin{enumerate}
\item Electric charge neutrality $Q=0$, given by
\begin{align}\label{QGP_Q}
 \frac{Q}{V}=n_{Q}\equiv\sum_f\,Q_f\,n_f(\mu_f,T)=0
\end{align}
where $Q_f$ is the charge and $n_{f}$ is the numerical density of each species $f$. $Q$ is a conserved quantity in the Standard Model under global $U(1)_{EM}$ symmetry. This is summed is over all particles present in the QGP epoch.
\item Baryon number and lepton number neutrality $B-L=0$, given by
\begin{align}\label{QGP_LB}
\frac{B-L}{V}=n_{B}-n_{L}\equiv\sum_f(B_f-L_f)n_f(\mu_f,T)=0
\end{align}
where $L_f$ and $B_f$ are the lepton and baryon number for the given species $f$. This condition is phenomenologically motivated by baryogenesis and is exactly conserved in the Standard Model under global $U(1)_{B-L}$ symmetry. We note many Beyond-Standard-Model (BSM) models also retain this as an exact symmetry though Majorana neutrinos do not.
\item The entropy-per-baryon density ratio $s/n_B$ is a constant and can be written as
\begin{align}\label{QGP_sB}
\frac{S}{B}=\frac{s}{n_B}=\frac{\sum_fs_f(\mu_f,T)}{\sum_fB_fn_f(\mu_f,T)}=\mathrm{const}
\end{align}
where $s_f$ is the entropy density of given species $f$. As the expanding Universe remains in thermal equilibrium, the entropy is conserved within a co-moving volume. The baryon number within a co-moving volume is also conserved. As both quantities dilute with $1/a(t)^{3}$ within a normal volume, the ratio of the two is constant. This constraint does not become broken until spatial inhomogeneities from gravitational attraction becomes significant, leading to increases in local entropy.
\end{enumerate}
At each temperature $T$, the above three conditions form a system of three coupled, nonlinear equations of the three chosen unknowns (here we have $\mu_d$, $\mu_e$, and $\mu_\nu$). In \rf{QGPchem1} we present numerical solutions to the conditions \req{QGP_Q}-\req{QGP_sB} and plot the chemical potentials as a function of time. As seen in the figure, the three potentials are in alignment during the QGP phase until the hadronization epoch where the down quark chemical potential diverges from the leptonic chemical potentials before reaching an asymptotic value at late times. This asymptotic value is given as approximately $\mu_{q}\approx m_{N}/3$ the mass of the nucleons and represents the confinement of the quarks into the protons and neutrons at the end of hadronization.

\begin{figure}[ht]
 \centering
 \includegraphics[width=\textwidth]{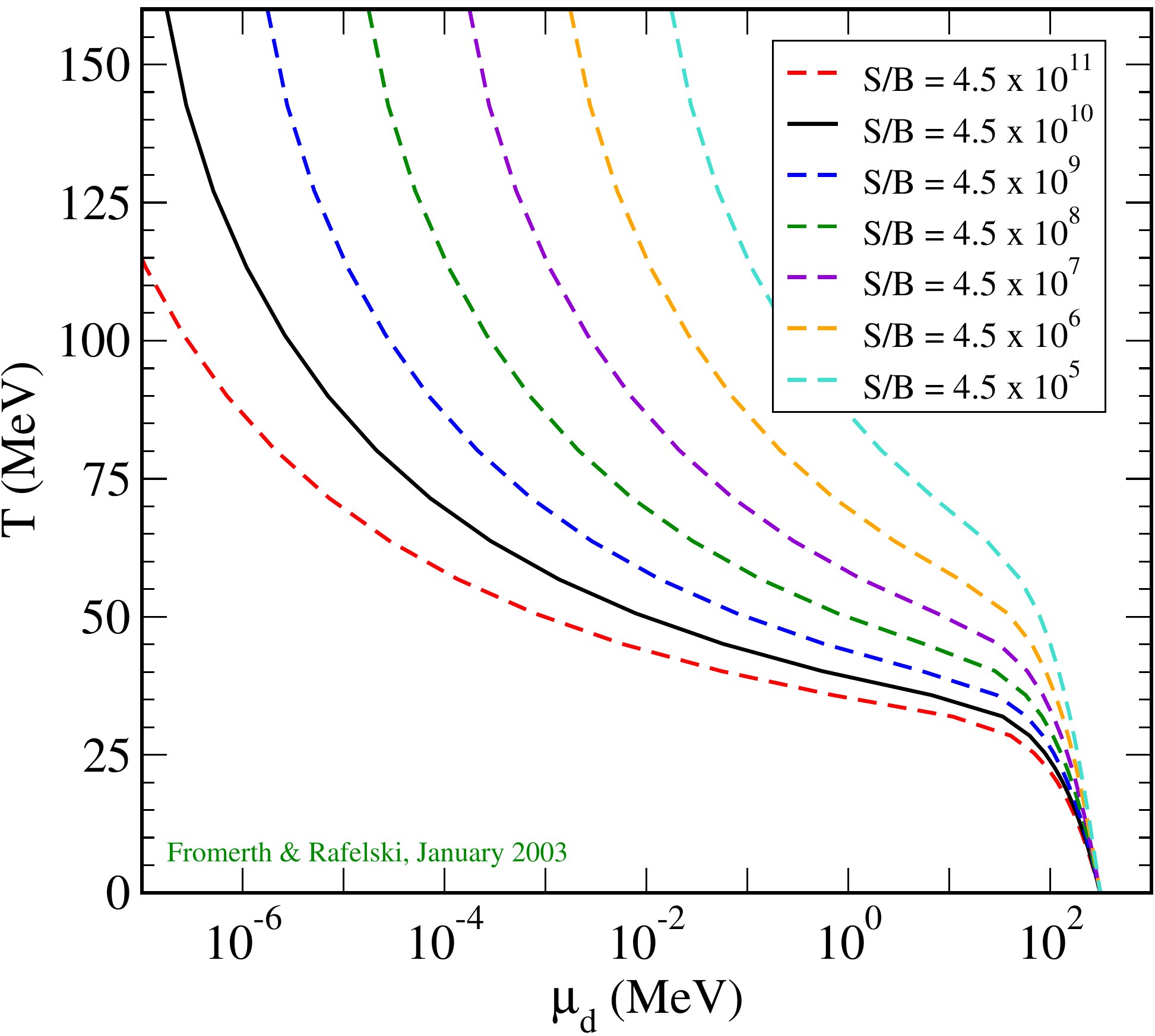}
 \caption{Plot of the down quark chemical potential $\mu_{d}$ as a function of temperature for differing values of entropy-per-baryon $S/B$ ratios. (2003 unpublished, Fromerth \& Rafelski~\cite{Rafelski:2019twp})}
 \label{QGPchem2} 
\end{figure}

This asymptotic limit is also shown in \rf{QGPchem2} where we present the down quark chemical potential for different values of the entropy-to-baryon ratio. While the $s/n_{B}$ ratio has large consequences for the plasma at high temperatures, the chemical potential is insensitive to this parameter at low temperatures the degrees of freedom are dominated by the remaining baryon number rather than the thermal degrees of freedom of the individual quarks. Therefore the entropy to baryon value today greatly controls the quark content when the Universe was very hot. We note that the distribution of quarks in the QGP plasma does not remain fixed to the Fermi-Dirac distribution for thermal and entropic equilibrium. The quark partition function is instead
\begin{align} \label{QuarkDistribution}\ln\mathcal{Z}_{\mathrm{quarks}}=\sum_{q}\ln\left(1+\Upsilon_{q}(t)e^{-\beta E_{q}}\right)\,,\qquad\Upsilon_{q}(t)=\gamma_{q}(t)\lambda_{q}\,\qquad q=u,d,c,s,t,b,
\end{align}
which is summed over all quarks and their quantum numbers. In \req{QuarkDistribution}, $\lambda_{q}$ is the quark fugacity while $\gamma_{q}(t)$ is the temporal inhomogeneity of the population distribution~\cite{Rafelski:2019twp}. The product of the two $\Upsilon_{q}(t)=\gamma_{q}(t)\lambda_{q}$ is then defined as the generalized fugacity for the species. Because of nuclear reactions, these distributions populate and depopulate over time which pulls the gas off entropic equilibrium while retaining temperature $T$ with the rest of the Universe~\cite{Letessier:2002ony}. When $\gamma\neq1$, the entropy of the quarks is no longer minimized. As entropy in the cosmic expansion is conserved overall, this means the entropy gain or loss is then related to the entropy moving between the quarks or its products.

In practice, the generalized fugacity is $\Upsilon=1$  during the QGP epoch as the quarks in early Universe remained in both thermal and entropic equilibrium. This is because the Universe's expansion was many orders of magnitude slower than the process reaction and decay timescales~\cite{Letessier:2002ony}. However near the hadronization temperature, heavy quarks abundance and deviations from chemical equilibrium have not yet been studied in great detail. We show in \rsec{sec:BottomCharm} and~\cite{Yang:2020nne} that the bottom quarks can deviate from chemical equilibrium $\gamma\neq1$ by breaking the detailed balance between reactions of the quarks.

\subsection{Heavy flavor: Bottom and charm in QGP}\label{sec:BottomCharm}
\noindent In the QGP epoch, up and down $(u,d)$ (anti)quarks are effectively massless and remain in equilibrium via quark-gluon fusion. Strange $(s)$ (anti)quarks are in equilibrium via weak, electromagnetic, and strong interactions until $T\sim12\MeV$~\cite{Yang:2021bko}. In this section, we focus on the heavier charm and bottom $(c,b)$ (anti)quarks. In primordial QGP, the bottom and charm quarks can be produced from strong interactions via quark-gluon pair fusion processes and disappear via weak interaction decays. For production, we have the following processes
\begin{align}
 q+q&\longrightarrow b+\bar b,\qquad q+q\longrightarrow c+\bar c,\\
 g+g&\longrightarrow b+\bar b,\qquad g+g\longrightarrow c+\bar c,
\end{align}
for bottom and charm and 
\begin{align}
 &b\longrightarrow c+l+\nu_l, \qquad b\longrightarrow c+q+\bar{q}\\
&c\longrightarrow s+l+\nu_l,\qquad c\longrightarrow s+q+\bar{q}
\end{align}
for their decay. A detailed calculation of production and decay rate can be found in~\cite{Yang:2020nne}.

\begin{figure} 
    \centering
    \includegraphics[width=0.9\textwidth]{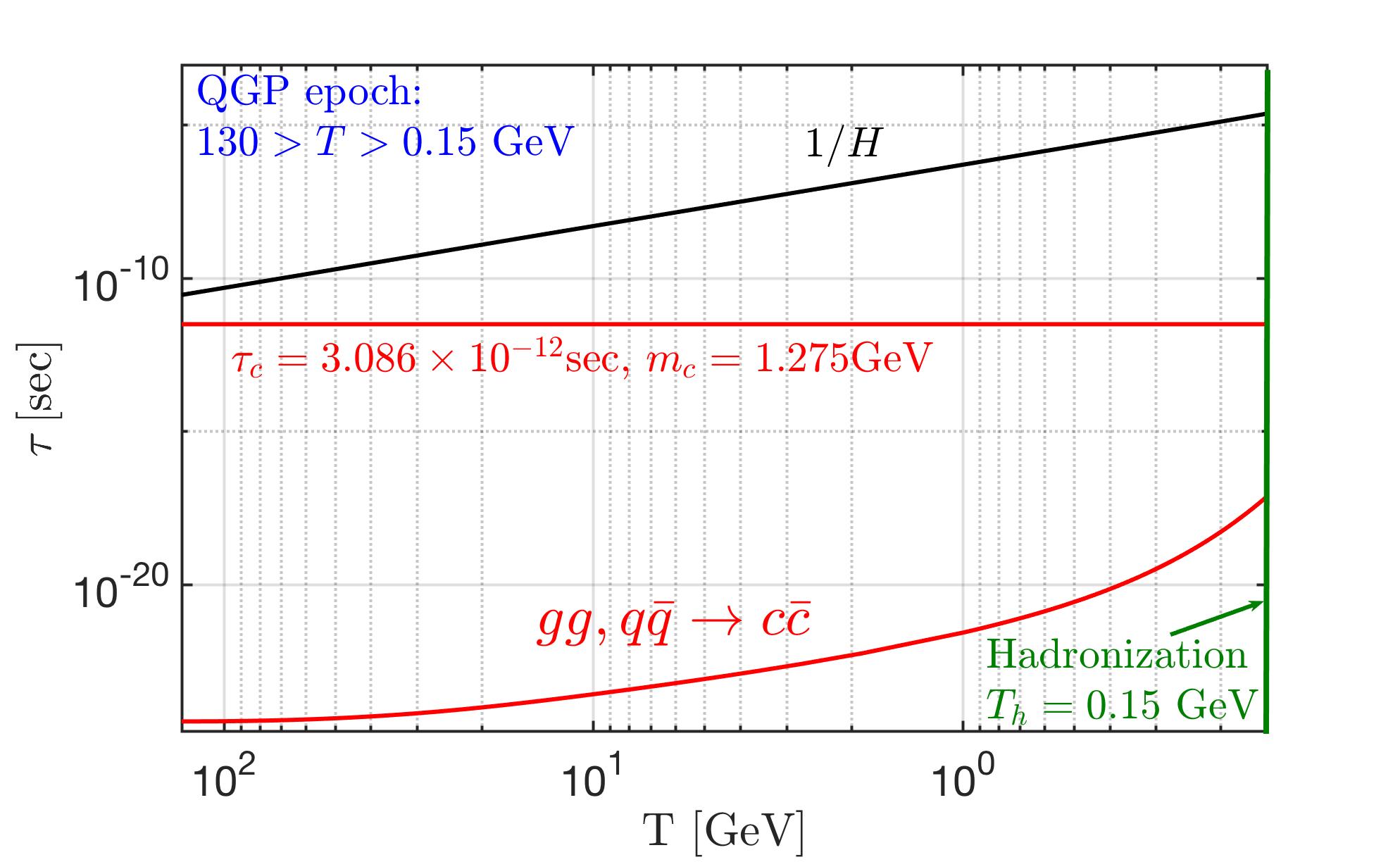}
    \includegraphics[width=0.9\textwidth]{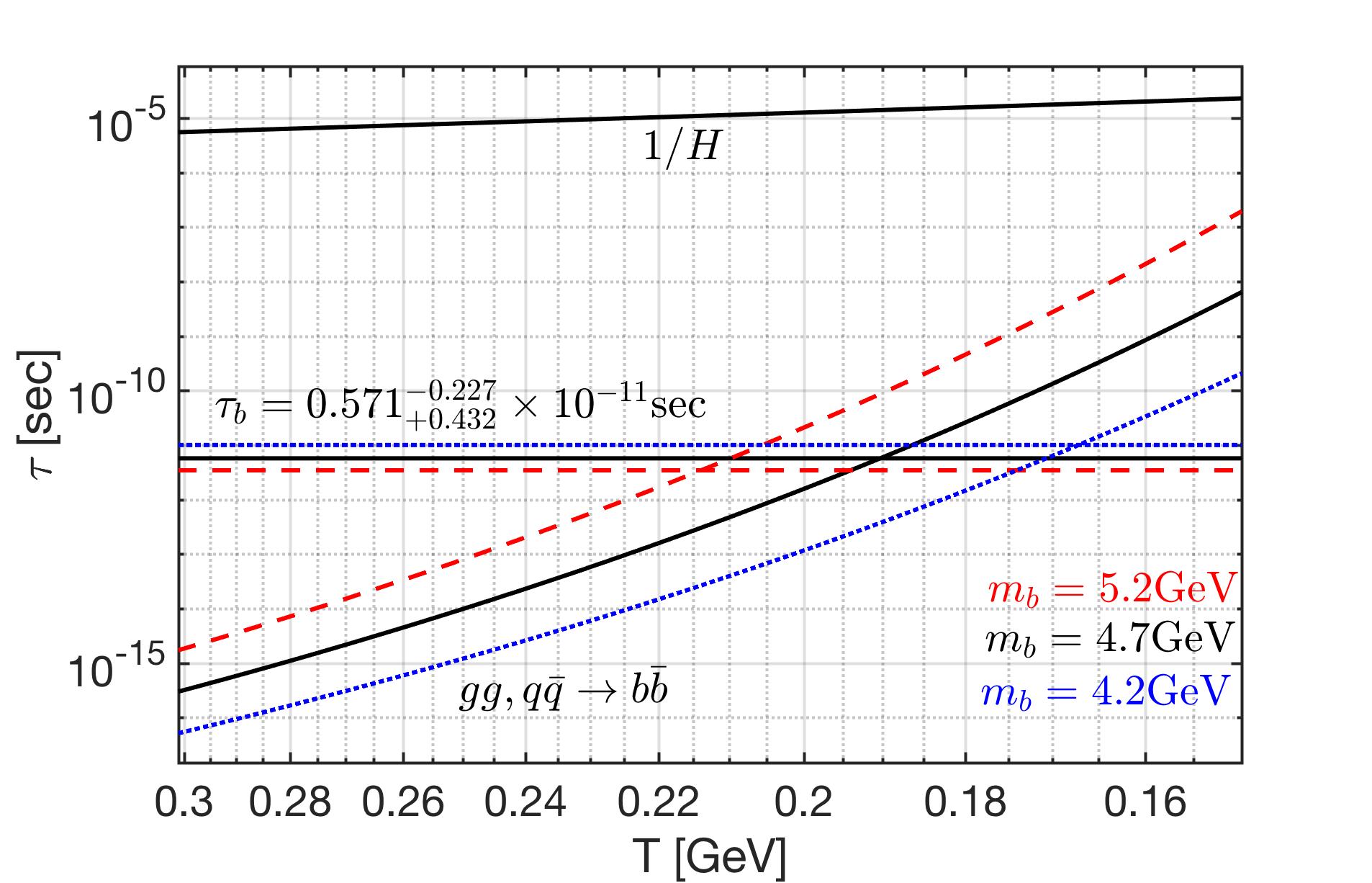}
    \caption{Comparison of Hubble time $1/H$, quark lifespan $\tau_{q}$, and characteristic time for production via quark-gluon pair fusion for (top figure) charm and (bottom figure) bottom quarks as a function of temperature. Both figures end at approximately the hadronization temperature of $T_{h}\approx150\MeV$. Three different masses $m_{b}=4.2\GeV$ (blue short dashes), $4.7\GeV$, (solid black), $5.2\GeV$ (red long dashes) for bottom quarks are plotted to account for its decay width.}
\label{BCreaction_fig}
\end{figure}

In the early Universe within the temperature range $130\, \mathrm{GeV}>T>150\, \mathrm{MeV}$ we have the following particles: photons, $8_c$-gluons, $W^\pm$, $Z^0$, three generations of $3_c$-quarks and leptons in the primordial QGP. The Hubble parameter can be written as the sum of particle energy densities $\rho_i$ for each species
\begin{align}
H^2=\frac{8\pi G_{N}}{3}\left(\rho_\gamma+\rho_{\mathrm{lepton}}+\rho_{\mathrm{quark}}+\rho_{g,{W^\pm},{Z^0}}\right),
\end{align}
where $G_{N}$ is Newton's constant of gravitation. Ultra-relativistic particles (which are effectively massless) and radiation dominate the speed of expansion.

The Universe's characteristic expansion time constant $1/H$ is seen in \rf{BCreaction_fig} (both top and bottom figures). The (top) figure plots the relaxation time for the production and decay of charm quarks as a function of temperature. For the entire duration of QGP, the Hubble time is larger than the decay lifespan and production times of the charm quark. Therefore, the heavy charm quark remains in equilibrium as its processes occur faster than the expansion of the Universe. Additionally, the charm quark production time is faster than the charm quark decay. The faster quark-gluon pair fusion keeps the charm in chemical equilibrium up until hadronization. After hadronization, charm quarks form heavy mesons that decay into multi-particles quickly. Charm content then disappears from the Universe's particle inventory.

In \rf{BCreaction_fig} (bottom) we plot the relaxation time for production and decay of the bottom quark with different masses as a function of temperature. It shows that both production and decay are faster than the Hubble time $1/H$ for the duration of QGP. Unlike charm quarks however, the relaxation time for bottom quark production intersects with bottom quark decay at a temperatures dependant on the mass of the bottom. This means that the bottom quark decouples from the primordial plasma before hadronization as the production process slows down at low temperatures. The speed of weak interaction decays then dilutes bottom quark content of the QGP plasma pulling the distribution off equilibrium with $\Upsilon\neq1$ (see \req{QuarkDistribution}) in the temperature domain below the crossing point, but before hadronization. All of this occurs with rates faster than Hubble expansion and thus as the Universe expands, the system departs from a detailed chemical balance rather than thermal freezeout.

\begin{figure}[t]
\centering
\includegraphics[width=\textwidth]{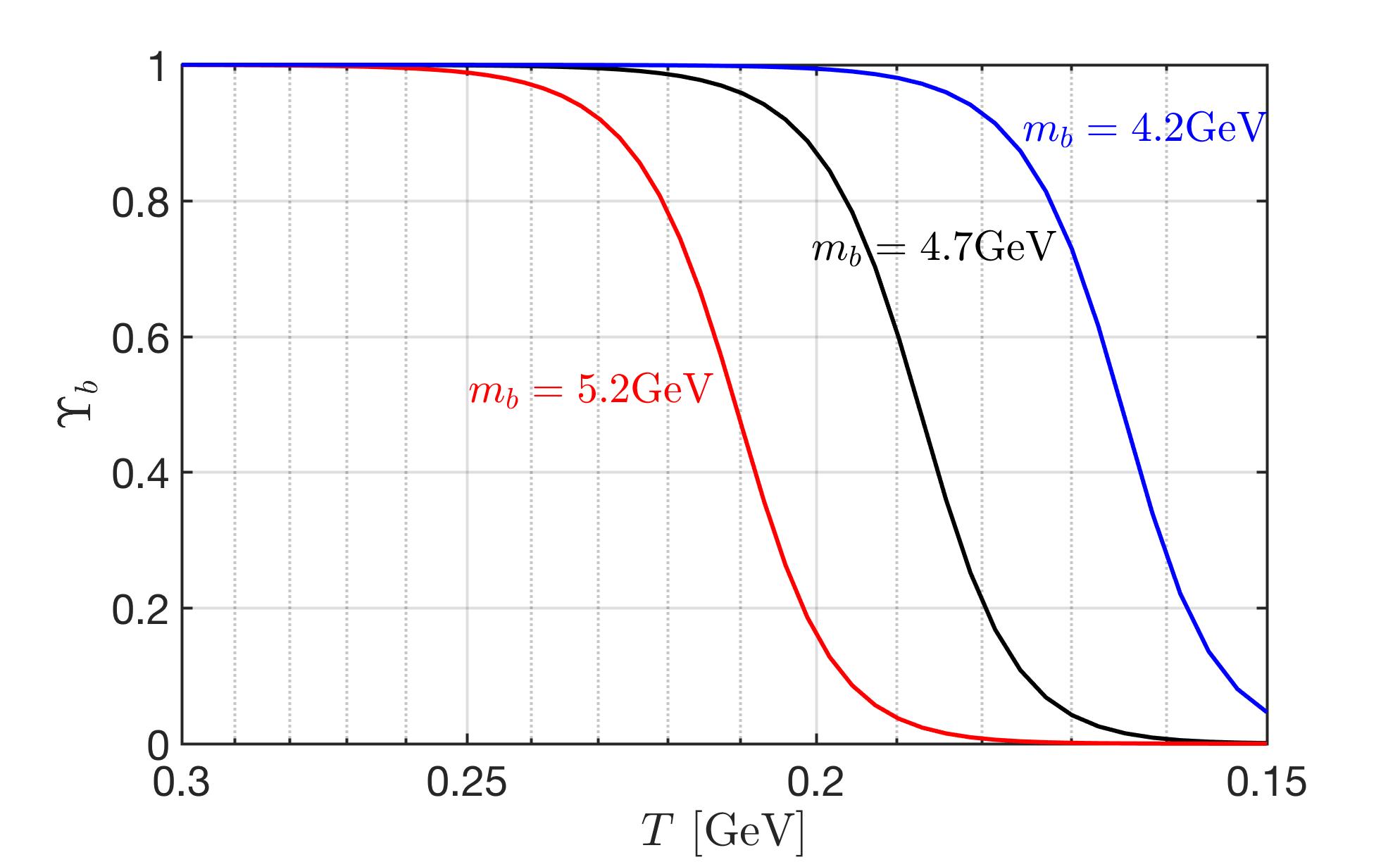}
    \caption{The generalized fugacity $\Upsilon_{b}$ of free unconfined bottom quark as a function of temperature in QGP up to the hadronization temperature of $T_{h}\approx150\MeV$ for three different bottom masses $m_{b}=4.2\GeV$ (solid blue), $4.7\GeV$, (solid black), $5.2\GeV$ (solid red).}
\label{UpsilonBottom_fig}
\end{figure}

Let us describe the dynamical non-equilibrium of bottom quark abundance in QGP in more detail. The competition between decay and production reaction rates for bottom quarks in the early Universe can be written as
\begin{align}
\label{Bquark_eq}
\frac{1}{V}\frac{dN_b}{dt}=\big(\,1-\Upsilon^2_{b}\,\big)\,R^{\mathrm{Source}}_{b}-\Upsilon_b\,R^{\mathrm{Decay}}_{b}\;,
\end{align}
where $N_b$ is the bottom quark abundance, $\Upsilon_b$ is the general fugacity of bottom quarks, and $R^{\mathrm{Source}}_{b}$ and $R^{\mathrm{Decay}}_{b}$ are the thermal reaction rates per volume of production and decay of bottom quark, respectively~\cite{Yang:2020nne}. The bottom source rate is controlled by quark-gluon pair fusion rate which vanishes upon hadronization. The decay rate depends on whether the bottom quarks are unconfined and free or bound within B-mesons which is controlled by the plasma temperature. Under the adiabatic approximation 
, we solve for the generalized bottom fugacity $\Upsilon_{b}$ in \req{Bquark_eq} yielding
\begin{align}
 \Upsilon_{b}=\frac{R^{\mathrm{Decay}}_{b}}{2R^{\mathrm{Source}}_{b}}\left[\sqrt{1+\left(2R^{\mathrm{Source}}_{b}/R^{\mathrm{Decay}}_{b}\right)^2}-1\right]\,.
\end{align}
In \rf{UpsilonBottom_fig} we show the fugacity of the bottom quarks as a function of temperature $T=0.3\sim0.15\GeV$ for different masses of bottom quarks. In all cases, we have prolonged non-equilibrium $\Upsilon_{b}\neq1$ because the decay and production rates of bottom quarks are of comparable temporal size to one another. The bottom content of QGP is exhausted as $\Upsilon_{b}\rightarrow0$ as the Universe cools in temperature. For smaller masses, some bottom quark content is preserved up until hadronization as the strong interaction formation rate slows the depletion from weak decay near the QGP to HG phase transformation.

As demonstrated above, the bottom quark flavor is capable to imprint arrow in time on physical processes being out of chemical equilibrium during the epoch $T = 0.3 \sim0.15\GeV$. This is one of the required Sakharov condition (see \rsec{sec:Timeline}) for baryogenesis. Our results provide a strong motivation to explore the physics of baryon non-conservation involving the bottom quarks and bound $b\bar b$ bottonium states in a thermal environment. Given that the non-equilibrium of bottom flavor arises at a relatively low QGP temperature allows for the baryogenesis to occur across primordial QGP hadronization epoch~\cite{Yang:2020nne}. This result establishes the temperature era for the non-equilibrium abundance of bottom quarks.

\section{Hadronic Epoch}\label{sec:Hadrons}
\subsection{The formation of matter}\label{sec:Creation}
\noindent It is in this epoch that the matter of the Universe, including all the baryons which make up visible matter today, was created~\cite{Fromerth:2002wb,Rafelski:2019twp}. Unlike the fundamental particles, such as the quarks or W and Z, the mass of these hadrons is not due to the Higgs mechanism, but rather from the condensation of the QCD vacuum~\cite{Rafelski:2015cxa,Roberts:2021xnz,Roberts:2022rxm}. The quarks from which protons and neutrons are made have a mass more than 100 times smaller than these nucleons. The dominant matter mass-giving mechanism arises from quark confinement~\cite{Hagedorn:1984hz}. Light quarks are compressed by the quantum vacuum structure into a small space domain a hundred times smaller than their natural `size'. A heuristic argument can be made by considering the variance in valance quark momentum $\Delta p$ required by the Heisenberg uncertainty principle by confining them to a space of order $\Delta x\approx1\ \mathrm{fm}$ and the energy density of the attractive gluon field required to balance that outward pressure. That energy cost then manifests as the majority of the nucleon mass. The remaining few percent of mass is then due to the fact that quarks also have inertial mass provided by the Higgs mechanism as well as the electromagnetic mass for particles with charge.

The QGP-hadronization transformation is not instantaneous and involves a transitory period containing both hadrons and QGP~\cite{Rafelski:2019twp}. Therefore the conservation laws outlined in \req{QGP_Q} - \req{QGP_sB} can be violated in one phase as long as it is equally compensated in the other phase. This means the partition function during hadronization, and thus the formation of matter, should be parameterized between the hadron gas (HG) component and QGP component as
\begin{align}\label{QGP_trans}
 \ln\mathcal{Z}_{tot}=f_{HG}(T)\ln\mathcal{Z}_{HG}+\left[1-f_{HG}(T)\right]\ln\mathcal{Z}_{QGP}\,,
\end{align}
where $f_{HG}(T)$ is the proportion of the phase space occupied by the hadron gas with values between $0<f_{HG}<1$. The charge neutrality condition \req{QGP_Q} is then modified to be
\begin{align}\label{QGP_transQ}
 n_{Q,HG+QGP}=f_{HG}(T)n_{HG,Q}+\left[1-f_{HG}(T)\right]n_{QGP,Q}=0\,.
\end{align}
At a temperature of $T_{h}\approx150\MeV$, the quarks and gluons become confined and condense into hadrons (both baryons and mesons). During this period, the number of baryon-antibaryon pairs is sufficiently high that the asymmetry (of $\sim1$ in $10^{9}$) would be essentially invisible until a temperature of between $40-50\MeV$. We note that CPT symmetry is protected by the lack of asymmetry in normal Standard Model reactions to some large factor by the accumulation of scattering events through the majority of the Universe's evolution. CPT-violation is similarly restricted by possible mass difference in the Kaons~\cite{Fromerth:2002dv} via the hypothetical difference in strange-antistrange quark masses which are expected to be small if not identically zero.

\begin{figure}[ht]
\centering
\includegraphics[width=\textwidth]{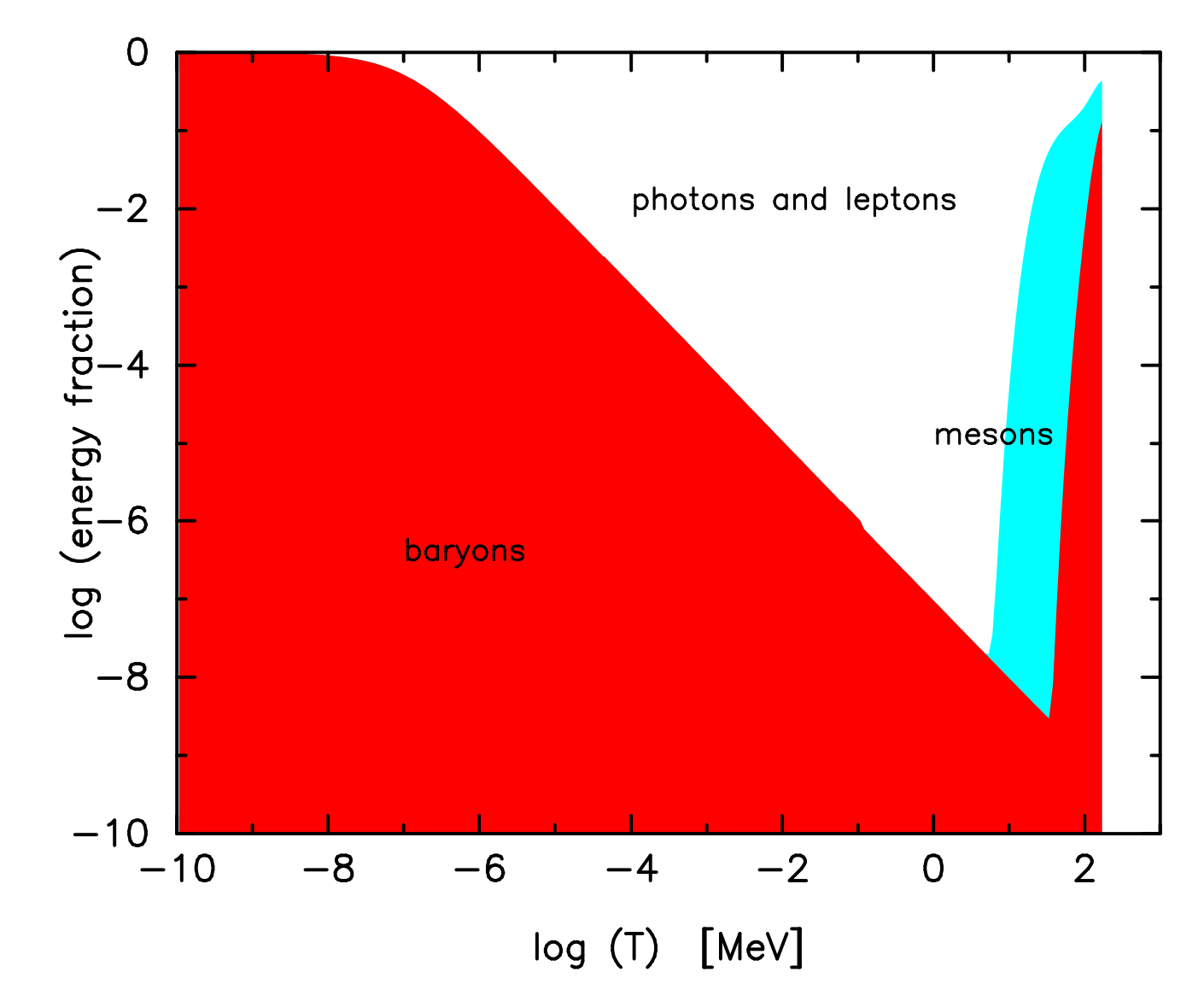}
\caption{The fractional energy density of the luminous Universe (photons and leptons (white), mesons (blue), and hadrons (red)) as a function of the temperature of the Universe from hadronization to the contemporary era. This figure is a companion figure to \rf{CosmicFraction}. (2003 unpublished, Fromerth \& Rafelski~\cite{Rafelski:2019twp})}
\label{hadron_content}
\end{figure}

In \rf{hadron_content}, we present the fraction of visible radiation and matter split between the baryons, mesons, and photons and leptons. {\xblue For a brief early Universe period after QGP hadronization  when the large amount of antimatter found in antiquarks converted into the dense gas of hadrons, their contribution to the energy density of the Universe competed  with that of radiation and leptons~\cite{Rafelski:2019twp}. Mass of matter will not emerge again} until the late Universe after recombination though by that point dark matter would become the dominant form of matter in the cosmos.

The chemical potential of baryons after hadronization can be determined by the conserved baryon-per-entropy ratio under adiabatic expansion. Considering the net baryon density in the early Universe with temperature range $150\,\mathrm{MeV}>T>5\MeV$~\cite{Yang:2021bko} we write
\begin{align}\label{Baryon_ChemicalPotential}
\frac{\left(n_B-n_{\overline{B}}\right)}{s}&=\frac{1}{s}\left[\left(n_p-n_{\overline{p}}\right)+\left(n_n-n_{\overline{n}}\right)+\left(n_Y-n_{\overline{Y}}\right)\right]\notag\\
&=\frac{45}{2\pi^4g^s_\ast}\sinh\left[\frac{\mu_B}{T}\right]F_N\left[1+\frac{F_Y}{F_N}\sqrt{\frac{1+e^{-\mu_B/T}\,F_Y/F_K}{1+e^{\mu_B/T}\,F_Y/F_K}}\right].
\end{align}
where $\mu_{B}$ is the baryon chemical potential, $g_{*}^{s}$ represents the effective entropic degrees of freedom, and we employ phase-space functions $F_i$ for the set of nucleon $N$, kaon $K$, and hyperon $Y$ particles. These functions are defined in Section 11.4 of~\cite{Letessier:2002ony} and given by
\begin{align}
&F_N=\sum_{N_i}\,g_{N_i}W(m_{N_i}/T)\;, \quad N_i=n, p, \Delta(1232),\\
&F_K=\sum_{K_i}\,g_{K_i}W(m_{K_i}/T)\;, \quad K_i=K^0, \overline{K^0}, K^\pm, K^\ast(892),\\
&F_Y=\sum_{Y_i}\,g_{Y_i}W(m_{Y_i}/T)\;, \quad Y_i=\Lambda, \Sigma^0,\Sigma^\pm, \Sigma(1385),
\end{align}
where $g_{N_i,K_i,Y_i}$ is the degeneracy of each baryonic species. We define the function $W(x)=x^2K^\mathrm{B}_2(x)$ where $K^\mathrm{B}_2$ is the modified Bessel functions of integer order \lq\lq$2$\rq\rq. 

The net baryon-per-entropy-ratio can be obtained from the present-day measurement of the net baryon-per-photon ratio $\left(n_{B}-n_{\overline{B}}\right)/n_{\gamma}$, where $n_{\gamma}$ is the contemporary photon number density from the CMB~\cite{Yang:2021bko}. This value is determined to be
\begin{align}\label{BdS}
\frac{n_B-n_{\overline{B}}}{s}= \left.\frac{n_B-n_{\overline{B}}}{s}\right|_{t_0}=(0.865\pm0.008)\times10^{-10} \;.
\end{align}
We arrive at this ratio from considering the observed baryon-per-photon ratio~\cite{ParticleDataGroup:2018ovx} of
\begin{align}
 \frac{n_B-n_{\overline{B}}}{n_\gamma}= (0.609\pm0.006)\times10^{-9}\,, 
\end{align}
as well as the entropy-per-particle~\cite{Fromerth:2012fe} for massless bosons and fermions 
\begin{align}
s/n|_\mathrm{boson}\approx 3.60\,,\qquad
s/n|_\mathrm{fermion}\approx 4.20\,.
\end{align}

\begin{figure}[h]
\centering
\includegraphics[width=\textwidth]{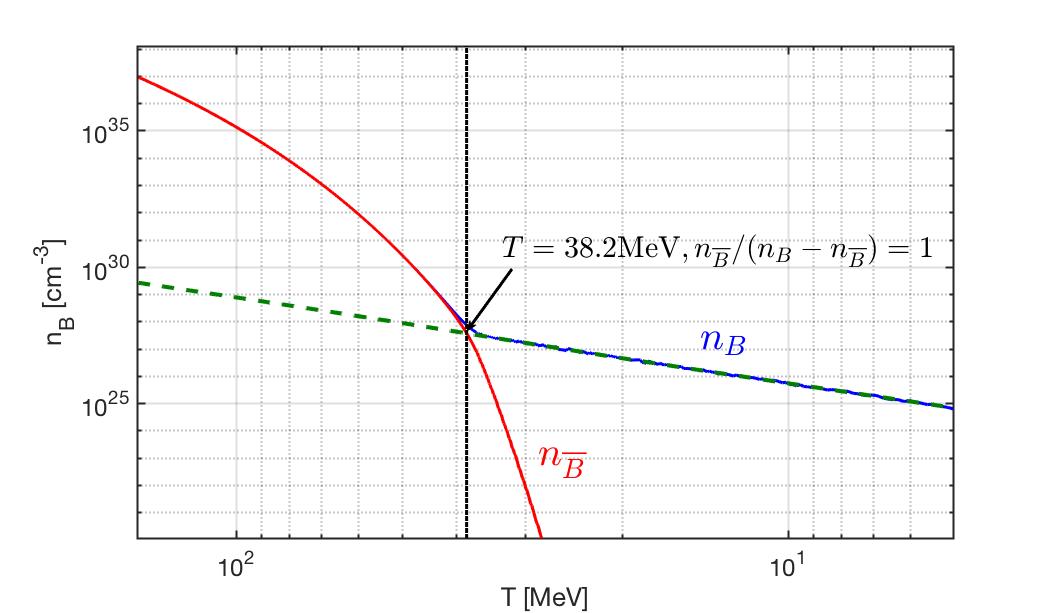}
\caption{The baryon (blue solid line) and antibaryon (red solid line) number density as a function of temperature in the range $150\MeV>T>5\MeV$. The green dashed line is the extrapolated value for baryon density. The temperature $T=38.2\MeV$ (black dashed vertical line) is denoted when the ratio $n_{\overline B}/(n_B-n_{\overline B})=1$ which define the condition where antibaryons disappear from the Universe.}
\label{Baryon_fig}
\end{figure}

Considering the inventory of strange mesons and baryons in the cosmos after hadronization, we evaluated the temperature of the net baryon disappearance in \rf{Baryon_fig}. In solving \req{Baryon_ChemicalPotential} numerically, we plot the baryon and antibaryon number density as a function of temperature in the range $150\,\mathrm{MeV}>T>5\,\mathrm{MeV}$. The temperature where antibaryons disappear from the Universe inventory can be defined when the ratio $n_{\overline B}/(n_B-n_{\overline B})=1$. This condition was reached at temperature $T=38.2\MeV$ which is in agreement with the qualitative result in Kolb and Turner~\cite{kolb1990early}. After this temperature, the net baryon density dilutes with a residual co-moving conserved quantity determined by the baryon asymmetry.

The antibaryon disappearance temperature does not depend on baryon and lepton number neutrality $L=B$. Rather, it depends only on the baryon-per-entropy ratio which is assumed to be constant during the Universe's evolution, a condition which is maintained well after the plasmas discussed here vanish. The assumption of co-moving baryon number conservation is justified by the wealth of particle physics experiments, and the co-moving entropy conservation in an adiabatic evolving Universe is a common assumption.

\subsection{Strangeness abundance}\label{sec:Strangeness}
\noindent As the energy contained in QGP is used up {\xblue to create mesons, that is massive particles containing matter and antimatter}, the high abundance of (anti)strange $(s,\bar{s})$ quark pairs present in the plasma is preserved. A smaller abundance of  (anti)charm $(c,\bar{c})$ can combine with abundant strange quarks to form `exotic' heavy mesons. With time, charmness and later strangeness decay away as these flavors are heavier than the light $(u,d)$ quarks and antiquarks. Unlike charm, which disappears from the particle inventory relatively quickly, strangeness can still persist~\cite{Yang:2021bko} in the Universe until $T\approx\mathcal{O}(10\MeV)$. {\xblue As already noted, the meson sector is of particular interest in our work since mesons carry antimatter in form of their antiquark component. After the loss of  antibaryons at $T=38.2\MeV$, \rf{Baryon_fig}, the remaining light mesons then act as a proxy for the hadronic antimatter evolution.}

We illustrate this by considering an unstable strange particle $S$ decaying into two particles $1$ and $2$ which themselves have no strangeness content. In a dense and high-temperature plasma with particles $1$ and $2$ in thermal equilibrium, the inverse reaction populates the system with particle $S$. This is written schematically as
\begin{align}
 S\Longleftrightarrow1+2,\qquad \mathrm{Example}: K^0\Longleftrightarrow\pi+\pi\,.
\end{align}
The natural decay of the daughter particles provides the intrinsic strength of the inverse strangeness production reaction rate. As long as both decay and production reactions are possible, particle $S$ abundance remains in thermal equilibrium. This balance between production and decay rates is called a detailed balance. The thermal reaction rate per time and volume for two-to-one particle reactions $1+2\rightarrow 3$ has been presented before~\cite{Kuznetsova:2008jt,Kuznetsova:2010pi}. In full kinetic and chemical equilibrium, the reaction rate per time per volume is given by~\cite{Kuznetsova:2010pi} :
\begin{align}
&R_{12\to 3}=\frac{g_3}{(2\pi)^2}\,\frac{m_3}{\tau^0_3}\,\int^\infty_0\frac{p^2_3dp_3}{E_3}\frac{e^{E_3/T}}{e^{E_3/T}\pm1}\Phi(p_3)\;,
\end{align}
where $\tau^0_3$ is the vacuum lifetime of particle $3$. The positive sign $``+"$ is for the case when particle $3$ is a boson, while it is negative $``-"$ for fermions. The function $\Phi(p_3)$ in the non-relativistic limit $m_3\gg p_3,T$ can be written as 
\begin{align}\label{photonfusion}
\Phi(p_3\to0)=2\frac{1}{(e^{E_1/T}\pm1)(e^{E_2/T}\pm1)}.
\end{align}
When back-reactions are faster than the Universe expansion, a condition we characterize in the following, we can explore the Universe composition assuming both kinetic and particle abundance equilibrium (chemical equilibrium). In \rf{EquilibPartRatiosFig} we numerically solve for the chemical potential of strangeness and show the chemical equilibrium particle abundance ratios~\cite{Yang:2021bko} for various mesons, the baryons, and their antiparticles. In the temperature range $150\MeV>T>40\MeV$ the Universe is rich in physics phenomena involving strange mesons and (anti)baryons including (anti)hyperon abundances. While antibaryons vanish after temperature $T\approx40\MeV$, kaons persist compared to baryons until $T=20\MeV$. For temperatures $T<20\MeV$, the Universe becomes light-quark baryons  dominant. Pions $\pi(q\bar q)$ persist the longest of the mesons (a feature explored in \rsec{sec:Pions}) until $T=5.6\MeV$. Pions are the most abundant hadrons in this period because of their low mass and the inverse decay reaction $\gamma+\gamma\rightarrow\pi^0$ which assures chemical equilibrium~\cite{Kuznetsova:2008jt}.

Below $T=5.6\MeV$, we have $n_\pi/n_B<1$ and the number density of pion become sub-dominate compared to the remaining baryons. It is important to realize that hadrons always are a part of the evolving Universe, a point we wish to see emphasized more in literature. For temperatures $150\MeV>T>20\MeV$ the Universe is meson-dominant with (anti)strangeness well represented in the meson sector with $s=\bar s$. Below temperature $T<13\MeV$, strangeness inventory is mostly found in the hyperons as we have $(s -\bar s)\ne 0$. We note that hyperons never exceed baryon content throughout the hadron epoch. This period of meson physics ends the stage of the Universe where antimatter was dominant in the quark sector.

\begin{figure}[bt]
\centering
\includegraphics[width=\textwidth]{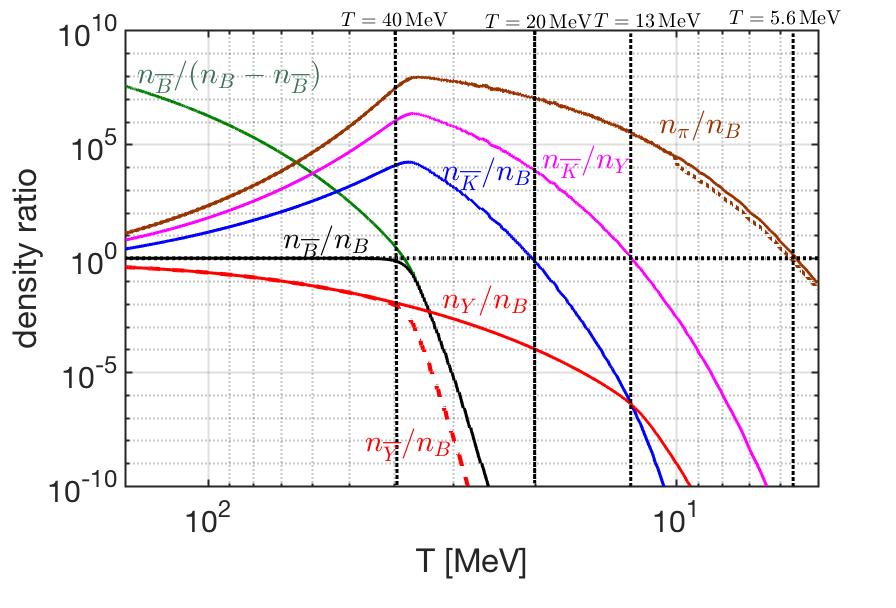}
\caption{Ratios of hadronic particle number densities as a function of temperature $150\MeV>T>5\MeV$ in the early Universe with baryon $B$ yields: Pions $\pi(q\bar q)$ (brown line), kaons $K( q\bar s)$ (blue line), antibaryon $\overline B$ (black line), hyperon $Y$ (red line) and antihyperons $\overline Y$ (dashed red line). Also shown is the $\overline K/Y$ ratio (purple line) and the $\bar B$ to asymmetry $B-\bar B$ ratio (green line). Temperature crossings are included (as vertical dashed black lines) at $T=40\MeV,\ 20\MeV,\ 13\MeV,\ 5.6\MeV$ as different abundances become sub-dominate compared to other species. The dashed brown line represents the drop in overall pion $\pi$ abundance when the vanishing of the charged pions $\pi^{\pm}$ from the particle inventory is taken into account.}
\label{EquilibPartRatiosFig}
\end{figure}

In \rf{Strangeness_map2} we schematically show important source reactions for strange quark abundance in baryons and mesons considering both open and hidden strangeness ($s\bar s$-content). The important strangeness processes (involving both the quark and lepton sectors) are 
\begin{align}
 l^-+l^+&\leftrightarrow\phi\,,\qquad\rho+\pi\leftrightarrow\phi\,,\nonumber\\ 
 \pi+\pi&\leftrightarrow K\,,\qquad
 \Lambda \leftrightarrow \pi+ N\,,\qquad\mu^\pm+\nu\leftrightarrow K^\pm\,.
\end{align}
Muons and pions are coupled through electromagnetic reactions 
\begin{align}
 \mu^++\mu^-\leftrightarrow\gamma+\gamma\,,\qquad \pi^0\leftrightarrow\gamma+\gamma\,,
\end{align}
to the photon background and retain their chemical equilibrium respectively~\cite{Rafelski:2021aey,Kuznetsova:2008jt}. The large $\phi\leftrightarrow K+K$ rate assures $\phi$ and $K$ are in relative chemical equilibrium. 

\begin{figure}[ht]
\centering
\includegraphics[width=0.8\linewidth]{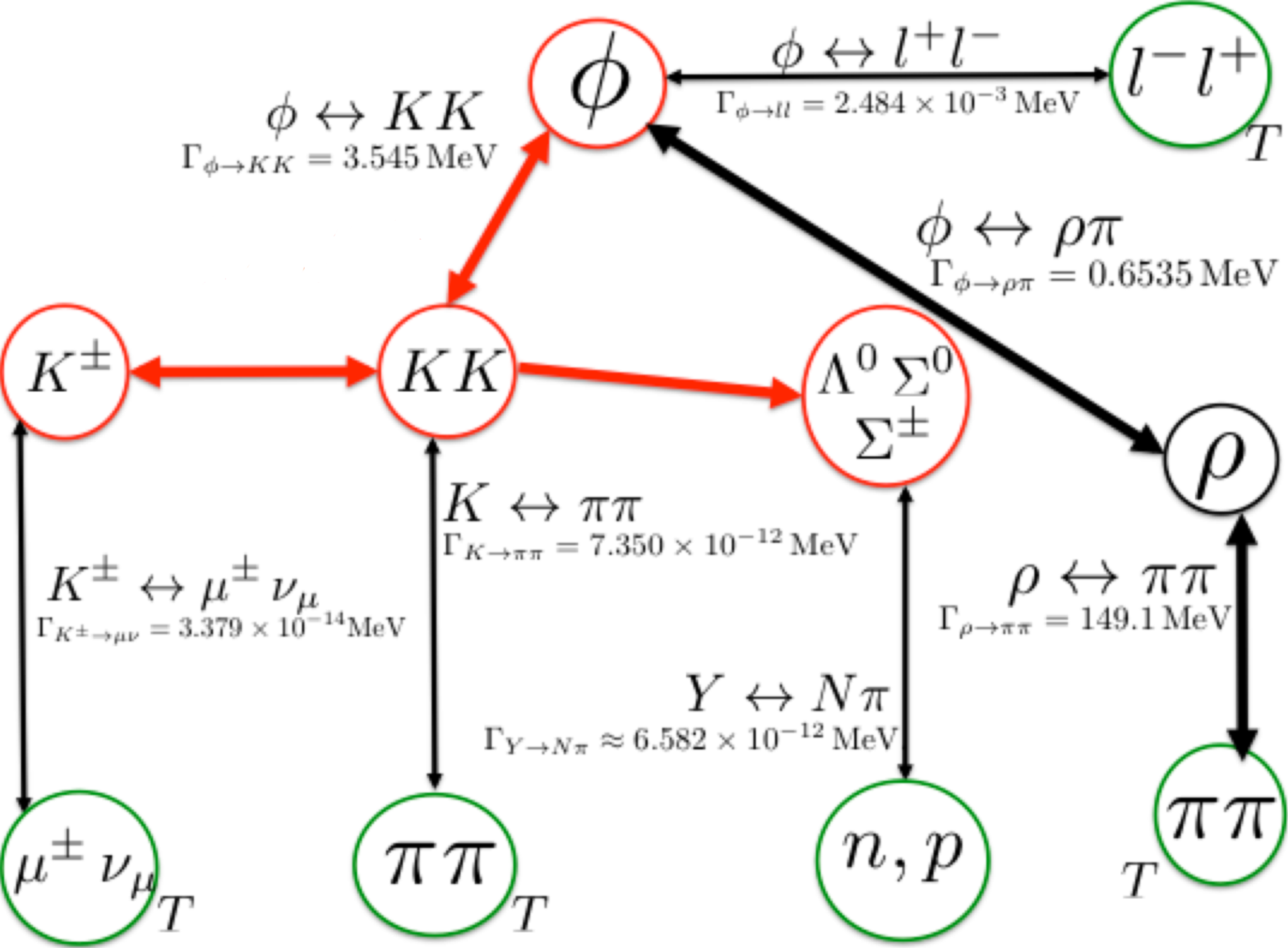}
\caption{The strangeness abundance changing reactions in the primordial Universe. Red circles show strangeness carrying hadronic particles and thick red lines denote effectively instantaneous reactions. Thick black lines show relatively strong hadronic reactions.}
\label{Strangeness_map2}
\end{figure}

\begin{figure}[ht]
\centering
\includegraphics[width=1.0\linewidth]{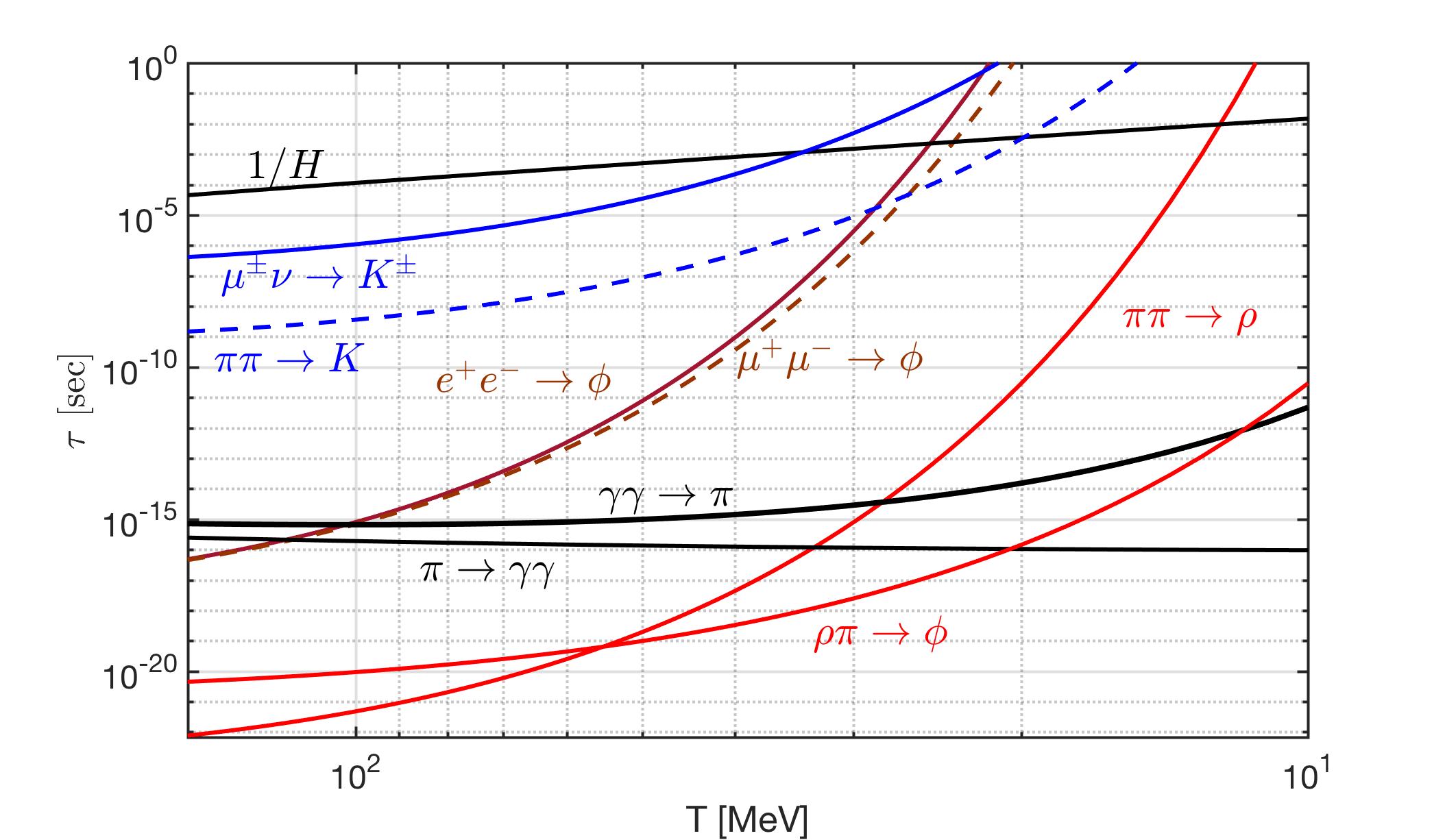}
\caption{The hadronic reaction relaxation times $\tau_{i}$ in the meson sector as a function of temperature compared to Hubble time $1/H$ (black solid line). The following processes are presented: The leptonic (solid blue line) and strong (dashed blue line) kaon $K$ processes, the electronic (solid dark red line) and muonic (dashed dark red line) phi meson $\phi$ processes, the forward and backward (thick black lines) electromagnetic pion $\pi$ processes, and the strong (red lines) rho meson $\rho$ processes.}
\label{reaction_time_tot}
\end{figure}

Once the primordial Universe expansion rate (given as the inverse of the Hubble parameter $1/H$) overwhelms the strongly temperature-dependent back-reaction, the decay $S\rightarrow 1+2$ occurs out of balance and particle $S$ disappears from the Universe. In order to determine where exactly strangeness disappears from the Universe inventory we explore the magnitudes of a relatively large number of different rates of production and decay processes and compare these with the Hubble time constant~\cite{{Yang:2021bko}}. Strangeness then primarily resides in two domains:
\begin{itemize}
 \item Strangeness in the mesons
 \item Strangeness in the (anti)hyperons
\end{itemize}
In the meson domain, the relevant interaction rates competing with Hubble time are the reactions
\begin{align}
 \pi+\pi&\leftrightarrow K\,,\qquad\mu^\pm+\nu\leftrightarrow K^\pm\,,\nonumber\\
 l^++l^-&\leftrightarrow\phi\,,\qquad \rho+\pi\leftrightarrow\phi\,,\qquad \pi+\pi\leftrightarrow\rho\,.
\end{align}
The relaxation times $\tau_{i}$ for these processes are compared with Hubble time in \rf{reaction_time_tot}. The criteria for a detailed reaction balance is broken once a process crosses above the Hubble time $1/H$ and thus can no longer be considered as subject to adiabatic evolution. As the Universe cools, these various processes freeze out as they cross this threshold. In Table~\ref{FreezeoutTemperature_table} we show the characteristic strangeness reactions and their freeze-out temperatures in the hadronic epoch. 

\begin{table}[ht]
\caption{The characteristic strangeness reaction, their freeze-out temperature, and temperature width in the hadronic epoch.}
\label{FreezeoutTemperature_table} 
\centering
\begin{tabular}{c| c| c}
\hline\hline
Reactions &freeze-out Temperature (MeV) & {$\Delta T_f$\,(MeV)} \\
\hline
$\mu^\pm\nu\rightarrow K^\pm$ & $T_f=33.8$\,MeV & {$3.5$ \,MeV}\\ 
\hline
$e^+e^-\rightarrow \phi$ & $T_f=24.9$\,MeV &{$0.6$\,MeV}\\
$\mu^+\mu^-\rightarrow\phi$ & $T_f=23.5$\,MeV &{$0.6$\,MeV}\\
\hline
 $\pi\pi\rightarrow K$ & $T_f=19.8$\,MeV&{$1.2$\,MeV}\\
\hline
$\pi\pi\rightarrow\rho$ & $T_f=12.3$\,MeV&{$0.2$\,MeV}\\
\hline\hline
\end{tabular}
\end{table}

Once freeze-out occurs and the corresponding detailed balance is broken, the inverse decay reactions act like a \lq\lq hole\rq\rq\ in the strangeness abundance siphoning strangeness out of the Universe's particle inventory. The first freeze-out reaction is the weak interaction kaon production process
\begin{align}
    \mu^\pm+\nu_{\mu}\rightarrow K^\pm\,,\qquad T_f^{K^\pm}=33.8\MeV\,,
\end{align}
which is followed by the electromagnetic $\phi$ meson production process
\begin{align}
    l^-+l^+\rightarrow\phi\,,\qquad T_f^\phi=23\sim25\MeV\,.
\end{align}
Hadronic kaon production via pions follows next in the freeze-out process
\begin{align}
    \pi+\pi\rightarrow K\,,\qquad T_f^K=19.8\MeV\,.
\end{align}
as it becomes slower than the Hubble expansion. The reactions
\begin{align}
    \gamma+\gamma\leftrightarrow\pi\,,\qquad\rho+\pi\leftrightarrow\phi
\end{align}
remain faster compared to $1/H$ for the duration of the hadronic plasma epoch. Most $\rho$ meson decays are faster~\cite{ParticleDataGroup:2018ovx} than $\rho$ meson producing processes and cannot contribute to the strangeness creation in the meson sector. Below the temperature $T<20\MeV$, all the detail balances in the strange meson sector are broken by freeze-out and the strangeness inventory in meson sector disappears rapidly.

\begin{figure}[ht]
\centering
\includegraphics[width=\linewidth]{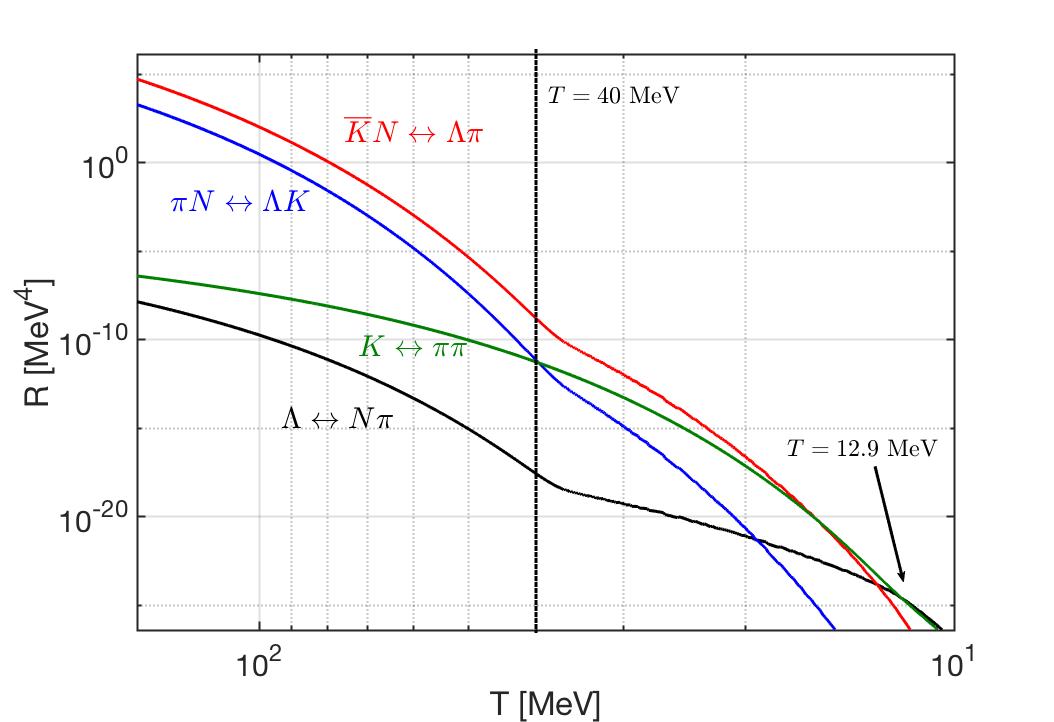}
\caption{Thermal reaction rate $R$ per volume and time for important hadronic strangeness production, exchange and decay processes as a function of temperature $150\,\mathrm{MeV}> T>10\,\mathrm{MeV}$. The following processes are presented: $\Lambda\leftrightarrow N\pi$ (solid black line), $K\leftrightarrow\pi\pi$ (solid green line), $\pi N\leftrightarrow\Lambda K$ (solid blue line), $\bar K N\leftrightarrow\Lambda\pi$ (solid red line). Two temperature crossings are denoted at $T=40\MeV,\ 12.9\MeV$.}
\label{Lambda_Rate_volume.fig}
\end{figure}

Were it not for the small number of baryons present, strangeness would entirely vanish with the loss of the mesons. In order to understand strangeness in hyperons in the baryonic domain, we evaluated the reactions 
\begin{align}
 \pi +N\leftrightarrow K+\Lambda\,,\qquad \overline{K}+N\leftrightarrow \Lambda+\pi\,,\qquad \Lambda\leftrightarrow N+\pi\,,
\end{align}
for strangeness production, exchange, and decay respectively in detail. The general form for thermal reaction rate per volume is discussed in Ch.~17 of~\cite{Letessier:2002ony}. In \rf{Lambda_Rate_volume.fig} we show that for $T<20\MeV$, the reactions for the hyperon $\Lambda$ production is dominated by $\overline{K}+N\leftrightarrow\Lambda+\pi$. Both strangeness and antistrangeness disappear from the Universe via the reactions
\begin{align}
    \Lambda\rightarrow N+\pi\,,\qquad K\to\pi+\pi\,,
\end{align}
which conserves $s=\bar s$. Beginning with $T=12.9\MeV$, the dominant reaction is $\Lambda\leftrightarrow N+\pi$, which shows that at lower temperatures strangeness content resides in the $\Lambda$ baryon. This behavior is seen explicitly in \rf{EquilibPartRatiosFig} where the hyperon abundance (of which the $\Lambda$ baryon is a member) exceeds the rapidly diminishing kaon abundance as the Universe cools. While hyperons never form a dominant component of the hadronic content of the Universe, it is an important life-boat for strangeness persisting after the more transitory mesons. In this case, the strangeness abundance becomes asymmetric and we have $s\gg\bar{s}$ at temperatures $T<12.9\MeV$. Hence, strange hyperons and antihyperons could enter into dynamic non-equilibrium condition including $\langle s-\bar s\rangle \ne 0$. The primary conclusion of the study of strangeness production and content in the early Universe, following on QGP hadronization, is that the relevant temperature domains indicate a complex interplay between baryon and meson (strange and non-strange) abundances and non-trivial decoupling from equilibrium for strange and non-strange mesons.

\subsection{Pion abundance}\label{sec:Pions}
\noindent Pions ($q\bar q, q\in u,d$), the lightest hadrons, are the dominant hadrons in the hadronic era and the most abundant hadron family well into the leptonic epoch (see \rsec{sec:Leptonic}). The neutral pion $\pi^0$ vacuum lifespan of $\tau_{\pi^0}^0=(8.52\pm0.18)\times10^{-17}$ seconds~\cite{ParticleDataGroup:2018ovx} is far shorter compared to the Hubble expansion time of $1/H=(10^{-3}\sim10^{-4})$ seconds within this epoch as depicted in~\rf{reaction_time_tot}.

At seeing such a large discrepancy in characteristic times, one is tempted to presume that the decay process dominates and that $\pi^0$ disappears quickly in the hadronic gas. However, in the high temperature $T=\mathcal{O}(100\MeV)\sim\mathcal{O}(10\MeV)$ thermal bath of this era, the inverse decay reaction forms neutral pions $\pi^0$ at rate corresponding to the decay process maintaining the abundance of the species (see \rf{EquilibPartRatiosFig}). In general, $\pi^0$ is produced in the QED plasma predominantly by thermal two-photon fusion:
\begin{align}
\gamma+\gamma \rightarrow \pi^{0}. 
\end{align}
This formation process is simply the inverse of the dominant decay process. While we do not address it in detail here, the $\pi^{\pm}$ charged pions are also in thermal equilibrium with the other pions species via hadronic and electromagnetic reactions
\begin{align}
    \pi^{0}+\pi^{0}\leftrightarrow\pi^{+}+\pi^{-}\,\qquad l^{+}+l^{-}\leftrightarrow\pi^{+}+\pi^{-}\,,\qquad\gamma+\gamma\leftrightarrow\pi^{+}+\pi^{-}\,.
\end{align}
Of these, the hadronic interaction is the fastest and controls the charged pion abundance most directly~\cite{Kuznetsova:2009xh,Fromerth:2012fe} such that the condition
\begin{align}
    \rho_{\pi^{0}}\sim\rho_{\pi^{\pm}}\,,
\end{align}
where $\rho$ is the energy density of the species and is maintained for most of the hadronic era. We point out that the in the late (colder) hadronic era, the charged pions will scatter off the remaining baryons with asymmetric reactions due to the lack of antibaryons. The smallness of the electronic $e^{+}e^{-}$ formation of $\pi^0$ is characterized by its small branching ratio in $\pi^0$ decay $B=\Gamma_{ee}/\Gamma_{\gamma\gamma}=6.2\pm 0.5\times10^{-8}$~\cite{ParticleDataGroup:2018ovx} which can be neglected compared to photon fusion. The general form for invariant production rates and relaxation time is discussed in~\cite{Kuznetsova:2008jt} where we have for the photon fusion process
\begin{align}
R_{\gamma\gamma\to\pi^0}=&\int\frac{d^{3}{p_{\pi}}}{(2\pi\ )^32E_{\pi}}
 \int\frac{d^{3} {p_{2\,\gamma}}}{(2\pi\ )^32E_{2\,\gamma}}
 \int\frac{d^{3}{p_{1\,\gamma}}} {(2\pi\ )^32E_{1\,\gamma} }\left(2\pi\right)^{4}
 \delta^{4}\left(p_{1\,\gamma}+p_{2\,\gamma}-p_{\pi}\right)\times \nonumber\\ &
 \sum_{spin}\left|\langle p_{1\,\gamma}p_{2\,\gamma}\left| M\right|p_{\pi}\rangle\right|^{2}
 f_{\pi}(p_{\pi})f_{\gamma}(p_{1\,\gamma})f_{\gamma}(p_{2\,\gamma})
 \Upsilon^{-2}_{\gamma}\Upsilon_{\pi^{0}}^{-1}e^{u \cdot p_{\pi}/T}\,, \label{pi0pr}
 \end{align}
where $\Upsilon_i$ is the fugacity and $f_{i}$ is the Bose-Einstein distribution of particle $i$, and $M$ is the matrix element for the process. Since the $\gamma+\gamma\to \pi^0$ is the dominant mechanism of pion production, we can omit all sub-dominant processes, and the dynamic equation of $\pi^0$ abundance can be written as~\cite{Fromerth:2012fe}:
\begin{align}
\frac{d}{dt}\Upsilon_{\pi^0}=\frac{1}{\tau_T}\Upsilon_{\pi^0}+\frac{1}{\tau_S}\Upsilon_{\pi^0}+\frac{1}{\tau_{\pi^0}}\left(\Upsilon^2_\gamma-\Upsilon_{\pi^0}\right)\,,
\end{align}
where $\tau_T$ and $\tau_S$ are the kinematic relaxation times for temperature and entropy evolution and $\tau_{\pi^0}$ is the chemical relaxation time for $\pi^0$. We have
\begin{align}
\frac{1}{\tau_T}&\equiv -T^3g^*\frac{d (n_{\pi}/(\Upsilon_3
g^*T^3))/dT}{dn_{\pi}/d{\Upsilon_3}}{\dot T},\label{tauT} \nonumber\\
\frac{1}{\tau_{S}}&\equiv
-\frac{n_{\pi}/\Upsilon_3}{dn_{\pi}/d{\Upsilon_3}}\frac{d\ln (g^*VT^3)}{dT}
\dot{T},\nonumber \\
\tau_{\pi^0}&\equiv \frac{dn_{\pi^0}/d\Upsilon_{\pi^0}}{R_{\pi^0}}\,,
\end{align}
Where $n_{\pi^0}$ is the number density of pions. A minus sign is introduced in the above expressions to maintain $\tau_T$, $\tau_S>0$. Since entropy is conserved within the radiation-dominated epoch, we have $T^{3}V={\rm constant}$ thus $d(T^3V(T))/dT=0$. This implies the entropic relaxation time is infinite yielding $1/\tau_S=0$. The effect of Universe expansion and dilution of number density is described by $1/\tau_T$. Comparing $\tau_T$ to the chemical relaxation time $\tau_{\pi^0}$ can provide the quantitative condition for freeze-out from chemical equilibrium. In the case of pion mass being much larger than the temperature, $m_{\pi}\gg T$, we have~\cite{Kuznetsova:2009xh}
\begin{align}
\tau_T\approx\frac{T}{m_{\pi}H}.
\end{align}
 In \rf{reaction_time_tot} we compare the relaxation time of $\tau_{\pi^0}$ to the Hubble time $1/H$ which shows that $\tau_{\pi^0}\ll 1/H$. In such a case, the yield of $\pi^0$ is expected to remain in chemical equilibrium (even as its thermal number density gradually decreases) with no freeze-out temperature occurring. This makes pions distinct from all other meson species. This phenomenon can be attributed to the high population of photons as in such an environment, it remains sufficiently probable to find high-energy photons to fuse back into neutral pions $\pi^0$~\cite{Fromerth:2012fe} for the duration of large pion abundance. As shown in \rf{EquilibPartRatiosFig},  {\xblue pions remained as proxy for hadronic matter and antimatter down to} $T=5.6\MeV$.

\section{Leptonic Epoch} \label{sec:Leptonic}
\subsection{Thermal degrees of freedom}\label{sec:Freedom}
\noindent The leptonic epoch, dominated by photons and both charged and neutral leptons, is notable for being the last time where neutrinos played an active role in the Universe's thermal dynamics before decoupling and becoming free-streaming. In the early stage of this plasma after the hadronization era ended $T\approx\mathcal{O}(10\MeV)$, neutrinos represented the highest energy density followed by the light charged leptons and then finally the photons. The differing relativistic limit energy densities can be related by
\begin{align}
\rho_{e^{\pm}}\approx\left(2\times\frac{7}{8}\right)\rho_{\gamma}\,,\qquad\rho_{\nu}\approx\left(3\times\frac{7}{8}\right)\rho_{\gamma}\,.
\end{align}
The reason for this hierarchy is because of the degrees of freedom~\cite{Letessier:2002ony,Rafelski:2013yka} available in each species in thermal equilibrium; the factor $7/8$ arises from the difference in pressure contribution between bosons and fermions. 

While photons only exhibit two polarization degrees of freedom, the charged light leptons could manifest as both matter (electrons), antimatter (positrons) and as well as two polarizations yielding $2\times2=4$. The neutral leptons made up of the neutrinos however had three thermally active species $3\times2=6$ boosting their energy density in that period to more than any other contribution. The muon-antimuon energy density was also controlled by its degrees of freedom matching that of $e^{\pm}$ until $T\approx\mathcal{O}(100\MeV)$, still well within the hadronic epoch, when the heavier lepton no longer satisfied the ultra-relativistic (and thus massless) limit. This separation of the two lighter charge lepton dynamics is seen in \rf{CosmicFraction} after hadronization.

The known cosmic degrees of freedom require that if and when neutrinos are Dirac-like and have chiral right-handed (matter) components, then  these right handed components must not  drive the neutrino effective degrees of freedom $N^{\nu}_{\mathrm{eff}}$ away from three. In a more general context the non-interacting sterile neutrinos could also inflate $N_{\mathrm{eff}}^{\nu}$ during this epoch for the same reasoning~\cite{Kopp:2011qd,Hamann:2011ge,Kopp:2013vaa,Lello:2014yha,Birrell:2014qna} or have a connection to dark matter~\cite{Weinberg:2013kea,Giusarma:2014zza}. The neutrino degrees of freedom will be more fully discussed in \rsec{sec:EffectiveNeutrino}.

\subsection{Muon abundance} \label{sec:Muons}
\noindent {\xblue As seen in \rsec{sec:Strangeness}, muon abundance and their associated reactions are integral to the understanding of the strangeness and antistrangeness content of the primordial Universe~\cite{Yang:2021bko}. Therefore we determine to what extent and temperature (anti)muons remained in chemical abundance equilibrium. Without a clear boundary separating the hadronic epoch from the leptonic epoch, there is complete overlap in the hadronic and leptonic species dynamics in the period $T=\mathcal{O}(10\MeV)\sim\mathcal{O}(1\MeV)$.}

In the cosmic plasma, muons can be produced by predominately electromagnetic and weak interaction processes
\begin{align} 
&\gamma+\gamma\longrightarrow\mu^++\mu^-,\qquad & e^++e^-\longrightarrow \mu^++\mu^-\;,\\
&\pi^-\longrightarrow\mu^-+\bar{\nu}_\mu,\qquad & \pi^+\longrightarrow\mu^++\nu_\mu\;.
\end{align}
{\xblue Provided that all particles shown on the left-hand side of each reaction (namely the photons, electrons(positrons) and charged pions) exist in chemical equilibrium, the back-reaction for each of the above processes occurs in detailed balance.}

The scattering angle averaged thermal reaction rate per volume for the reaction $a\overline{a}\rightarrow b\overline{b}$ in Boltzmann approximation is given by~\cite{Letessier:2002ony}
\begin{align}\label{pairR}
R_{a\overline{a}\rightarrow b\overline{b}}=\frac{g_ag_{\overline{a}}}{1+I}\frac{T}{32\pi^4}\int_{s_{th}}^\infty ds\frac{s(s-4m^2_a)}{\sqrt{s}}\sigma_{a\overline{a}\rightarrow b\overline{b}} K_1(\sqrt{s}/T),
\end{align}
where $s_{th}$ is the threshold energy for the reaction, $\sigma_{a\overline{a}\rightarrow b\overline{b}}$ is the cross section for the given reaction. We introduce the factor $1/(1+I)$ to avoid the double counting of indistinguishable pairs of particles where $I=1$ for an identical pair and $I=0$ for a distinguishable pair. 

{\xblue The muon weak decay processes are} 
\begin{equation}
\mu^-\rightarrow\nu_\mu+e^-+\bar{\nu}_e,\qquad \mu^+\rightarrow\bar{\nu}_\mu+e^++\nu_e\,,
\end{equation} 
with the vacuum life time $\tau_{\mu}=2.197 \times 10^{-6}$ seconds {\xblue producing (anti)neutrino pairs of differing flavor and electrons(positrons). We recall the considerable shorter vacuum lifetime of pions  $\tau_{\pi^\pm}=2.6033\times10^{-8}$ seconds.} The thermal decay rate per volume in the Boltzmann limit is~\cite{Kuznetsova:2008jt}
\begin{align}
&R_i=\frac{g_i}{2\pi^2}\left(\frac{T^3}{\tau_i}\right)\left(\frac{m_i}{T}\right)^2K_1(m_i/T) 
\end{align}
where $\tau_i$ is the vacuum lifespan of a given particle $i$.

\begin{figure}[ht]
\centering
\includegraphics[width=0.9\columnwidth]{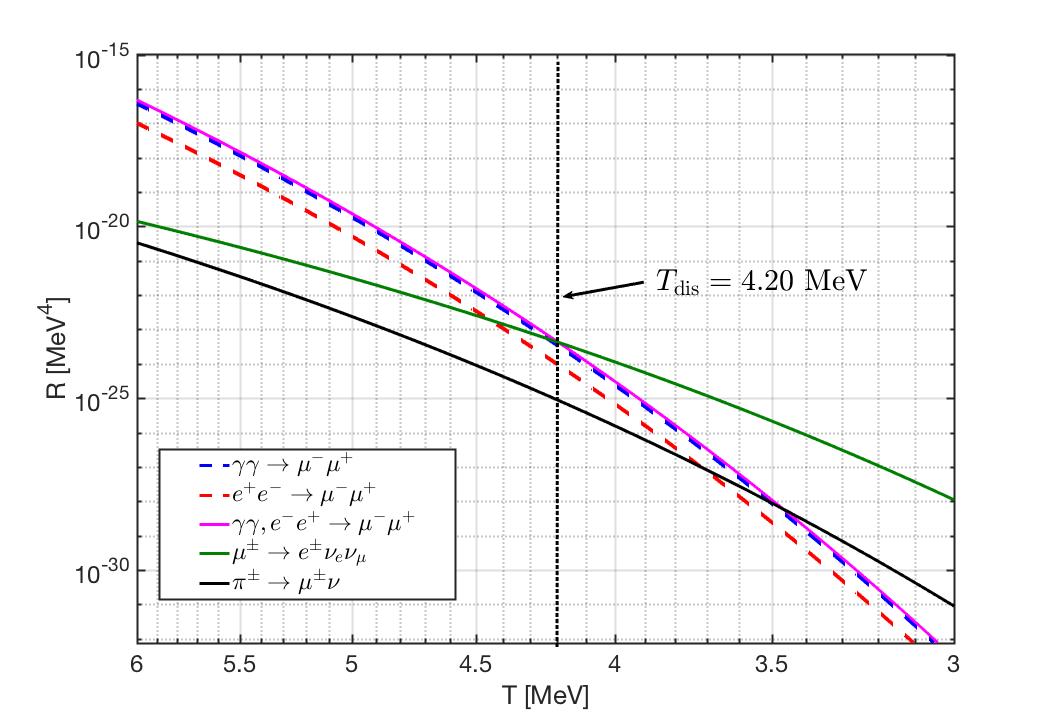}
\caption{The thermal reaction rate per volume for muon related reactions as a function of temperature adapted from~\cite{Rafelski:2021aey}. {\xblue The dominant reaction rates for $\mu^\pm$ production are: The $\gamma\gamma$ channel (blue dashed line), $e^{\pm}$ (red dashed line), these two combined as the total electromagnetic rate (pink solid line), and the charged pion decay feed channel (black solid line). The muon decay rate is also shown (green solid line). The crossing point between the electromagnetic production processes and the muonic decay rate is seen as the dashed vertical black line at $T_{\rm dis}=4.2\MeV$.}}
\label{muon_fig} 
\end{figure}

{\xblue These production and decay rates for muonic processes are evaluated in~\cite{Rafelski:2021aey}. From this, we can determine the temperature when muons rather suddenly disappear from the particle inventory of the Universe which occurs when their decay rate exceeds their production rate.} In \rf{muon_fig} we show the invariant thermal reaction rates per volume and time for the relevant muon reactions.  {\xblue As the temperature decreases in the expanding Universe, the initially dominant production rates become rapidly smaller due to the mass threshold effect. This is allowing the production and decay rates to become equal.  The characteristic times are much faster than the Hubble time (not shown in \rf{muon_fig}). Muon abundance therefore disappears just when the decay rate overwhelms production at the temperature $T_\mathrm{dis} = 4.20\MeV$.}

In \rf{muonRatio_fig} we show that the number density ratio of muons to baryons $n_{\mu^{\pm}}/n_{B}$ at the muon disappearance temperature $T_\mathrm{dis} = 4.20\MeV$ is $n_{\mu^\pm}/n_\mathrm{B}\approx0.91$~\cite{Yang:2021bko}. {\xblue  Interestingly, this means that the muon abundance may still be able to influence baryon evolution up to this point because their number density is comparable to that of baryons (there are no antibaryons). This coincidence of abundance offers a novel and tantalizing model-building opportunity for  both baryon-antibaryon separation models and/or strangelet formation models.}

\begin{figure}[ht]
\centering
\includegraphics[width=0.9\columnwidth]{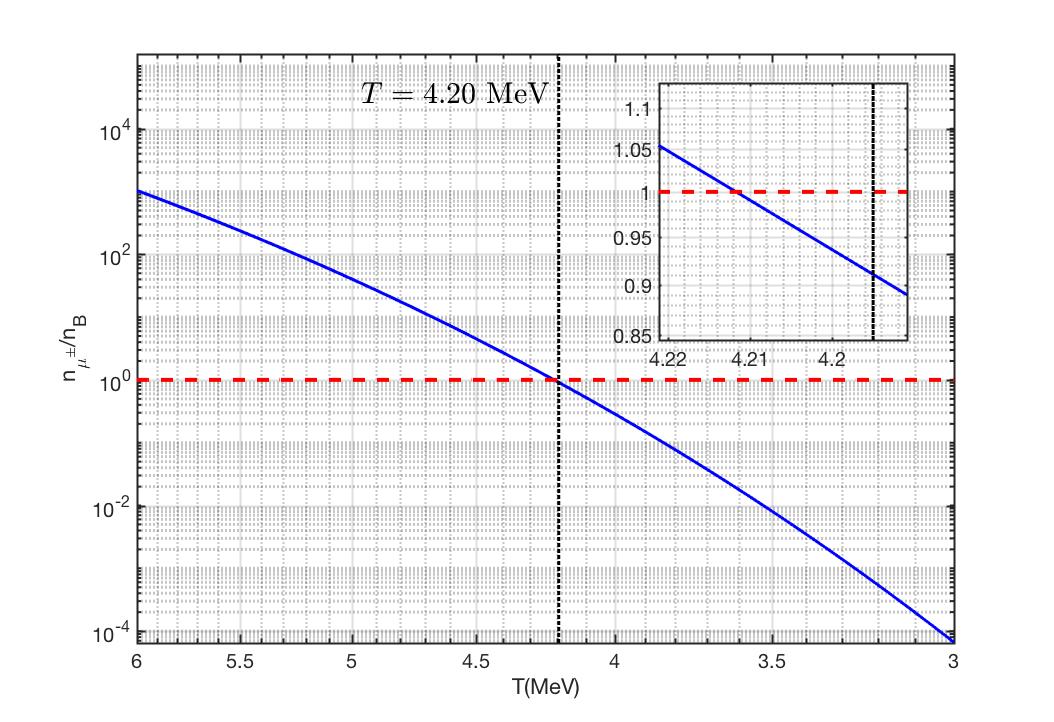}
\caption{The density ratio between $\mu^\pm$ and baryons $n_{\mu^{\pm}}/n_{B}$ (blue solid line) is plotted as a function of temperature. The red dashed line indicates a density ratio value of $n_{\mu^{\pm}}/n_{B}=1$. The density ratio at the muon disappearance temperature (vertical black dashed line) is about $n_{\mu^\pm}/n_\mathrm{B}(T_\mathrm{dis})\approx0.911$.}
\label{muonRatio_fig} 
\end{figure}

\subsection{Neutrino masses and oscillation} \label{sec:Neutrinos}
\noindent Neutrinos are believed to have a small, but nonzero mass due to the phenomenon of flavor oscillation~\cite{SuperKamiokande:1998kpq,KamLAND:2002uet,Fogli:2005cq}. This is seen in the flux of neutrinos from the Sun, and also in terrestrial reactor experiments. In the Standard Model neutrinos are produced via weak charged current (mediated by the W boson) as flavor eigenstates. If the neutrino was truly massless, then whatever flavor was produced would be immutable as the propagating state. However, if neutrinos have mass, then they propagate through space as their mass-momentum eigenstates. 
Neutrino masses can be written in terms of an effective theory where the mass term contains various couplings between neutrino states determined by some BSM theory. The exact form of such a BSM theory is outside the scope of this work, we refer the reader to some standard references~\cite{Giunti:2014ixa,Fritzsch:1999ee,giunti2007fundamentals,Fritzsch:2015gxa}.

{\xblue Within the Standard Model keeping two degrees of freedom for each neutrino flavor the} Majorana fermion mass term is given by
\begin{alignat}{1}
	\label{Majorana} \mathcal{L}_{m}^{Maj.} = -\frac{1}{2}\bar{\nu}^{\alpha}_{L}M^{M}_{\alpha\beta}(\nu^{\beta}_{L})^{c}+\mathrm{h.c.}\,,
\end{alignat}
where $\nu^{c} = \hat{C}(\bar{\nu})^{T}$ is the charge conjugate of the neutrino field. The operator $\hat{C} = i\gamma^{2}\gamma^{0}$ is the charge conjugation operator. {\xblue An interesting consequence of neutrinos being Majorana particles is that they would be their own antiparticles like photons allowing for violations of total lepton number. Neutrinoless double beta decay is an important, yet undetected, evidence for Majorana nature of neutrinos~\cite{Dolinski:2019nrj}.} Majorana neutrinos with small masses can be generated from some high scale via the See-Saw mechanism~\cite{Arkani-Hamed:1998wuz,Ellis:1999my,Casas:2001sr} which ensures that the degrees of freedom separate into heavy neutrinos and light nearly massless Majorana neutrinos. The See-Saw mechanism then provides an explanation for the smallness of the neutrino masses as has been experimentally observed.

A flavor eigenstate $\nu^{\alpha}$ can be described as a superposition of mass eigenstates $\nu^{k}$ with coefficients given by the Pontecorvo-Maki-Nakagawa-Sakata (PMNS) mixing matrix~\cite{King:2013eh,FernandezMartinez:2016lgt} which are both in general  complex and unitary. This is given by
\begin{align}\label{NuFlavors}
	\nu^{\alpha}=\sum_k^nU^\ast_{\alpha k}\nu^{k}, \qquad\alpha=e,\mu,\tau,\qquad k=1,2,3,\dots,n
\end{align}
where $U$ is the PMNS mixing matrix. The PMNS matrix is the lepton equivalent to the CKM mixing matrix which describes the misalignment between the quark flavors and their masses. For Majorana neutrinos, there can be up to three complex phases $(\delta,\rho,\gamma)$ which are CP-violating~\cite{Pascoli:2006ci} which are present when the number of generations is $n\geq3$. For Dirac-like neutrinos, only the $\delta$ complex phase is required. In principle, the number of mass eigenstates can exceed three, but is restricted to three generations in most models. By standard convention~\cite{schwartz2014quantum} found in the literature we parameterize the rotation matrix $U$ as
	\begin{alignat}{1}
 	\label{PMNS} U =
		\begin{pmatrix}
			c_{12}c_{13} & s_{12}c_{13} & s_{13}e^{-i\delta}\\
			-s_{12}c_{23} - c_{12}s_{13}s_{23}e^{i\delta} & c_{12}c_{23} - s_{12}s_{13}s_{23}e^{i\delta} & c_{13}s_{23}\\
			s_{12}s_{23} - c_{12}s_{13}c_{23}e^{i\delta}& -c_{12}s_{23} - s_{12}s_{13}c_{23}e^{i\delta} & c_{13}c_{23}
		\end{pmatrix}\times
        \begin{pmatrix}
            1 & &\\
            & e^{i\rho} &\\
            & & e^{i\gamma}
        \end{pmatrix}\,,
	\end{alignat}
where $c_{ij} = \mathrm{cos}(\theta_{ij})$ and $s_{ij} = \mathrm{sin}(\theta_{ij})$. In this convention, the three mixing angles $(\theta_{12}, \theta_{13}, \theta_{23})$, are understood to be the Euler angles for generalized rotations.

The neutrino proper masses are generally considered to be small with values no more than $0.1\eV$. Because of this, neutrinos produced during fusion within the Sun or radioactive fission in terrestrial reactors on Earth propagate relativistically. Evaluating freely propagating plane waves in the relativistic limit yields the vacuum oscillation probability between flavors $\nu_\alpha$ and $\nu_\beta$ written as~\cite{ParticleDataGroup:2022pth}
\begin{align}\label{NuOscillation}
  P_{\alpha\rightarrow\beta}
 =&\delta_{\alpha\beta}-4\sum_{i<j}^n \mathrm{Re}\left[U_{\alpha i}U^\ast_{\beta i}U^\ast_{\alpha j}U_{\beta j}\right]\sin^2\!\!\left(\frac{\Delta m^2_{ij}L}{4E}\right)\notag\\
 &+2\sum_{i<j}^n \mathrm{Im}\left[U_{\alpha i}U^\ast_{\beta i}U^\ast_{\alpha j}U_{\beta j}\right]\sin\!\!\left(\frac{\Delta m^2_{ij}L}{2E}\right)
 ,\qquad\Delta m^2_{ij}\equiv{m^2_i-m^2_j}
\end{align}
where $L$ is the distance traveled by the neutrino between production and detection. The square mass difference $\Delta m^2_{ij}$ has been experimentally measured~\cite{ParticleDataGroup:2022pth}. As oscillation only restricts the differences in mass squares, the precise values of the masses cannot be determined from oscillation experiments alone. It is also unknown under what hierarchical scheme (normal or inverted)~\cite{Avignone:2007fu,Esteban:2020cvm} the masses are organized as two of the three neutrino proper masses are close together in value. 

It is important to point out that oscillation does not represent any physical interaction (except when neutrinos must travel through matter which modulates the $\nu_{e}$ flavor~\cite{NuSTEC:2017hzk,DUNE:2020ypp}) or change in the neutrino during propagation. Rather, for a given production energy, the superposition of mass eigenstates each have unique momentum and thus unique group velocities. This mismatch in the wave propagation leads to the oscillatory probability of flavor detection as a function of distance.

{\xblue We further note that non-interacting BSM so called}
sterile neutrinos of any mass have not yet been observed despite extensive searching. The existence of such neutrinos, if they were ever thermally active in the early cosmos would leave fingerprints on the Cosmic Neutrino Background (CNB) spectrum~\cite{Birrell:2014qna}. The presence of an abnormally large anomalous magnetic moment~\cite{Morgan:1981zy,Fukugita:1987uy,Vogel:1989iv,Elmfors:1997tt,Giunti:2008ve,Giunti:2014ixa,Canas:2015yoa} for the neutrino would also possibly leave traces in the evolution of the early Universe.
\subsection{Neutrino freeze-out}\label{sec:Freezeout}
\noindent The relic neutrino background (or CNB) is believed to be a well-preserved probe of a Universe only a second old which at some future time may become experimentally accessible. The properties of the neutrino background are influenced by the details of the freeze-out or decoupling process at a temperature $T=\mathcal{O}(2\MeV)$. The freeze-out process, whereby a particle species stops interacting and decouples from the photon background, involves several steps that lead to the species being described by the free-streaming momentum distribution. We outline freeze-out properties, including what distinguishes it from the equilibrium distributions~\cite{Birrell:2012gg}.

Chemical freeze-out of a particle species occurs at the temperature, $T_{ch}$, when particle number changing processes slow down and the particle abundance can no longer be maintained at an equilibrium level. Prior to the chemical freeze-out temperature, number changing processes are significant and keep the particle in chemical (and thermal) equilibrium, implying that the distribution function has the Fermi-Dirac form, obtained by maximizing entropy at fixed energy {\xblue (parameter $1/T$) and  particle number (parameter $\lambda$)}
\begin{equation}\label{equilibrium}
f_{c}(t,E)=\frac{1}{\lambda\exp(E/T)+1}, \text{ for } T(t)> T_{ch}.
\end{equation}
Kinetic freeze-out occurs at the temperature, $T_f$, when momentum exchanging interactions no longer occur rapidly enough to maintain an equilibrium momentum distribution. When $T_f<T(t)<T_{ch}$, the number-changing process no longer occurs rapidly enough to keep the distribution in chemical equilibrium but there is still sufficient momentum exchange to keep the distribution in thermal equilibrium. The distribution function is therefore obtained by maximizing entropy, with fixed energy, particle number, and antiparticle number separately. This implies that the distribution function has the form
\begin{equation}\label{kinetic_equilib}
f_k(t,E)=\frac{1}{\Upsilon^{-1}\exp(E/T)+1}, \text{ for }T_f< T(t)< T_{ch}.
\end{equation}
{\xblue The time dependent generalized} fugacity $\Upsilon(t)$ controls the occupancy of phase space and is necessary once $T(t)<T_{ch}$ in order to conserve particle number.
 
For $T(t)<T_f$ there are no longer any significant interactions that couple the particle species of interest and so they begin to free-stream through the Universe, i.e. travel on geodesics without scattering. The Einstein-Vlasov equation can be solved, see~\cite{choquet2008general}, to yield the free-streaming momentum distribution
\begin{equation}\label{free_stream_dist}
f(t,E)=\frac{1}{\Upsilon^{-1}e^{\sqrt{p^2/T^2+m^2 /T_f^2}}+ 1}
\end{equation}
where the free-streaming effective temperature
\begin{equation}\label{T_freestream_dist}
T(t)=\frac{T_fa(t_k)}{a(t)}
\end{equation}
is obtained by redshifting the temperature at kinetic freeze-out. The corresponding free-streaming energy density, pressure, and number densities are given by
\begin{align}
\rho&=\frac{d}{2\pi^2}\!\int_0^\infty\!\!\!\frac{\left(m^2+p^2\right)^{1/2}p^2dp }{\Upsilon^{-1}e^{\sqrt{p^2/T^2+m^2/T_f^2}}+ 1},\label{freestream_rho}\\[0.2cm]
P&=\frac{d}{6\pi^2}\!\int_0^\infty\!\!\!\frac{\left(m^2+p^2\right)^{-1/2}p^4dp }{\Upsilon^{-1} e^{\sqrt{p^2/T^2+m^2/T_f^2}}+ 1},\label{freestream_P}\\[0.2cm]
n&=\frac{d}{2\pi^2}\!\int_0^\infty\!\!\!\frac{p^2dp }{\Upsilon^{-1}e^{\sqrt{p^2/T^2+m^2/T_f^2}}+ 1},
\label{num_density}
\end{align}
where $d$ is the degeneracy of the particle species. These differ from the corresponding expressions for an equilibrium distribution in Minkowski space by the replacement $m\rightarrow m T(t)/T_f$ {\em only} in the exponential. 

The separation of the freeze-out process into these three regimes is of course only an approximation. In principle, there is a smooth transition between them. However, it is a very useful approximation in cosmology. See~\cite{Mangano:2005cc,Birrell:2014gea} for methods capable of resolving these smooth transitions.

\begin{figure}[ht]
\centerline{\includegraphics[width=0.47\columnwidth]{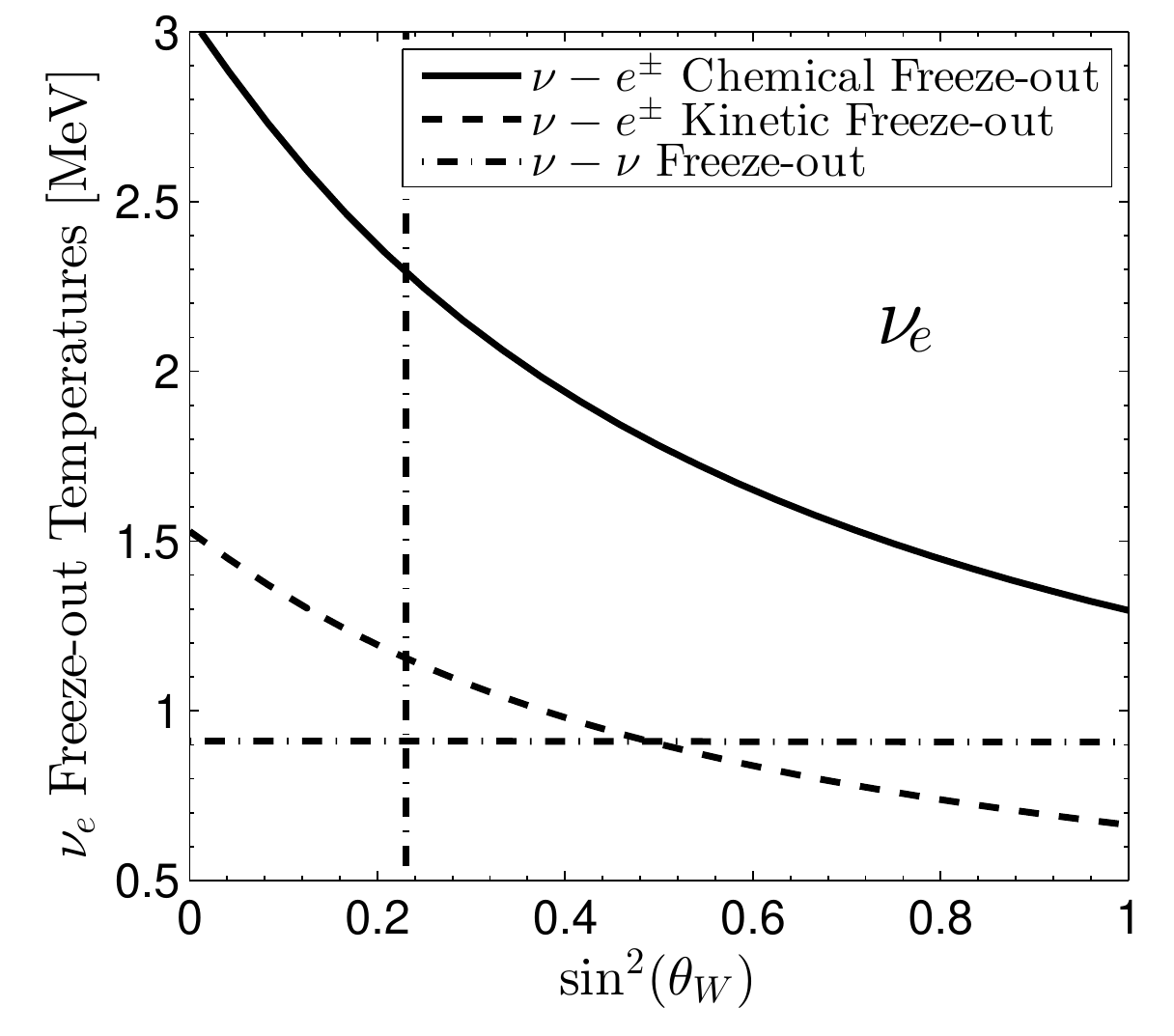}
\hspace{1mm}\includegraphics[width=0.47\columnwidth]{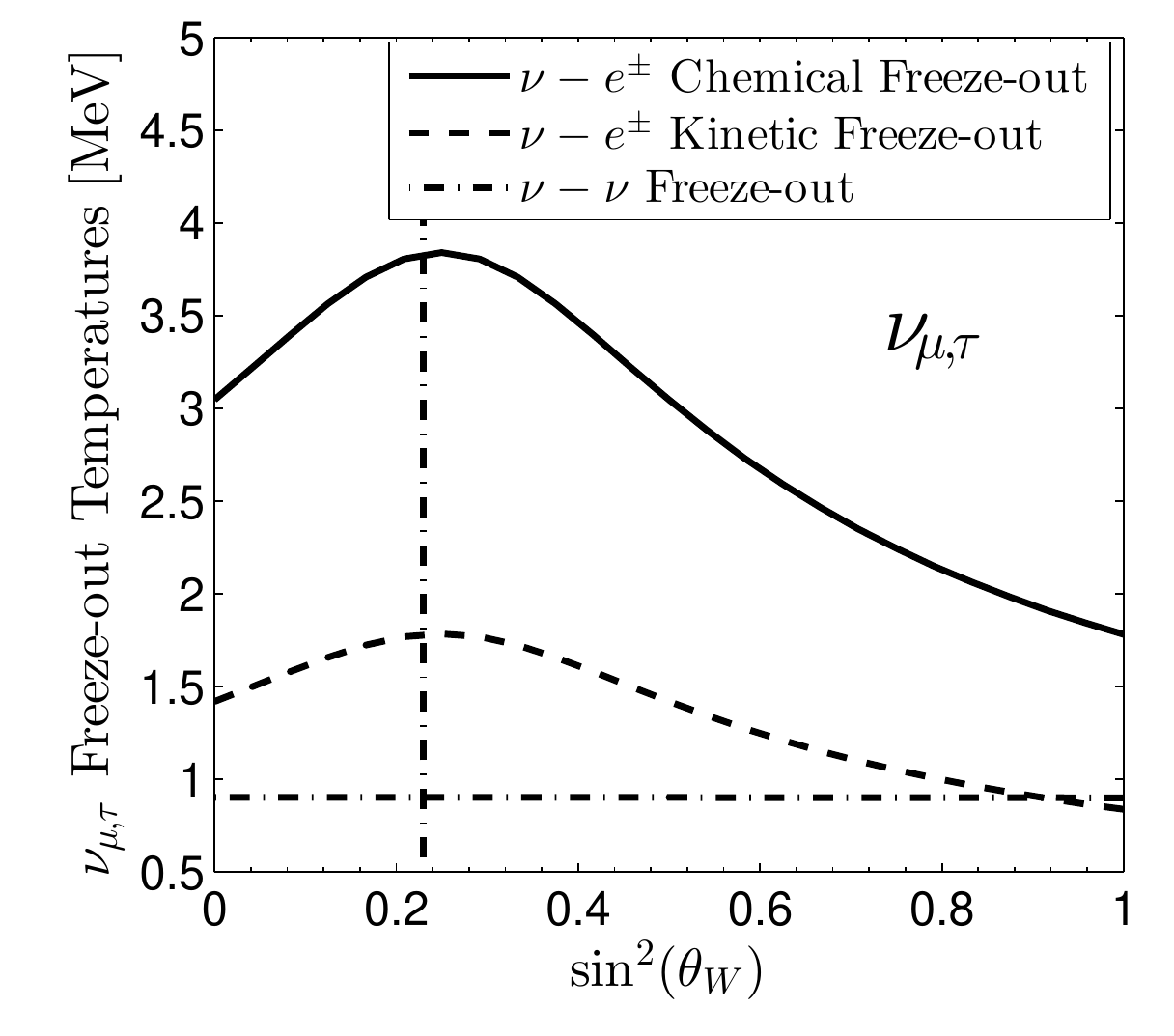}}
\centerline{\includegraphics[width=0.47\columnwidth]{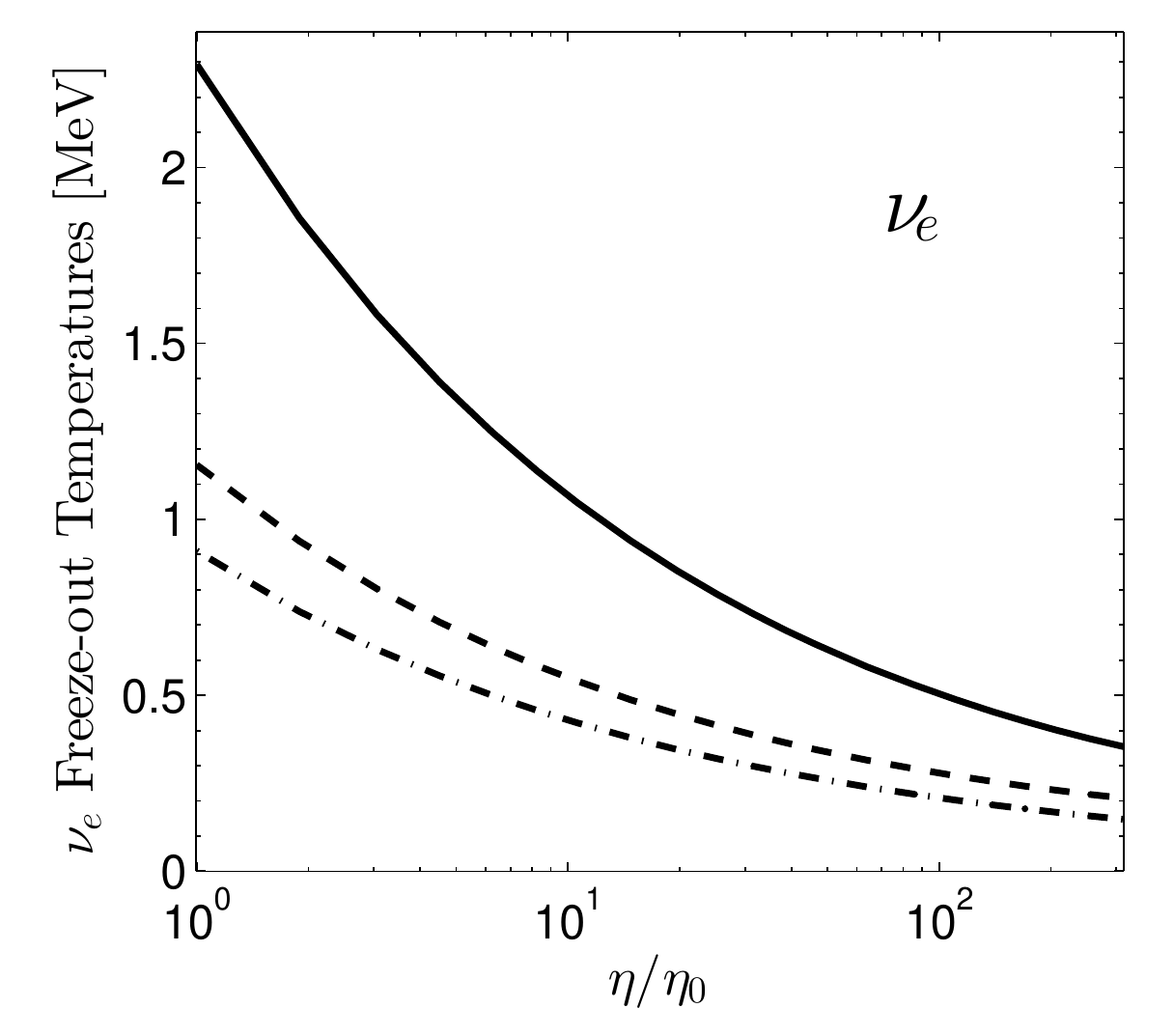}
\hspace{1mm}\includegraphics[width=0.47\columnwidth]{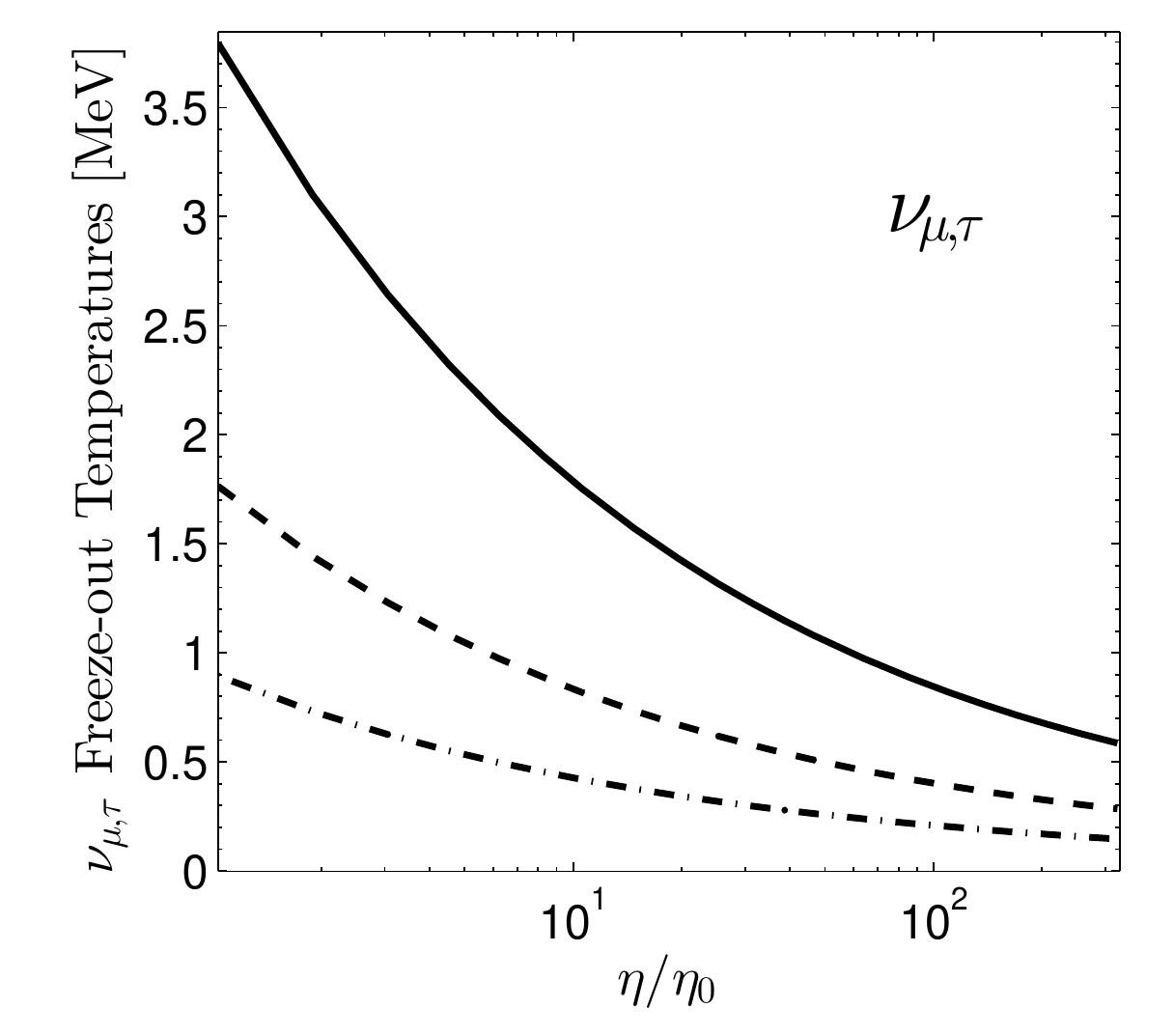}}
\caption{Freeze-out temperatures for electron neutrinos (left) and $\mu$, $\tau$ neutrinos (right) for the three types of freeze-out processes adapted from paper~\cite{Birrell:2014uka}. Top panels print temperature curves as a function of $\sin^2\theta_W$ for $\eta=\eta_0$, the vertical dashed line is $\sin^2\theta_W=0.23$; bottom panels are printed as a function of relative change in interaction strength $\eta/\eta_0$ obtained for $\sin^2\theta_W=0.23$.}
    \label{fig:freezeoutT}
 \end{figure}

 To estimate the freeze-out temperature we need to solve the Boltzmann equation with different types of collision terms. In~\cite{Birrell:2014uka} we detail a new method for analytically simplifying the collision integrals and show that the neutrino freeze-out temperature is controlled by standard model (SM) parameters. The freeze-out temperature depends only on the magnitude of the Weinberg angle in the form $\sin^2\theta_W$ , and a dimensionless relative interaction strength parameter $\eta$,
\begin{align}
\eta\equiv M_p m_e^3 G_F^2, \qquad M_p^2\equiv \frac{1}{8\pi G_N}, \end{align}
a combination of the electron mass $m_e$, Newton constant $G_N$ (expressed above in terms of Planck mass $M_p$), and the Fermi constant $G_F$. The dimensionless interaction strength parameter $\eta$ in the present-day vacuum has the value
\begin{align}
\eta_0\equiv \left.M_p m_e^3 G_F^2\right|_0 = 0.04421 .
\end{align}

The magnitude of $\sin^2\theta_W$ is not fixed within the SM and could be subject to variation as a function of time or temperature. In \rf{fig:freezeoutT} we show the dependence of neutrino freeze-out temperatures for $\nu_e$ and $\nu_{\mu,\tau}$ on SM model parameters $\sin^2\theta_W$ and $\eta$ in detail. The impact of SM parameter values on neutrino freeze-out and the discussion of the implications and connections of this work to other areas of physics, namely Big Bang Nucleosynthesis and dark radiation can be found in detail in~\cite{Dreiner:2011fp,Boehm:2012gr,Blennow:2012de,Birrell:2014uka}.

After neutrinos freeze-out, the neutrino co-moving entropy is independently conserved. However, the presence of electron-positron rich plasma until $T=20\keV$ provides the reaction $\gamma\gamma\to e^-e^+\to\nu\bar{\nu}$ to occur even after neutrinos decouple from the cosmic plasma. This suggests the small amount of $e^\pm$ entropy can still transfer to neutrinos until temperature $T=20\keV$ and can modify free streaming distribution and the effective number of neutrinos.

We expect that incorporating oscillations into the freeze-out calculation would yield a smaller freeze-out temperature difference between neutrino flavors as oscillation provides a mechanism in which the heavier flavors remain thermally active despite their direct production becoming suppressed. In work by Mangano et. al.~\cite{Mangano:2005cc}, neutrino freeze-out including flavor oscillations is shown to be a negligible effect.

\subsection{Effective number of neutrinos}\label{sec:EffectiveNeutrino}
\noindent The population of each flavor of neutrino is not a fixed quantity throughout the evolution of the Universe. In the earlier hot Universe, the population of neutrinos is controlled thermally and to maximize entropy, each flavor is equally filled. As the expansion factor $a(t)$ is radiation dominated for much of this period (see \rf{CosmicFraction}), the CMB is ultimately sensitive to the total energy density within the neutrino sector (which is sometimes referred to as the dark radiation contribution). 
This is described by the effective number of neutrinos $N_{\nu}^{\mathrm{eff}}$ which captures the number of relativistic degrees of freedom for neutrinos as well as any reheating that occurred in the sector after freeze-out. This quantity is related to the total energy density in the neutrino sector as well as the photon background temperature of the Universe via
\begin{align}\label{Neff}
N_\nu^{\mathrm{eff}}\equiv\frac{\rho^{\mathrm{tot}}_\nu}{\frac{7\pi^2}{120}\left(\frac{4}{11}\right)^{4/3}T_\gamma^4}\;,
\end{align}
where $\rho_\nu^{\mathrm{tot}}$ is the total energy density in neutrinos and $T_\gamma$ is the photon temperature. 

$N_\nu^{\mathrm{eff}}$ is defined such that three neutrino flavors with zero participation of neutrinos in reheating during $e^\pm$ annihilation results in $N_\nu^{\mathrm{eff}}=3$. The factor of $\left(4/11\right)^{1/3}$ relates the photon temperature to the (effective) temperature of the free-streaming neutrinos after $e^\pm$ annihilation, under the assumption of zero neutrino reheating. Strictly speaking, the number of true degrees of freedom is exactly determined by the number of neutrino families and available quantum numbers, therefore deviations of $N_{\nu}^{{\rm eff}}>3$ are to be understood as reheating which goes into the neutrino energy density $\rho_{\nu}^{{\rm tot}}$.

Experimentally, $N_\mathrm{eff}$ has been determined from CMB data by the Planck collaboration~\cite{Planck:2018vyg} in their 2018 analysis yielding $N_{\nu, exp}^{\mathrm{eff}}=2.99\pm0.17$ though this value has evolved substantially since their 2013 and 2015 analyses~\cite{Planck:2013pxb,Planck:2015fie}. Precise study of neutrino decoupling (as outlined in \rsec{sec:Freezeout}) and thus freeze-out can improve the predictions for the value of $N_\nu^{\mathrm{eff}}$. Many studies focus on improving the calculation of decoupling through various means such as
\begin{enumerate}
    \item Determining the dependence of freeze-out on the natural constants found in the Standard Model of particle physics~\cite{Birrell:2014uka,Birrell:2014ona}.
    \item The entropy transfer from electron-positron annihilation and finite temperature correction at neutrino decoupling~\cite{Dicus:1982bz,Heckler:1994tv,Fornengo:1997wa}.
    \item Neutrino decoupling with flavor oscillations~\cite{Mangano:2001iu,Mangano:2005cc}. Nonstandard neutrino interactions have been investigated, including neutrino electromagnetic~\cite{Morgan:1981zy,Fukugita:1987uy,Elmfors:1997tt,Vogel:1989iv,Mangano:2006ar,Giunti:2008ve} and nonstandard neutrino electron coupling~\cite{Mangano:2006ar}.
\end{enumerate}
As $N_\nu^\mathrm{eff}$ is only a measure of the relativistic energy density leading up to photon decoupling, a natural alternative mechanism for obtaining $N_\nu^\mathrm{eff}>3$ is the introduction of additional, presently not discovered, weakly interacting massless particles~\cite{Anchordoqui:2011nh,Abazajian:2012ys,Anchordoqui:2012qu,Steigman:2013yua,Giusarma:2014zza}. Alternatively, theories outside conventional freeze-out considerations have been proposed to explain the tension in $N_\mathrm{eff}$ including: QGP as the possible source of $N_\mathrm{eff}$ or connection between lepton asymmetry $L$ and $N_\nu^{\mathrm{eff}}$.

The natural consistency of the reported CMB range of $N_\nu^{\mathrm{eff}}$ with the range of QGP hadronization temperatures, motivates the exploration of a connection between $N_\nu^{\mathrm{eff}}$ and the decoupling of sterile particles at and below the QGP phase transition~\cite{Birrell:2014cja}. This demonstrates that that $N_\nu^{\mathrm{eff}}>3.05$ can be associated with the appearance of several light particles at QGP hadronization in the early Universe that either are weakly interacting in the entire space or is only allowed to interact within the deconfined domain, in which case their coupling would be strong. Such particles could leave a clear dark radiation experimental signature in relativistic heavy-ion experiments that produce the deconfined QGP phase.

In standard $\Lambda$CDM, the asymmetry between leptons and antileptons $L\equiv [N_\mathrm{L}-N_{\overline{\mathrm{L}}}] /N_\gamma $ (normalized with the photon number) is generally assumed to be small (nano-scale) such that the net normalized lepton number equals the net baryon number $L=B$ where $B=[N_\mathrm{B}-N_{\overline{\mathrm{B}}}]/N_\gamma $. Barenboim, Kinney, and Park~\cite{Barenboim:2016shh,Barenboim:2017dfq} note that the lepton asymmetry of the Universe is one of the most weakly constrained parameters is cosmology and they propose that models with leptogenesis are able to accommodate a large lepton number asymmetry surviving up to today. 

{\xblue If lepton number is grossly broken, this could provide a connection between cosmic neutrino properties and the baryon-antibaryon asymmetry present in the Universe today~\cite{Barenboim:2017dfq}.} We quantify in~\cite{Yang:2018oqg} the impact of large lepton asymmetry on Universe expansion and show that there is another \lq natural\rq\ choice $L\simeq 1$, making the net lepton number and net photon number in the Universe similar. Thus because $N_{\nu}^{{\rm eff}}$ can be understood as a characterization of the relativistic dark radiation energy content in the early Universe, independent of its source, there still remains ambiguity in regard to measurements of $N_{\nu}^{\rm eff}$.

\section{Electron-Positron Epoch}\label{sec:ElectronPositron}
\subsection{The last bastion of antimatter}\label{sec:ElectronPositronDensity}
\noindent The electron-positron epoch of the early Universe was home to Big Bang Nucleosynthesis (BBN), the annihilation of most electrons and positrons reheating both the photon and neutrino fields, as well as setting the stage for the eventual recombination period which would generate the cosmic microwave background (CMB). The properties of the electron-positron $e^{\pm}$ plasma in the early Universe has not received appropriate attention in an era of precision BBN studies~\cite{Pitrou:2018cgg}. The presence of $e^{\pm}$ pairs before and during BBN has been acknowledged by Wang, Bertulani and Balantekin~\cite{Wang:2010px,Hwang:2021kno} over a decade ago. This however was before necessary tools were developed to explore the connection between electron and neutrino plasmas~\cite{Mangano:2005cc,Birrell:2012gg,Birrell:2014uka}.

\begin{figure}[ht]
\centerline{\includegraphics[width=\textwidth]{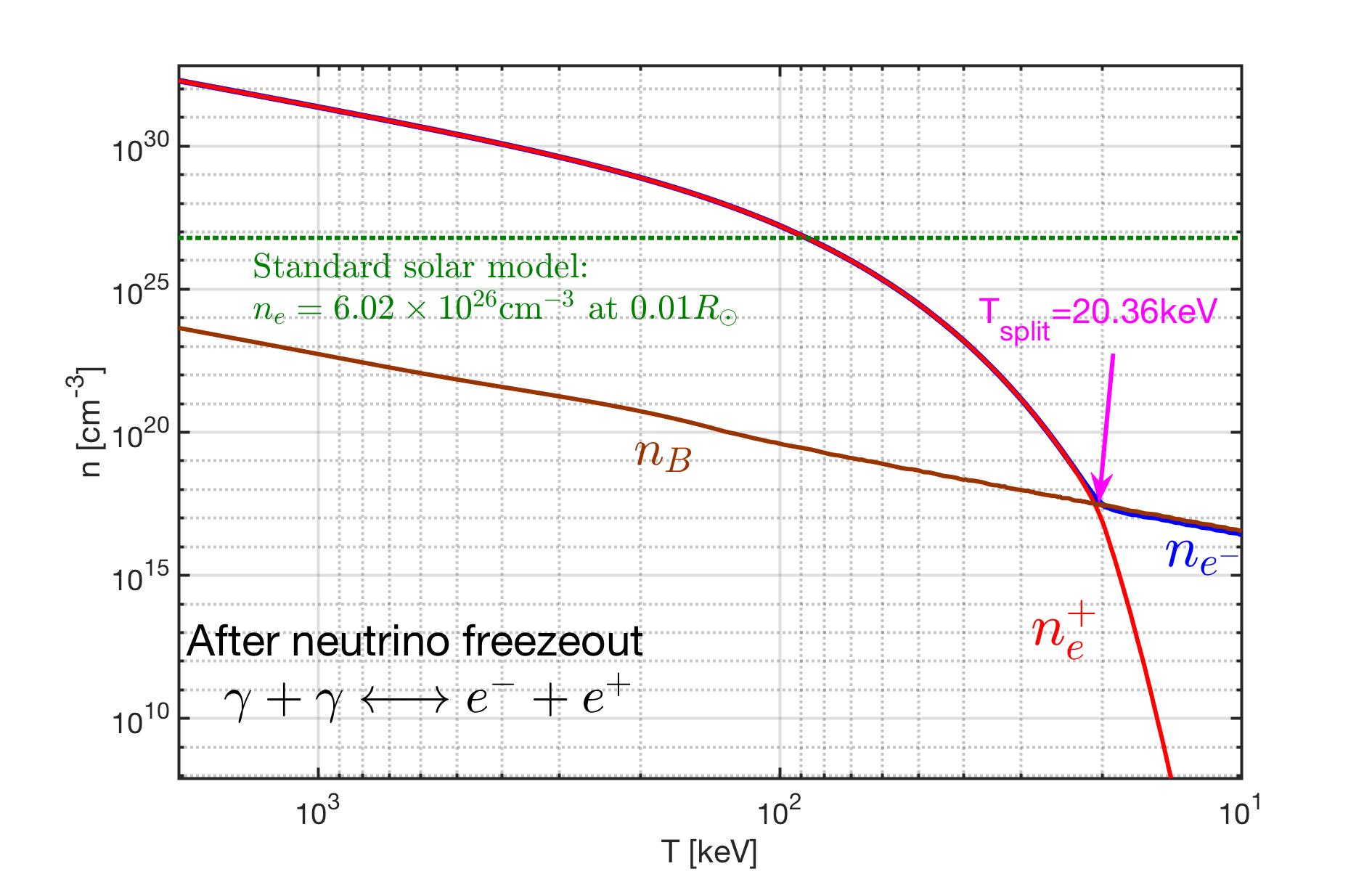}}
\caption{The $e^{\pm}$ number densities as a function of temperature in the range $2\,\mathrm{MeV}>T>10\,\mathrm{keV}$. The blue solid line is the electron density $n_{e^{-}}$, the red solid line is the positron density $n_{e^{+}}$, and the brown solid line is the baryon density $n_{B}$. For comparison, we also show the green dotted line as the solar electron density within the solar core~\cite{Bahcall:2000nu}.}
\label{Density_fig} 
\end{figure}

During the late stages of the $e^{\pm}$ epoch where BBN occurred, the matter content of the Universe was still mostly dominated by the light charged leptons by many orders of magnitude even though the Hubble parameter was still mostly governed by the radiation behavior of the neutrinos and photons. In \rf{Density_fig} we show that the dense $e^{\pm}$ plasma in the early Universe under the hypothesis charge neutrality and entropy conservation as a function of temperature $2\,\mathrm{MeV}>T>10\,\mathrm{keV}$~\cite{Chris:2023abc}. The plasma is electron-positron rich, i.e, $n_{e^\pm}\gg n_B$ in the early Universe until leptonic annihilation at $T_{\mathrm{split}} = 20.36\ \mathrm{keV}$. For $T<T_{\mathrm{split}}$ the positron density $n_{e^{+}}$ quickly vanishes because of annihilation leaving only a residual electron density as required by charge conservation.

\begin{figure}[ht]
\centerline{\includegraphics[width=0.9\textwidth]{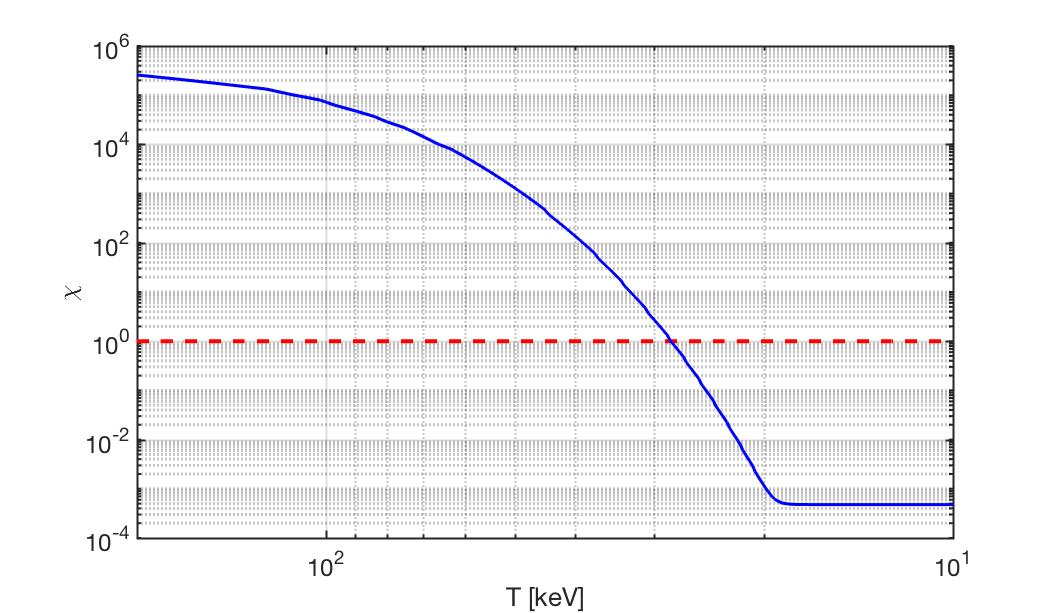}}
\caption{The energy density ratio $\chi$ (solid blue line) between $e^{\pm}$ and baryons as a function of temperature from $10\,\mathrm{keV}< T<200\,\mathrm{keV}$. The dashed red line crossing point represents where the baryon density exceeds that of the electron-positron pairs.}
\label{ratio_fig} 
\end{figure}

The temperatures during this epoch were also cool enough that the electrons and positrons could be described as partially non-relativistic to fairly good approximation while also still being as energy dense as the Solar core making it a relatively unique plasma environment not present elsewhere in cosmology. Considering the energy density between non-relativistic $e^{\pm}$ and baryons, we can write the ratio of energy densities as
\begin{align}\label{Eq_ratio}
\chi&\equiv\frac{\rho_e+\rho_{\bar e}}{\rho_p+\rho_n} \nonumber \\
&=\frac{m_e(n_e+n_{\bar e})}{m_pn_p+m_n n_n}=\frac{m_e(n_e+n_{\bar e})}{n_B(m_pX_p+m_nX_n)}=\left(\frac{n_e+n_{\bar e}}{n_B}\right)\,\left(\frac{m_e}{m_pX_p+{m_n X_\alpha}/2}\right)\,,
\end{align}
where we consider all neutrons as bound in $^4H_e$ after BBN. Species ratios $X_p=n_p/n_B$ and $X_\alpha=n_\alpha/n_B$ are given by the PDG~\cite{ParticleDataGroup:2022pth} as
\begin{align}
X_p=0.878,\qquad X_\alpha=0.245\,,
\end{align}
with masses
\begin{align}
m_e=0.511\,\mathrm{MeV}, \qquad m_p=938.272\,\mathrm{MeV},\qquad m_n=939.565\,\mathrm{MeV}\,.
\end{align}
In \rf{ratio_fig} we plot the energy density ratio \req{Eq_ratio} as a function of temperature $10\,\mathrm{keV}< T<200\,\mathrm{keV}$. This figure shows that the energy density of electron and positron is dominant until $T=28.2\keV$, i.e., at higher temperatures we have $\rho_{e}\gg\rho_B$. Until around $T\approx85\keV$, the $e^{\pm}$ number density remained higher than that of the solar core, though notably at a much higher temperature than the Sun's core of $T_{\odot}=1.36\keV$~\cite{Castellani:1996cm}. After $T=28.2\keV$, where $\rho_{e}\ll\rho_B$, the ratio becomes constant around $T=20\keV$ because of positron annihilation and charge neutrality.

\subsection{Cosmic magnetism}\label{sec:Energy}
\noindent The Universe today filled with magnetic fields~\cite{Kronberg:1993vk} at various scales and strengths both within galaxies and in deep extra-galactic space far and away from matter sources. Extra-galactic magnetic fields (EGMF) are not well constrained today, but are required by observation to be non-zero~\cite{Anchordoqui:2001bs,Widrow:2002ud} with a magnitude between $10^{-12}\ \mathrm{T}>B_{EGMF}>10^{-20}\ \mathrm{T}$ over Mpc coherent length scales. The upper bound is constrained from the characteristics of the CMB while the lower bound is constrained by non-observation of ultra-energetic photons from blazars~\cite{Neronov:2010gir}. There are generally considered two possible origins~\cite{Widrow:2011hs,Vazza:2021vwy} for extra-galactic magnetic fields: (a) matter-induced dynamo processes involving Amperian currents and (b) primordial (or relic) seed magnetic fields whose origins may go as far back as the Big Bang itself. It is currently unknown which origin accounts for extra-galactic magnetic fields today or if it some combination of the two models. Even if magnetic fields in the Universe today are primarily driven via amplification through Amperian matter currents, {\xblue such models could still benefit from the presence of primordial fields to act as catalyst. The purpose of this section is then to consider the magnetization properties of the $e^{\pm}$ plasma period due to spin which has not yet been considered. 

While matter (and thus electrons) are relatively dilute today, the early Universe plasmas contained relatively large quantity of both matter $(e^-)$ and antimatter $(e^+)$. We explore here the spin response of the electron-positron plasma to external and self-magnetization fields thus developing methods for future detailed study.}

\begin{figure}[htbp]
 \centering \includegraphics[trim=110 50 120 40,clip,width=\textwidth]{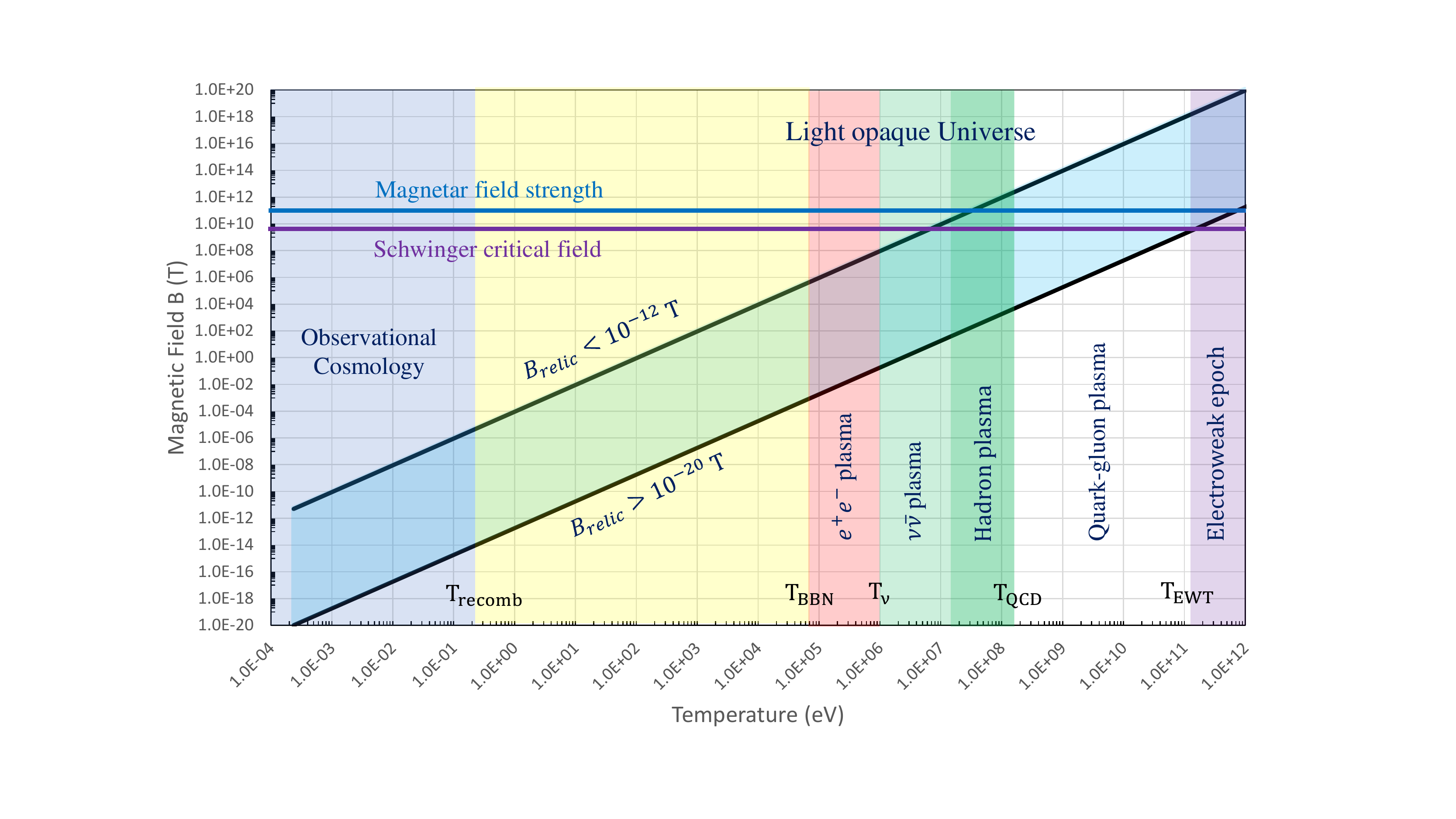}
 \caption{Qualitative value of the primordial magnetic field over the evolutionary lifespan of the Universe. The upper and lower black lines represent extrapolation of the EGMF bounds into the past. The major phases of the Universe are indicated with shaded regions. The values of the Schwinger critical field (purple line) and the upper bound of surface magnetar field strength (blue line) are included for scale.\label{relic_plot}}
\end{figure}

As magnetic flux is conserved over co-moving surfaces, we see in \rf{relic_plot} that the primordial relic field is expected to dilute as $B\propto1/a(t)^{2}$. This means the contemporary small bounded values of $5\times10^{-12}\ \mathrm{T}>B_{relic}>10^{-20}\ \mathrm{T}$ (coherent over $\mathcal{O}(1\ \mathrm{Mpc})$ distances) may have once represented large magnetic fields in the early Universe. Therefore, correctly describing the dynamics of this $e^{\pm}$ plasma is of interest when considering modern cosmic mysteries such as the origin of extra-galactic magnetic fields~\cite{Anchordoqui:2001bs,Neronov:2010gir}. While most approaches tackle magnetized plasmas from the perspective of classical or semi-classical magneto-hydrodynamics (MHD)~\cite{berezhiani1992influence,berezhiani1995large,Schlickeiser:2018hzq,melrose2008quantum}, our perspective is to demonstrate that fundamental quantum statistical analysis can lead to further insights on the behavior of magnetized plasmas.

As a starting point, we consider the energy eigenvalues of charged fermions within a homogeneous magnetic field. Here, we have several choices: We could assume the typical Dirac energy eigenvalues with gyro-magnetic g-factor set to $g=2$. But as electrons, positrons and most plasma species have anomalous magnetic moments (AMM), we require a more complete model. {\xblue Particle dynamics of classical particles with AMM are explored in~\cite{Rafelski:2017hce,Formanek:2017mbv,Formanek:2020zwc,Formanek:2021mcp}.} Another option would be to modify the Dirac equation with a Pauli term~\cite{thaller2013dirac}, often called the Dirac-Pauli (DP) approach, via
\begin{align}
 \label{Pauli} \hat{H}_{\mathrm{AMM}} = -a\frac{e}{2m_{e}}\frac{\sigma_{\mu\nu}F^{\mu\nu}}{2}\,,
\end{align}
where $\sigma_{\mu\nu}$ is the spin tensor proportional to the commutator of the gamma matrices and $F^{\mu\nu}$ is the EM field tensor. For the duration of this section, we will remain in natural units $(\hbar=c=k_{B}=1)$ unless explicitly stated otherwise. The AMM is defined via g-factor as
\begin{align}
 \label{AMM} \frac{g}{2}=1+a\,.
\end{align}
This approach, while straightforward, would complicate the energies making analytic understanding and clarity difficult without a clear benefit. Modifying the Dirac equation with \req{Pauli} yields the following eigen-energies
\begin{align}
 \label{DPEnergy} E_{n}^{s}\vert_{DP}=\sqrt{\left(\sqrt{m_{e}^{2}+2eB\left(n+\frac{1}{2}-s\right)}-\frac{eB}{2m}(g-2)s\right)^{2}+p_{z}^{2}}
\end{align}
This model for the electron-positron plasma of the early Universe has been used in work such as Strickland et. al.~\cite{Strickland:2012vu}. Our work in this section is then in part a companion piece which compares and contrasts the DP model of fermions to our preferred model for the AMM via the Klein-Gordon-Pauli (KGP) equation given by
\begin{alignat}{1}
 \label{KGP} \left(\left(i\partial_{\mu}-eA_{\mu}\right)^{2}-m_{e}^{2}-e\frac{g}{2}\frac{\sigma_{\mu\nu}F^{\mu\mu}}{2}\right)\Psi=0\,.
\end{alignat}
We wish to emphasize, that each of the three above models (Dirac, DP, KGP) are distinct and have differing physical consequences and are not interchangeable which we explored in the context of hydrogen-like atoms in~\cite{Steinmetz:2018ryf}. Recent work done in~\cite{rafelski2023study} discuss the benefits of KGP over other approaches for $g\neq2$ from a quantum field theory perspective. Exploring the statistical behavior of KGP in a cosmological context can lead to new insights in magnetization which may be distinguished from pure $g=2$ behavior of the Dirac equation or the \emph{ad hoc} modification imposed by the Pauli term in DP. One major improvement of the KGP approach over the more standard DP approach is that the energies take eigenvalues which are mathematically similar to the Dirac energies. Considering the $e^\pm$ plasma in a uniform magnetic field $B$ pointing along the $z$-axis, the energy of $e^\pm$ fermions can be written as
\begin{align}
 \label{KGPEnergy} 
 E_{n}^{s}&=\sqrt{p^2_z+\tilde{m}^2+2eBn},\\ \nonumber
 \tilde{m}^2&=m^2_e+eB\left(1-gs\right),\qquad s=\pm\frac{1}{2},\qquad n=0,1,2,3,\dots
\end{align}
where $n$ is the principle quantum number for the Landau levels and $s$ is the spin quantum number. Here we introduce a notion of effective mass $\tilde{m}$ which inherits the spin-specific part of the energy adding them to the mass. This convention is also generalizable to further non-minimal electromagnetic models with more exotic energy contributions such that we write a general replacement as
\begin{align}
 \label{MagMass} m_{e}^{2}\rightarrow\tilde{m}^2(B)\,.
\end{align}
This definition also pulls out the ground state Landau energy separating it from the remainder of the Landau tower of states. One restriction is that the effective mass must remain positive definite in our analysis thus we require
\begin{align}
 \label{MassLimit} \tilde{m}^2(B)=m^2_e+eB\left(1-gs\right)>0\,.
\end{align}
This condition fails under ultra-strong magnetic fields of order
\begin{align}
 \label{MagMassFail} B_{\mathrm{crit}}=\frac{m_{e}^{2}}{ea}=\frac{\mathcal{B}_{S}}{a}\approx3.8\times10^{12}\ \mathrm{T}\,,
\end{align}
where $\mathcal{B}_{S}$ is the Schwinger critical field strength. For electrons, this field strength is well above the window of magnetic field strengths of interest during the late $e^{\pm}$ epoch.

\subsection{Landau eigen-energies in cosmology}\label{sec:Landau}
\noindent There is another natural scale for the magnetic field besides \req{MagMassFail} when considering the consequences of FLRW expansion on the $e^{\pm}$ gas. As the Universe expands, different terms in the energies and thus partition function evolve as a function of the scale factor $a(t)$ which arises in the FLRW metric. We can consider the expansion to be an adiabatic process which results in a smooth shifting of the relevant dynamical quantities. From the conservation of magnetic flux through a co-moving surface, the magnetic field under expansion starting at some initial time $t_{0}$ is given by
\begin{alignat}{1}
 \label{BScale} B(t) = B(t_{0})\frac{a(t_{0})^{2}}{a(t)^{2}}\,.
\end{alignat}
As the Universe expands, the temperature also cools as the cosmological redshift reduces the momenta of particles in the Universe lowering their contribution to the energy content of the Universe. This cosmological redshift is written as
\begin{alignat}{1}
 \label{Redshift} p_{i}(t) = p_{i}(t_{0})\frac{a(t_{0})}{a(t)}\,,\qquad T(t) = T(t_{0})\frac{a(t_{0})}{a(t)}\,.
\end{alignat}
The momenta scale with the same factor as temperature as it is the origin of cosmological redshift. The energy of massive free particles in the Universe scales differently based on their momentum (and thus temperature). When hot and relativistic, particle energy scales with inverse scale factors like radiation. However as particles transition to non-relativistic momenta, their energies scale with the inverse square of the scale factor like magnetic flux.
\begin{alignat}{1}
 \label{EScale} E(t) = E(t_{0})\frac{a(t_{0})}{a(t)}\xrightarrow{\mathrm{NR}}\ E(t_{0})\frac{a(t_{0})^{2}}{a(t)^{2}}\,.
\end{alignat}
This occurs because of the functional dependence of energy on momentum in the relativistic versus non-relativistic cases. The argument in the Boltzmann statistical factor is given by
\begin{alignat}{1}
 \label{Boltz} X_{n}^{s}\equiv\frac{E_{n}^{s}}{T}\,.
\end{alignat}
We can explore this relationship for the magnetized system explicitly by writing out \req{Boltz} using the KGP eigen-energies as
\begin{alignat}{1}
 \label{XExplicit} X_{n}^{s} = \sqrt{\frac{m_{e}^{2}}{T^{2}}+\frac{p_{z}^{2}}{T^{2}}+\frac{2eB}{T^{2}}\left(n+\frac{1}{2}-\frac{gs}{2}\right)}\,,
\end{alignat}
where we now introduce the expansion scale factor via \req{BScale} - \req{Redshift}. The Boltzmann factor can then be written as
\begin{alignat}{1}
 \label{XScale} X_{n}^{s}[a(t)] = \sqrt{\frac{m_{e}^{2}}{T^{2}(t_{0})}\frac{a(t)^{2}}{a(t_{0})^{2}}+\frac{p_{z}^{2}(t_{0})}{T^{2}(t_{0})}+\frac{2eB(t_{0})}{T^{2}(t_{0})}\left(n+\frac{1}{2}-\frac{gs}{2}\right)}\,.
\end{alignat}
This reveals that only the mass contribution is dynamic over cosmological time. For any given eigen-state, the mass term increases driving the state into the non-relativistic limit while the momenta and magnetic contributions are frozen by initial conditions.

Following reasoning outlined in~\cite{rafelski2023study} and~\cite{Steinmetz:2018ryf} we will proceed using the KGP eigen-energies. Motivated by \req{XScale}, we can introduce a dimensionless cosmic magnetic scale which is frozen in the homogeneous case as
\begin{alignat}{1}
 \label{Bo} b_{0}\equiv\frac{eB}{T^{2}}=\frac{eB\hbar c^{2}}{(k_{B}T)^{2}}\ \mathrm{(S.I)}\,,
\end{alignat}
where we've included the expression explicitly in full SI units. We can estimate the value of $b_{0}$ from the bounds of the extra-galactic magnetic field strength and the temperature of the Universe today. If the origin of deep space extra-galactic magnetic fields are relic fields from the early Universe, which today are expected to exist between $5\times10^{-12}\ \mathrm{T}>B_{relic}>10^{-20}\ \mathrm{T}$, then at temperature $T=2.7\ \mathrm{K}$, the value of the cosmic magnetic scale is between
\begin{alignat}{1}
 \label{BoScale} 5.5\times10^{-3}>b_{0}>1.1\times10^{-11}\,.
\end{alignat}
This should remain constant in the Universe at-large up to the last epoch the Universe was sufficiently magnetized to disturb this value. As the electron-proton $(e^{-}p)$ plasma which generated the CMB was relatively dilute over its duration, it was unlikely sufficiently magnetized to significantly alter this value over extra-galactic scales. Rather, the first candidate plasma, going backwards in time, to have been sufficiently magnetized and dense to have set the relic field magnetic scale would have been the electron-positron plasma which existed during the duration of Big Bang Nucleosynthesis (BBN) and beforehand.

Higher order non-minimal magnetic contributions which can be introduced via \req{MagMass} to the eigen-energies like $\approx\mu_{B}^{2}B^{2}/T^{2}$ are even more suppressed over cosmological time which drives the system into minimal electromagnetic coupling with the exception of the anomalous magnetic moment in the KGP eigenenergies. It is interesting to note that cosmological expansion serves to \lq\lq smooth out\rq\rq\ the characteristics of more complex BSM electrodynamics erasing them from a statistical perspective in favor of the minimal or minimal-like dynamics. As $b_0$ is a constant of expansion, assuming the electron-proton plasma between the CMB and electron-positron annihilation did not greatly disturbed it, we can calculate the remnant values at the temperature $T=50\ \mathrm{keV}$ (which takes place in the middle of BBN) with the expression
\begin{align}
 \label{BBNFields} B(T)=\frac{b_{0}}{e}T^{2}\,,
\end{align}
yielding a range of field strengths
\begin{align}
 \label{BBNRange} 2.3\times10^{5}\ \mathrm{T}>B(T=50\ \mathrm{keV})>4.6\times10^{-4}\ \mathrm{T}\,,
\end{align}
during which the electron-positron plasma in the Universe had a number density comparable to that of the Solar core~\cite{Bahcall:2000nu}. We note that while the density of leptons is comparable to that of the solar core during this period, the temperature is not. The $e^{\pm}$ plasma during BBN was far hotter than the solar core's comparatively cool temperature of $T_{\odot}=1.37\keV$~\cite{Castellani:1996cm}.

\subsection{Electron-positron statistical physics}\label{sec:Partition}
\noindent We now turn our attention now to the statistical behavior of the $e^{\pm}$ system. We can utilize the general fermion partition function given by~\cite{Elze:1980er}
\begin{align}
 \label{PartFunc} \ln\mathcal{Z}=\sum_{\alpha}\ln\left(1+e^{-\beta(E-\eta)}\right)\,,
\end{align}
where $\beta=1/T$, $\alpha$ is the set of all quantum numbers in the system, and $\eta$ is the generalized chemical potential. The magnetized $e^{\pm}$ system should be considered a system of four quantum species: Particles and antiparticles, and spin aligned and anti-aligned. Taken together we consider a system where all electrons and positrons are spin aligned or anti-aligned with the magnetic field $B$ and the partition function of the system is written as
\begin{align}
 \label{PartFuncB}\ln\mathcal{Z}_{tot}=\frac{2eBV}{(2\pi)^2}\sum_{\sigma}^{\pm1}\sum_{s}^{\pm1/2}\sum_{n=0}^\infty\int^\infty_{0}dp_z\left[\ln\left(1+\Upsilon_{\sigma}^{s}(x)e^{-\beta E_{n}^{s}}\right)\right]\,,\\
 \label{Fugacity}\Upsilon_{\sigma}^{s}(x)=\gamma(x)\lambda_{\sigma}^{s}\,,\qquad\lambda_{\sigma}^{s}=e^{(\sigma\eta_{e}+s\eta_{s})/T}\,,
\end{align}
where $\eta_{e}$ is the electron chemical potential and $\eta_s$ is the spin chemical potential for the generalized fugacity $\lambda_{\sigma}^{s}$. The parameter $\gamma(x)$ is a spatial field which controls the distribution inhomogeneity of the Fermi gas. Inhomogeneities can arise from the influence of other forces on the gas such as gravitational forces. Deviations of $\gamma\neq1$ represent configurations of reduced entropy (maximum entropy yields the normal Fermi distribution itself with $\gamma=1$) without pulling the system off a thermal temperature.

This situation is similar to that of the quarks during QGP, but instead the deviation is spatial rather than in time. This is precisely the kind of behavior that may arise in the $e^{\pm}$ epoch as the dominant photon thermal bath keeps the Fermi gas in thermal equilibrium while spatial inhomogeneity could spontaneously develop. For the remainder of this work, we will retain $\gamma(x)=1$. The energy $E_{n}^\pm$ can be written as
\begin{align}
E_{n}^\pm&=\sqrt{p^2_z+\tilde m^2_\pm+2eBn},\qquad\tilde{m}^2_\pm=m^2_e+eB\left(1\mp\frac{g}{2}\right)\,,
\end{align}
where the $\pm$ script refers to spin aligned and anti-aligned eigenvalues. As we are interested in the temperature domain $T=50\ \mathrm{keV}$, we can consider a semi-relativistic approach obtained by the Boltzmann approximation. Taking the limit $m_{e}/T\ll1$, we obtain the first order Boltzmann approximation for semi-relativistic electrons and positrons. The Euler-Maclaurin formula is used to replace the sum over Landau levels with an integration which lets us split the partition function into three segments
\begin{align}
 \ln\mathcal{Z}_{tot}=\ln\mathcal{Z}_{free}+\ln\mathcal{Z}_B+\ln\mathcal{Z}_R\,,
\end{align}
where we define 
\begin{align}
    \label{FreePart}&\ln\mathcal{Z}_{free}=\frac{T^3V}{2\pi^2}\sum_{i=\pm}\left[2\cosh{\left(\frac{\eta_{e}^{i}}{T}\right)}\right]x_i^2K_2\left(x_i\right)\,,\qquad x_i=\frac{\tilde{m}_i}{T}\\
    \label{MagPart}&\ln\mathcal{Z}_B=\frac{eBTV}{2\pi^2}\sum_{i=\pm}\left[2\cosh{\left(\frac{\eta_{e}^{i}}{T}\right)}\right]\bigg[\frac{x_i}{2}K_1\left(x_i\right)+\frac{b_0}{12}K_0\left(x_i\right)\bigg]\,,\\
    \label{ErrorPart}&\ln\mathcal{Z}_R=\frac{eBTV}{\pi^2}\sum_{i=\pm}\left[2\cosh{\left(\frac{\eta_{e}^{i}}{T}\right)}\right]R.
\end{align}
The parameter $R$ is the error remainder which is defined by integrals over Bernoulli polynomials. {\xblue The parameter $\eta_{e}^{\pm}$ indicates that the chemical potential may be modified by the spin chemical potential and is in general non-zero as defined in \req{Fugacity}.}

While this would require further derivation to demonstrate explicitly, the benefit of the Euler-Maclaurin approach is if the error contribution remains finite or bound for the magnetized partition function, then a correspondence between the free Fermi partition function (with noticeably modified effective mass $\tilde{m}_{\pm}$) and the magnetized Fermi partition function can be established. The mismatch between the summation and integral in the Euler-Maclaurin formula would then encapsulate the immediate magnetic response and deviation from the free particle phase space.

While we label $\ln\mathcal{Z}_{free}$ in \req{FreePart} as the \lq\lq free\rq\rq\ partition function, this is not strictly true as this contribution to the overall partition function is a function of the effective mass we defined earlier in \req{MagMass}. When determining the magnetization of the quantum Fermi gas, derivatives of the magnetic field $B$ will not fully vanish on this first term which will resulting in an intrinsic magnetization which is distinct from the contribution from the ground state and mismatch between the quantized Landau levels and the continuum of the free momentum. Specifically, this free Fermi contribution represents the magnetization that uniquely arises from the spin magnetic energy rather than orbital contributions.

Assuming the error remainder $R$ is small and can be neglected, we can rewrite \req{FreePart} - \req{MagPart} obtaining
\begin{align}
 \label{lnZ}
 \ln\mathcal{Z}_{tot}=\frac{T^3V}{2\pi^2}\sum_{i=\pm}\left[2\cosh\left(\frac{\eta_{e}^{i}}{T}\right)\right]\left\{x_i^{2} K_2\left(x_i\right)+\frac{b_0}{2}x_iK_1\left(x_i\right)+\frac{b^2_0}{12}K_0\left(x_i\right)\right\}\,.
\end{align}
\req{lnZ} is a surprisingly compact expression containing only tractable functions and will be our working model for the remainder of the work. Note that the above does not take into consideration density inhomogeneities and is restricted to the domain where the plasma is well described as a Maxwell-Boltzmann distribution. With that said, we have not taken the non-relativistic expansion of the eigen-energies.

\subsection{Charge neutrality and chemical potential}\label{sec:ChargeNeutrality}
\noindent We explore the chemical potential of dense magnetized electron-positron plasma in the early Universe under the hypothesis of charge neutrality and entropy conservation. {\xblue To learn about orders of magnitude we set in the following $\eta_{e}=\eta_{e}^{+}=\eta_{e}^{-}$ and focus on the interval in the post-BBN temperature range $50\keV>T>20\keV$. We return to the full problem under separate cover.} The charge neutrality condition can be written as
\begin{align}
 \label{density_proton}
 \left(n_{e}-n_{\bar{e}}\right)=n_{p}=\left(\frac{n_{p}}{n_{B}}\right)\,\left(\frac{n_{B}}{s_{\gamma,\nu,e}}\right)\,s_{\gamma,\nu,e}= X_p\left(\frac{n_B}{s_{\gamma,\nu}}\right)\,s_{\gamma,\nu},\qquad X_p\equiv\frac{n_p}{n_B}\,,
\end{align}
where $n_{p}$ and $n_B$ is the number density of protons and baryons respectively.

{\xblue The radiation entropy component is given by $s_{\gamma,\nu}$.} The entropy density contribution of $e^\pm$ is negligible compared to the photon and neutrino entropy density at post-BBN temperatures $50\keV>T>20\keV$ because the low densities of $n_e\ll n_{\gamma,\nu}$ relative to the photon and neutrino gasses. The entropy density can be written as~\cite{kolb1990early}
\begin{align}
s=\frac{2\pi^2}{45}g_sT_\gamma^3,\qquad g_s=\sum_{i=boson}g_i\left(\frac{T_i}{T_\gamma}\right)^3+\frac{7}{8}\sum_{i=fermion}g_i\left(\frac{T_i}{T_\gamma}\right)^3\,,
\end{align}
where $g_s$ is the effective degree of freedom that contribute from boson and fermion species. The parameters $X_p$ and $(n_B/s)$ (see \req{BdS}) can be determined by the observation, yielding $X_p=0.878\pm0.015$~\cite{ParticleDataGroup:2022pth}. The net number density of electrons can be obtained by using the partition function of electron-positron plasma in the Boltzmann limit \req{lnZ} (with $g=2$) as follows:
\begin{align}\label{NetElectron}
\left(n_e-n_{\bar e}\right) 
&=\frac{T}{V}\frac{\partial}{\partial \eta_{e}}\ln\mathcal{Z}_{tot}\\ \nonumber
&=\frac{T^3}{2\pi^2}\left[2\sinh{(\eta_{e}/T)}\right]\sum_{i=\pm}\left[x_i^2K_2(x_i)+\frac{b_0}{2}x_i K_1(x_i)+\frac{b^2_0}{12}K_0(x_i)\right]\,.
\end{align}
Substituting \req{NetElectron} into the charge neutrality condition \req{density_proton} we can solve the chemical potential of electron $\eta_e/T$ yielding
\begin{align}\label{ChemicalPotential}
\sinh{(\eta_{e}/T)}&=\frac{2\pi^2}{2T^3}\,\frac{X_p(n_B/s_{\gamma,\nu})s_{\gamma,\nu}}{\sum_{i=\pm}\left[x_i^2K_2(x_i)+\frac{b_0}{2}x_i K_1(x_i)+\frac{b^2_0}{12}K_0(x_i)\right]}\,,\\
&\longrightarrow\frac{2\pi^2n_p}{2T^3}\,\frac{X_p(n_B/s_{\gamma,\nu})s_{\gamma,\nu}}{2x^2K_2(x)},\qquad x=m_e/T,\qquad \mathrm{for}\,\,b_0=0\label{ChemiticalPotential_000}\,.
\end{align}
We see in \req{ChemiticalPotential_000} that for the case $b_0=0$, the chemical potential agrees with the free particle result in~\cite{Chris:2023abc}. 

\subsection{Magnetization of the electron-positron plasma}\label{sec:Magnetization}
\noindent {\xblue We consider the electron-positron plasma in the mean field approximation where the external field is representative of the \lq\lq bulk\rq\rq\ internal magnetization of the gas. Each particle is therefore responding to the averaged magnetic flux generated by its neighbors as well as any global external field contribution. Considering the magnetized electron-positron partition function \req{lnZ} we introduce dimensionless magnetization in S.I units and the critical field as follows
\begin{align}
\label{Mdef}
\frac{M}{H_{c}}=\frac{1}{H_{c}}\frac{k_{B}T}{V}\frac{\partial \ln\mathcal{Z}_{tot}}{\partial B}\,,\qquad H_{c}&=\frac{B_{c}}{\mu_{0}}\qquad B_{c}=\frac{m_{e}^{2}c^{4}}{e\hbar c^{2}}\,.
\end{align}
Applying \req{Mdef} to \req{lnZ} we arrive at the expression
\begin{align}\label{Magnetization}
 M_{\pm}&=\frac{eT^{2}}{2\pi^2}\left[2\cosh\left(\frac{\eta_{e}}{T}\right)\right]\left\{c_{1}(x_{\pm})K_1(x_i)+c_{0}K_0(x_\pm)\right\}\,,\\
 c_{1}(x_{\pm}) &= \left[\frac{1}{2}-\left(\frac{1}{2}\pm\frac{g}{4}\right)\left(1+\frac{b^2_0}{12x^2_\pm}\right)\right]x_\pm\,,\qquad c_{0} = \left[\frac{1}{6}-\left(\frac{1}{4}\pm\frac{g}{8}\right)\right]b_0\,.
\end{align}
}

\begin{figure}[t]
\centering
\includegraphics[width=0.9\textwidth]{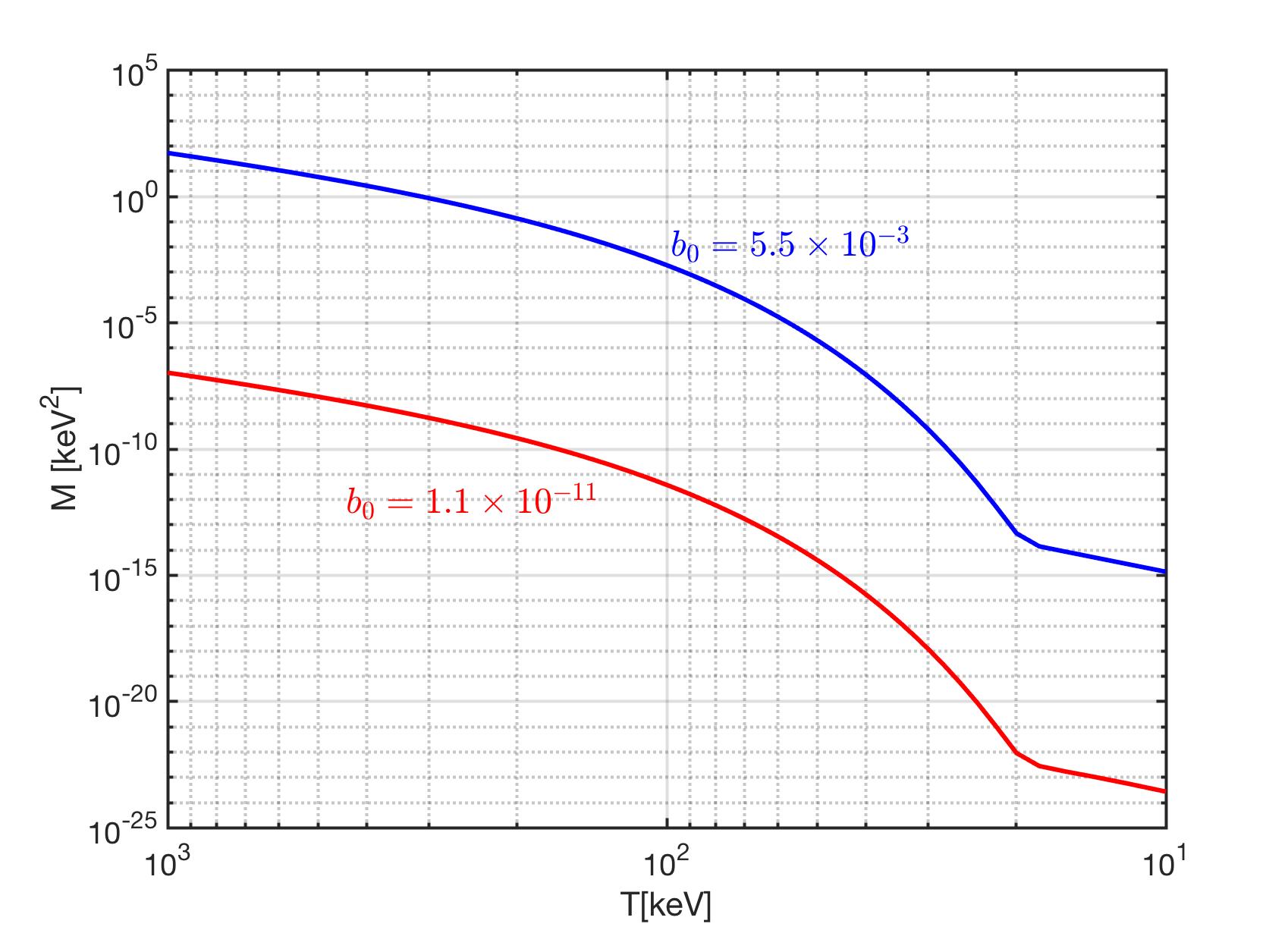}
\caption{{\xblue Estimate for the spin magnetization as a function of temperature in the range $10^{3}\keV>T>10\keV$, see text for detail.}}
\label{Case2_fig} 
\end{figure}

Substituting the chemical potential \req{ChemicalPotential} into \req{Magnetization} we can solve the magnetization $M$ numerically.
Considering the case $g=2$ the magnetization can be written as the sum of the aligned and anti-aligned polarizations {\xblue
\begin{align}
M=M_{+}+M_{-}\,,
\end{align}
}where the functions $M_\pm$ are defined as 
\begin{itemize}
\item[A.] The aligned polarized gas is described by $\tilde m_+=m_e$ and $x=\tilde m_+/T$. The magnetization of this contribution is therefore
\begin{align}\label{Magnetization_002}
M_{+}&=\frac{eT^{2}}{\pi^2}\sqrt{1+\sinh^2(\eta_e/T)}\left(\frac{1}{2}x_+K_1(x_+)+\frac{b_{0}}{6}K_0(x_+)\right)
\end{align}
\item[B.] The spin anti-aligned gas has effective masses $\tilde m_-=\sqrt{m^2_e+2eB}$, and $x_-=\tilde m_-/T$. This yields a magnetization contribution of
\begin{align}\label{Magnetization_001}
M_{-}&=-\frac{eT^{2}}{\pi^2}\sqrt{1+\sinh^2(\eta_e/T)}\left[\left(\frac{1}{2}+\frac{b_{0}^{2}}{12x_-^2}\right)x_-K_1(x_-)+\frac{b_{0}}{3}K_0(x_-)\right]
\end{align}
\end{itemize}
Using the cosmic magnetic scale parameter $b_0$ and chemical potential $\eta_e/T$ we solve the magnetization numerically. In \rf{Case2_fig}, {\xblue we present the outcome of this estimate. The solid lines (red for the lower bound of $b_{0}$ and blue for the higher bound of $b_{0}$) showing that the magnetization depends on the magnetic scale $b_0$.}

\section{Looking in the Cosmic Rear-view Mirror}\label{sec:Summary}
\noindent The present day Universe seems devoid of antimatter but the primordial Universe was nearly matter-antimatter symmetric. There was only a fractional nano-scale excess of matter which today makes up the visible matter we see around us. All that remains of the tremendous initial amounts of matter-antimatter from the Big Bang is now seen as background thermal entropy. The origin of this nano-matter excess remains to this day an unresolved puzzle. If matter asymmetry emerged along the path of the Universe's evolution, as most think, the previously discussed Sakharov conditions (see \rsec{sec:Timeline}) must be fulfilled.

We explored several major epochs in the Universe evolution where antimatter, in all its diverse forms, played a large roll. Emphasis was placed on understanding the thermal and chemical equilibria arising within the context of the Standard Model of particle physics. We highlighted that primordial quark-gluon plasma (QGP, which existed for $\approx 25\;\mu$sec) is an important antimatter laboratory with its gargantuan antimatter content. Study of the QGP fireballs created in heavy-ion collisions performed today informs our understanding of the early Universe and vice versa~\cite{Borsanyi:2016ksw,Rafelski:2013qeu,Petran:2013lja,Philipsen:2012nu}, even though the primordial quark-gluon plasma under cosmic expansion explores a location in the phase diagram of QCD inaccessible to relativistic collider experiments considering both net baryon density, see \rf{phaseQGP}, and longevity of the plasma. We described (see \rsec{sec:BottomCharm}) that the QGP epoch near to hadronization condition possessed bottom quarks in a non-equilibrium abundance: This novel QGP-Universe feature may be of interest in consideration of the QGP epoch as possible source for baryon asymmetry~\cite{Yang:2020nne}. 

Bottom non-equilibrium is one among a few interesting results presented bridging the temperature gap between QGP hadronization at temperature $T\simeq150\MeV$ and neutrino freeze-out. Specifically we shown {\bf persistence of:}
 \begin{itemize}
 \item Strangeness abundance, present beyond the loss of the antibaryons at $T=38.2\MeV$.
 \item Pions, which are equilibrated via photon production long after the other hadrons disappear; these lightest hadrons are also dominating the Universe baryon abundance down to $T=5.6\MeV$.
 \item Muons, disappearing at around $T=4.2\MeV$, the condition when their decay rate outpaces their production rate.
 \end{itemize}

At yet lower temperatures neutrinos make up the largest energy fraction in the Universe driving the radiation dominated cosmic expansion. Partway through this neutrino dominated Universe, in temperature  range $T\in 3.5-1\MeV$ (range spanning differing flavor freeze-out, chemical equilibria, and even variation in  standard natural constants; see \rf{fig:freezeoutT}), the neutrinos freeze-out and decouple from the rest of the thermally active matter in the Universe. We consider neutrino decoupling condition as a function of elementary constants: If these constants were not all ``constant'' or significantly temperature dependent, a noticeable entropy flow of annihilating $e^{\pm}$ plasma into neutrinos could be present, generating additional so-called neutrino degrees of freedom.

We presented a detailed study of the evolving disappearance of the lightest antimatter, the positrons; we quantify the magnitude of the large positron abundance during and after Big Bang Nucleosynthesis (BBN), see \rf{Density_fig}. In fact the energy density of electron-positron plasma exceeds greatly that of baryonic matter during and following the BBN period with the last positrons vanishing from the Universe near temperature $T=20\keV$, see \rf{ratio_fig}. 

Looking forward, we note that some of the topics we explored deserve a more intense followup work:
\begin{itemize}
 \item The study of matter baryogenesis in the context of bottom quarks chemical non-equilibrium persistence near to QGP hadronization;
 \item The impact of relatively dense $e^{\pm}$ plasma on BBN processes;
 \item Exploration of spatial inhomogeneities in dense $e^{\pm}$ plasma and eventual large scale structure formation and related spontaneous self magnetization process. 
 \item Appearance of a significant positron abundance at $T>25$ keV creates interest in understanding astrophysical object with core temperatures at, and beyond, this super-hot value; the high positron content enables in case of instability a rapid gamma ray formation akin to GRB events. 
\end{itemize}

GRBs are current knowledge frontier:  a tremendous amount of matter~\cite{Aksenov:2010vi} must be converted into gammas in a short time-span of a few seconds. Ruffini and collaborators~\cite{Ruffini:2001fe,Ruffini:2003yt,Ruffini:2009hg,Aksenov:2008zz,Ruffini:2012it,Han:2011er} suggests that strong field production of large amounts of antimatter which can be subsequently annihilated offers the most direct solution. This avoids the problem of excessive photon pressure needing to be balanced in super-hot objects where positron antimatter is already pre-existent. However, GRB events which lack classic  after-signature supernova~\cite{Burns:2023oxn,Levan:2023doz} could originate from novel super-hot stellar objects with primordial Universe properties which naturally possess, rather than create, larger amounts of positrons capable of rapid catalysis of gamma-rays upon gravitational collapse. 

In conclusion: We hope that this work provides to all interested parties a first glimpse at the very interesting epoch of Universe evolution involving in sequence numerous plasma phases made of all particles known today. In this work we provided a background and connection for more specific periods found in the comprehensive literature of observational cosmology~\cite{Davis:1985rj,Navarro:1995iw,Moore:1999nt,Springel:2005nw,Arbey:2021gdg}, the recombination period~\cite{Planck:2018vyg,Planck:2018nkj}, BBN~\cite{Steigman:2007xt,Cyburt:2015mya,Pitrou:2018cgg}, and baryon asymmetry~\cite{Kuzmin:1985mm,Canetti:2012zc,ParticleDataGroup:2022pth} or the origin of dark matter~\cite{Bertone:2004pz,Peccei:2006as,Wantz:2009it}. The Universe above temperatures $T>130 \GeV$ and the inflation era~\cite{Baumann:2009ds,Allahverdi:2020bys} was outside the purview of this work.\\

\noindent
{\bf Author Contributions:} Johann Rafelski: Conceived the article, supervised the development of research results, directed mode of presentation.  Jeremiah Birrell, Andrew Steinmetz, and Cheng Tao Yang contributed their research results in equal weight. MDPI  verbiage:
Conceptualization: JR; software: JB, AS, CTY; validation: all authors; formal analysis: all authors.; investigation: all authors; resources: all authors; data curation, all authors; writing---original draft preparation: JB, AS, CTY; writing---review and editing: JR+AS.; visualization: All authors; supervision: JR; project administration: JR; funding acquisition: NA.
All authors have read and agreed to the published version of the manuscript.\\

\noindent
{\bf Funding:} This research received no external funding.\\

\noindent
{\bf Data Availability Statement:} This is a theoretical review, all results were converted into graphic display at creation and were in this format shared with the reader.
\vfill\eject
\section*{Acknowledgements}
\vspace*{-1.88cm}
\noindent
{\bf \#\#\#\#\#\#\#\#\#\#\#\#\#\#\#\#\#\#\#\#\#\#\#\#\#\#\#\#\#\#\#\#\#\#\#\#\#\#\#\#\#\#\#\#\  https://arxiv.org/abs/2305.09055}
\vspace*{1.4cm}
\begin{figure}[ht]
    \centering
    \includegraphics[width=0.75\textwidth]{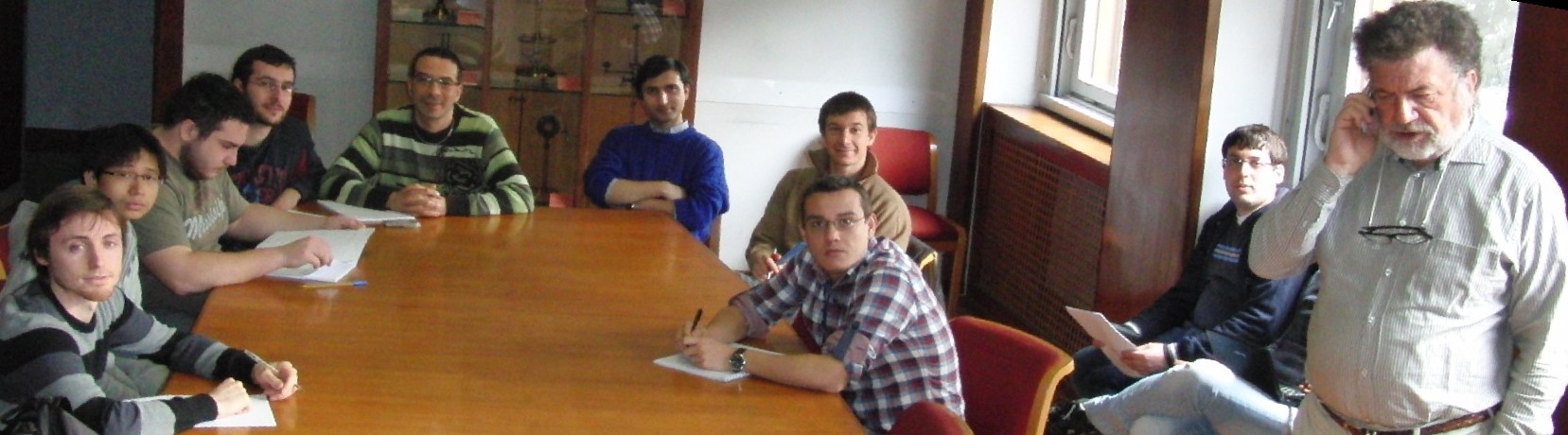}
    \caption{\noindent\href{https://www.icranet.org/}{ICRANet} group at work, Remo Ruffini on right. Photo by Johann Rafelski.\label{RemoAtWork}}
    \vspace{\floatsep}
    \includegraphics[width=0.75\textwidth]{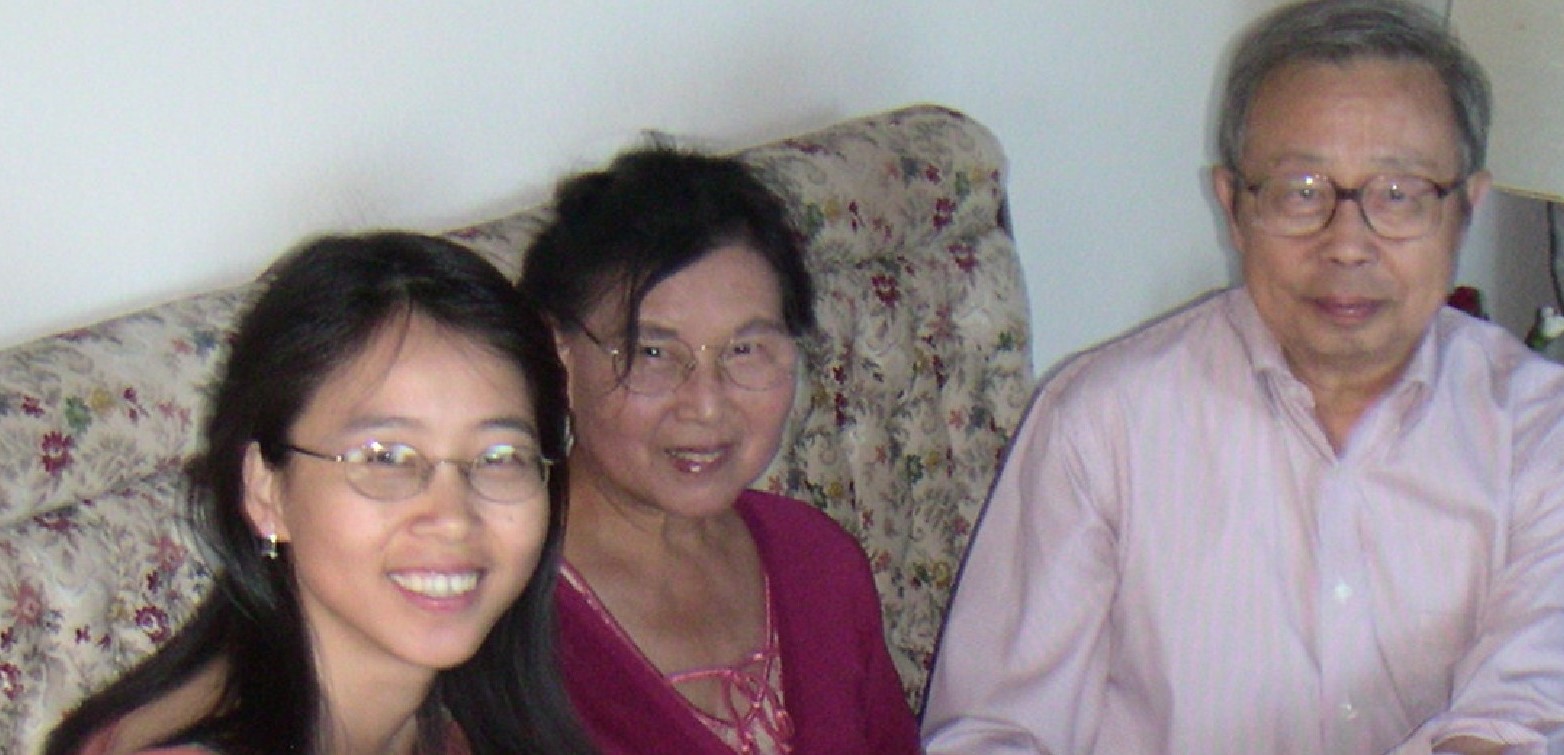}
    \caption{Lizhi Fang (on right), his wife Shuxian Li (center) and Shufang Su (Today: Physics Department Head at the University of Arizona) in April 2004. Photo taken by Johann Rafelski at his home in Tucson.\label{JRhomeFang}}
    \vspace{\floatsep}
    \includegraphics[width=0.75\textwidth]{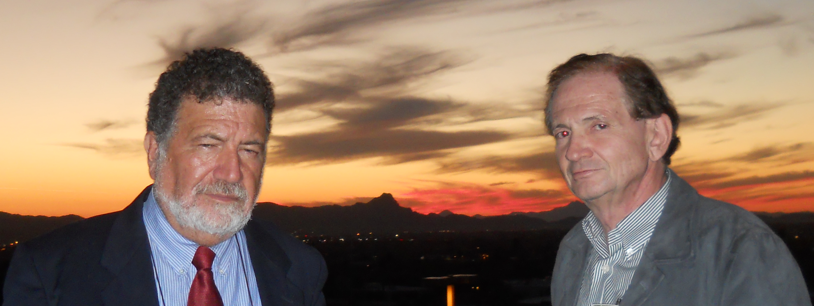}
    \caption{Remo Ruffini (on left) and Johann Rafelski beneath a sunset in Tucson, AZ on October 7th, 2012. The photo was taken by She Sheng Xue at a celebratory gathering honoring the life of Lizhi Fang.\label{remo_sunset}}
\end{figure}

This work was written in celebration of Professor Remo Ruffini's birthday, his contributions to astrophysics and cosmology and the large number of students and young scientists he mentored (see \rf{RemoAtWork}). To close, we also acknowledge our mentor and colleague in the Department of Physics at the University of Arizona, Lizhi Fang~\cite{fang1984cosmology,fang1985galaxies,fang1987quantum} (see \rf{JRhomeFang}) who passed away on April 6, 2012 at his home in Tucson, Arizona. Lizhi introduced Remo Ruffini to us, his career and life is remembered and celebrated (see \rf{remo_sunset}) as we continue to piece together the tapestry of the cosmos.\\

\noindent
{\bf  Conflicts of Interest:} The authors declare no conflict of interest.

\reftitle{References}
\bibliography{23june15refs-mdpi-final.bib}

\end{document}